\documentclass[aps,prb,twocolumn,superscriptaddress]{revtex4-1}

\usepackage{graphicx}
\usepackage{epsfig}
\usepackage{epstopdf}
\usepackage{float}
\usepackage{subfigure}
\usepackage{amsmath}
\usepackage[section]{placeins} 	
\usepackage{mathtools}
\usepackage{needspace}
\usepackage{grffile}
\usepackage{longtable}
\usepackage{adjustbox}
\usepackage{color}

\usepackage{needspace}
\usepackage{multirow}
\usepackage{bm}
\usepackage{url}
\usepackage{siunitx}
\DeclareSIUnit\b{b}
\usepackage{hyperref}

\graphicspath{{Figures/}}

\begin{document}

\title{Achieving DFT accuracy with a machine-learning interatomic potential: thermomechanics and defects in bcc ferromagnetic iron }

\author{Daniele Dragoni}
\affiliation{Theory and Simulation of Materials (THEOS), and National Centre for Computational Design and Discovery of Novel Materials (MARVEL), \'Ecole Polytechnique F\'ed\'erale de Lausanne, 1015 Lausanne, Switzerland}
\affiliation{Dipartimento di Scienza dei Materiali, Universit\`{a} di Milano-Bicocca, Via R. Cozzi 55, I-20125 Milano, Italy}
\author{Thomas D. Daff}
\affiliation{Engineering Laboratory, University of Cambridge, Trumpington Street, Cambridge, CB2 1PZ, United Kingdom}
\author{G\'abor Cs\'anyi}
\affiliation{Engineering Laboratory, University of Cambridge, Trumpington Street, Cambridge, CB2 1PZ, United Kingdom}
\author{Nicola Marzari}
\affiliation{Theory and Simulation of Materials (THEOS), and National Center for Computational Design and Discovery of Novel Materials (MARVEL), \'Ecole Polytechnique F\'ed\'erale de Lausanne, 1015 Lausanne, Switzerland}

\date{\today}

\begin{abstract}
We show that the Gaussian Approximation Potential machine learning framework can describe complex magnetic potential energy surfaces, taking ferromagnetic iron as a paradigmatic challenging case. The training database includes total energies, forces, and stresses obtained from density-functional theory in the generalized-gradient approximation, and comprises approximately 150,000 local atomic environments, ranging from pristine and defected bulk configurations to surfaces and generalized stacking faults with different crystallographic orientations. We find the structural, vibrational and thermodynamic properties of the GAP model to be in excellent agreement with those obtained directly from first-principles electronic-structure calculations. There is good transferability to quantities, such as Peierls energy barriers, which are determined to a large extent by atomic configurations that were not part of the training set. We observe the benefit and the need of using highly converged electronic-structure calculations to sample a target potential energy surface. The end result is a systematically improvable potential that can achieve the same accuracy of density-functional theory calculations, but at a fraction of the computational cost.
\end{abstract}
\pacs{}
\keywords{}

\maketitle

\section{Introduction}
Iron is the most abundant element at the Earth's core, it is responsible for the generation of the geomagnetic field, it is the main component of the most widely used structural engineering material (steel), and in its atomic form it's a component of e.g. oxygen-binding proteins. In its crystalline form it can host impurities that improve its mechanical properties and make it a formidably strong material suitable for many applications in the field of construction, automotive, machinery, and energy production. It is a metal with partially-filled $d$~electronic bands, and has a  complex phase diagram which presents  transformations driven by the interplay of magnetic, electronic and vibrational degrees of freedom. As a consequence, the modeling of iron is highly non-trivial. Density-functional theory (DFT)  provides a relatively good description of its zero temperature properties~\cite{moroni1997ultrasoft,alf1999melting,DalCorsodeGir,laio2000physics,shobhana,pozzo2012thermal}. However, even in this framework, theory shows discrepancies with respect to experimental data\cite{Dragoni}. 
Magnetic fluctuations as temperature approaches and crosses the Curie point~\cite{Koermann,Neugebauer-phonon,phonon-magnon} are crucial to describe the finite-temperature behavior of
thermodynamical quantities of the bulk crystal, although these can be well described up to a good fraction of the Curie point considering only 
vibrational effects~\cite{sha,Dragoni}.
In fact, despite the progress achieved in the past years, a satisfactory description of the thermodynamic phase transitions and of the paramagnetic phases of iron from first-principles remains a formidable task. Even more complex is the study of iron alloys and steels that, on top of the challenges mentioned above, requires in many cases the capability to deal with length and time scales which are beyond 
the reach of any ab-initio technique. 

For this reason, empirical interatomic potentials have been developed, fitted typically to a mixture of experimental and ab-initio data, that are capable of simulating systems containing thousands or millions of atoms for thousands or millions of time steps. These models allowed a detailed study of the microscopic processes at the origin of macroscopic mechanical properties of iron and iron alloys under different conditions. Embedded Atom Models (EAM)~\cite{EAM}, and other similar approaches such as the Finnis-Sinclair model~\cite{FS}, local volume potentials~\cite{Srolovitz} and the glue model~\cite{Glue}, have proved to be successful. In particular, the Mendelev family of parametrizations~\cite{mendelev,ackland2004development,malerba2010} are able to reproduce many fundamental properties of elemental bcc iron at zero temperature. These models however are not always fully satisfactory in reproducing the energetics of defective configurations such as self-interstitials~\cite{NPC-diSIA} and di-vacancies\cite{malerba2010}, the Peierls potentials associated with screw dislocations~\cite{malerba2010,ventelon2007core} or even fundamental bulk properties at finite temperature within the range of stability of the ferromagnetic $\alpha$-phase (below the Curie point)\cite{dragoni2016vibrational}. Additionally, due to their fixed functional form, these potentials are not easily generalizable to the modeling of bonds with mixed metallic and covalent character as can be found for example in Fe-C alloys.

More recently, new approaches such as the modified EAM~\cite{MEAM}, the (analytic) bond order potentials~\cite{BOP_iron,BOP_magnetic,BOP_iron_drautz}, magnetic EAM~\cite{magneticEAM}, or
metallic-covalent interatomic potentials~\cite{metallic-covalent-EAM} have been developed in order to overcome some of these limitations.

In this work we follow an alternative approach, generating a Gaussian Approximation Potential~\cite{Bartok_GAP} (GAP) for the $\alpha$-phase of iron. It is a highly flexible model fit directly and accurately first-principles potential energy surfaces (PES). Transferability is ensured by regular and smooth basis functions (kernels, in the language of machine learning), and an extended training database covering roughly 150,000 local atomic environments (LAEs). GAP is a machine learning framework, and similar approaches, such as neural networks, have been successful recently in modeling materials where previous, more empirical strategies have run out of steam~\cite{BehlerParr,seko2014sparse,artrith2016,Ramprasad2016,Kazutoshi,Oganov2016,shapeev,Goedecker,DeVita}. GAP uses Gaussian process regression~\cite{mackay,GPbook}, whose advantages are that (i)~its hyper-parameters (that control the kernel function and linear algebra regularisation) make physical sense and rarely need adjusting, (ii)~the fit itself is determined by simple linear algebra, rather than iterative nonlinear optimisation of a highly multimodal function as in the case of neural networks, (iii)~input data such as energies, forces and stresses are treated in a consistent manner, with appropriate error estimates that allow the inclusion of variable accuracy data. In the machine learning literature Gaussian process regression is often thought of as scaling badly (cubically) with the size of the input data, but we find that well known heuristics allow us to limit the number of basis functions to be much smaller than the number of input configurations, leading to training times of about a day on a single multi-core server and to prediction costs similar to that of neural network-based potentials. The key to the success of Gaussian process regression is an appropriate kernel function that captures the symmetries and describes the spatial correlation structure of the target function. We use the ``smooth overlap of atomic positions'' (SOAP) kernel~\cite{Bartok_envir} that has been shown previously to lead to excellent results for other materials\cite{Wojciech,Deringer_aC,PCCP-SOAP,Kermode}.

\section{method}
\label{sec:methods}
We start by assuming that the Born-Oppenheimer potential energy surface of a set of atoms is a smooth function of the atomic coordinates. As it is usually done when constructing interatomic potentials, we write the total energy as a sum of atomic contributions
\begin{equation}
E=\sum_i \epsilon (\boldsymbol{q}_i),
\end{equation}
where the short-ranged local atomic energy $\epsilon_i$ is assumed to depend explicitly on the positions of the atoms within a sphere of radius $r_{cut}$ centered on atom $i$.
The list of such atomic positions defines the local atomic environment of atom $i$ and is  represented by a suitable set of descriptors, here denoted by the vector $\boldsymbol{q}_i$. 
(Standard terms representing electrostatic and van der Waals interactions can be added as needed to account for long-range interactions.) Empirical interatomic potentials are designed using functional forms derived from physical intuition to approximate
$\epsilon (\boldsymbol{q}_i)$, and parameters are fitted to experimental or computational data. The moderate flexibility of these functional forms limits their scope to be systematically 
improved by increasing the fitting datasets; on the other hand, their qualitative description of the essential physical interactions ensures a good degree of transferability. 
In the GAP framework, Gaussian process regression is used instead to define a model for the local atomic energy function $\epsilon$ as a linear combination of non-linear kernel functions 
\begin{equation}
\label{eq:prediction}
\epsilon(\boldsymbol{q}^*)=\sum_s  \alpha_s K(\boldsymbol{q}_s,\boldsymbol{q}^*) \equiv \mathbf{K}(\boldsymbol{q}^*){^\mathrm{T}}\boldsymbol{\alpha}  ,
\end{equation}
where the sum runs over some representative subset $s$ of training configurations, usually far fewer than the total training set. 
The kernel function $K(\boldsymbol{q}_i,\boldsymbol{q}_j)$ of two local atomic environments, represented by their sets of descriptors $\boldsymbol{q}_i$ and $\boldsymbol{q}_j$, corresponds to the expected covariance of their respective local atomic energies $\epsilon(\boldsymbol{q}_i)$ and $\epsilon(\boldsymbol{q}_j)$, and can be interpreted  as a 
measure of similarity of the two local atomic environments. In the present work we use the ``smooth overlap of atomic positions'' (SOAP) kernel developed by Bart{\'o}k et al.~\cite{Bartok_envir} 
which is equivalent to choosing a polynomial kernel function 
\begin{equation}
\label{eq:SOAP}
K(\boldsymbol{q}_i,\boldsymbol{q}_j)=\sigma^2_w |\boldsymbol{\hat{q}}_i \cdot \boldsymbol{\hat{q}}_j|^{\xi},
\end{equation}
where the descriptor $\boldsymbol{\hat{q}}$ is the rotational power spectrum of the local atomic environment, which is a smooth and regular function, invariant to rotation and permutation of like atoms. All hyperparameters, including those inherent in the definition of the rotational power spectrum are shown in Table~\ref{tab:hyperparameters} and their role is extensively discussed in Ref.~\onlinecite{Wojciech}. The physically motivated hyperparameters include the energy scale $\sigma_w$  which roughly corresponds to the expected standard deviation of the the atomic energy, and the length scale $\sigma_\mathrm{atom}$ which controls the regularity of the potential. The power spectrum of the local environment includes a cutoff function that is zero for $r > r_\mathrm{cut}$, a parameter whose choice is governed by the decay of the force constant matrix, since the potential will give exactly zero force constants for $r > 2r_\mathrm{cut}$ by construction. 

The vector of coefficients $\bm{\alpha}$ is obtained by substituting the training data into Eq.~\ref{eq:prediction} and solving the linear system. We briefly outline the necessary steps, see Refs.~\cite{Wojciech,gap-tutorial} for further detail. Since the decomposition into atomic energies is not available from electronic structure calculations, the training data comprises total energies, and its derivatives (forces, and virial stresses) corresponding to collections of atoms. Let us define 
$\boldsymbol{y}$ as the vector with $D$ components containing the target data: all total energies, forces and virial stress components in the training database,  and $\boldsymbol{y}'$ as the vector with $N$ components containing the \textit{unknown}  atomic energies of the $N$ atomic environments in the database,  and  $\mathbf{L}$ as the linear differential operator of size $N\times D$  which connects $\boldsymbol{y}$ with $\boldsymbol{y}'$ such that $\boldsymbol{y}=\mathbf{L}{^\mathrm{T}}\boldsymbol{y}'$. After selecting $M$ representative atomic environments (with $M \ll N$), the expression for the coefficients in Eq.~\ref{eq:prediction} is given by\cite{sparseGP}
\begin{equation}
\label{eq:alphas}
    \boldsymbol{\alpha} = {\big[ \mathbf{K}_{MM} + \mathbf{K}_{MN} \mathbf{L} \mathbf{\Lambda}^{-1} \mathbf{L}{^\mathrm{T}} \mathbf{K}_{NM}  \big]}^{-1} \mathbf{K}_{MN} \mathbf{L} \mathbf{\Lambda}^{-1} \boldsymbol{y}   \,  ,
\end{equation}
where $K_{MM}$ is the covariance matrix between the $M$ representative atomic environments, and $K_{MN}$ is the covariance between matrix them and all $N$ environments in the training data.  (In the Gaussian process literature, using a subset of the data to construct the basis is called \emph{sparsification}.) While taking $\mathbf{\Lambda}=\sigma_{\nu}^2 \mathbf{I}$ as the \emph{regularization} matrix would be sufficient to solve the linear system (corresponding to simple $L_2$ regularisation), the probabilistic interpretation of Gaussian process regression suggests that the elements of  $\mathbf{\Lambda}$ are the tolerances, or expected errors in components of the training data vector $\boldsymbol{y}$, with even different units for different types of input data. Note that the expected errors are not just due to lack of numerical convergence in the electronic structure calculations, but also include the \emph{model error} of the GAP representation, e.g.\ due to the finite cutoff of the local environment. Our informed choices for these parameters are reported in Table~\ref{tab:hyperparameters}. The representative local environments are chosen by the CUR matrix decomposition procedure~\cite{CUR} applied to the matrix of descriptor vectors in the input dataset which essentially finds a subset of the atomic environments that would lead to a good low-rank approximation of the full covariance matrix.  The upshot of using only a small number of representative atomic environments is that the computational cost to train the model scales as $O(NM^2)$ rather than $O(N^3)$, and the cost of evaluating a single local atomic energy scales as $O(M)$ rather than $O(N)$. Typically we find that $M < 10\,000$ is sufficient (in the sense that prediction results do not improve when a larger $M$ is used) even when $N > 150\,000$. 

We trained the GAP model using the QUIP software code, which is publically available\cite{github-quip}, and the full set of command line parameters as follows,
\begin{verbatim}
at_file=data.xyz gap={soap l_max=12 n_max=12
cutoff=5.0 cutoff_transition_width=1.0 delta=1.0
atom_sigma=0.5 zeta=4 config_type_n_sparse=
{slice_sample_high:500:phonons_54_high:500:
phonons_128_high:500:default:3000}
sparse_method=cur_points
covariance_type=dot_product}
sparse_jitter=1e-12 default_sigma={0.005 0.2
1.0 0.0} config_type_sigma={slice_sample_high:
0.0001:0.01:0.01:0.0:phonons_54_high:0.001:
0.05:1.0:0.0:phonons_128_high:0.001:0.05:
1.0:0.0}
\end{verbatim}

\begin{table}[h]
\centering
\setlength{\tabcolsep}{12pt}
\renewcommand{\arraystretch}{1.4}
\begin{tabular}{@{}l l @{}}
\toprule
Atomic environment kernel & SOAP   \\
$r_{cut}$                 &  5.0 \AA         \\ 
$r_{\Delta}$              &  1.0 \AA         \\
\colrule
${\sigma_{\nu}^\mathrm{energy}}_\mathrm{DB1}$       &   $\num{1.0 e-04}$ eV/atom    \\
${\sigma_{\nu}^\mathrm{energy}}_\mathrm{DB2}$       &   $\num{1.0 E-03}$ eV/atom     \\
${\sigma_{\nu}^\mathrm{energy}}_\mathrm{default}$   &   $\num{5.0 E-03}$ eV/atom     \\
${\sigma_{\nu}^\mathrm{force}}_\mathrm{DB1}$        &   $\num{1.0 E-02}$ eV/\AA        \\
${\sigma_{\nu}^\mathrm{force}}_\mathrm{DB2}$        &   $\num{5.0 E-02}$ eV/\AA        \\ 
${\sigma_{\nu}^\mathrm{force}}_\mathrm{default}$    &   $\num{2.0 E-01}$ eV/\AA        \\ 
$\sigma_{\nu}^\mathrm{virial}$               &   $\num{1.0 E-02}$ eV/atom \\ 
\colrule
$\sigma_{w}$              & 1.0  eV          \\ 
$\sigma_\mathrm{atom}$           & 0.5  \AA         \\ 
$\xi$                     &  4               \\
$n_{\max}$                 & 12               \\
$l_{\max}$                 & 12               \\
GAP software version      &\textsf{1469201250}\\
\colrule
Represenative environments             & 4500             \\
sparse method             & CUR   \\
\botrule
\end{tabular}
\caption{Hyper-parameters for the SOAP kernel and the GAP model.}
\label{tab:hyperparameters}
\end{table}

\section{Database}
\label{sec:DB}
A large training database of electronic structure calculations is
required in order to ensure transferability of flexible GAP models to
a wide range of atomic environments.  In what follows we discuss the
details of how we generated such database.

\subsection{Generation protocol}
We choose to include in the database only first-principles
data. Although computationally costly, this approach allows for a
direct control and propagation of the accuracy and the degree of
convergence of the data entering the training procedure.  The database
generation protocol that we adopt can be rationalized as follows.  (1)
We start by selecting the physical properties that we require to be
well reproduced or predicted by our model. For each material property
of interest, we select a number of representative \emph{small} periodic
configurations (with varying cell parameters and atomic positions)
that are amenable for first-principles calculations and covers the
relevant local atomic environments needed for the potential to
reproduce that property.  (2) We sample the configurational space
associated to each unit cell selected in (1) by means of Monte Carlo
or molecular dynamics techniques using density functional theory
calculations that are configured to have only a moderate level of 
convergence.  (3) From each sampling run, we extract
a weakly correlated subset of
configurations.  Each of these subdatabases is
denoted as DB$x$.  (4)
Finally, we recompute total energies, forces and stresses for
each configuration in each subdatabase using highly
converged parameters in order to minimize the stochastic and
systematic errors due to the finite $k$-point sampling and plane wave
cutoff. Even so, it is not possible to use the same (consistent) $k$-point sampling 
across the entire database due to resource limitations, and the resulting
varying degrees of convergence are used to inform the magnitude 
of the regularisation terms corresponding to each subdatabase, as shown 
in Table~\ref{tab:hyperparameters}.

\subsection{Training configurations}
\begin{figure}[H]
\centering
  \includegraphics[trim=0mm 0mm 0mm 0mm, clip, width=0.2\textwidth]{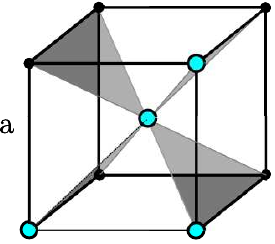} 
  \includegraphics[trim=0mm 0mm 0mm 0mm, clip, width=0.324\textwidth]{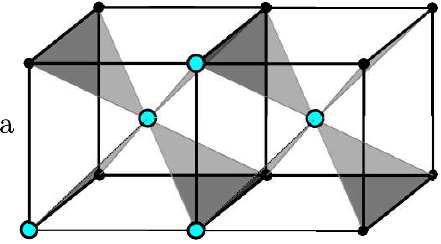} 
  \includegraphics[trim=0mm 0mm 0mm 0mm, clip, width=0.2\textwidth]{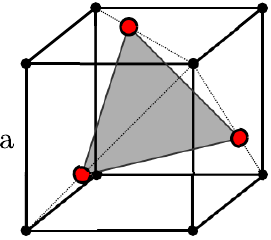} 
  \caption{From top to bottom we show the schematics of tetra-, penta-vacancy and non-parallel di-interstitials  of DB5/DB6. Azure circles represent the missing atoms in an otherwise perfect bcc
  structure. Red circles schematically represent the atomic arrangement of a non-parallel di-interstitial defect.}
  \label{fig:tetrapentaNPCschematics}
\end{figure}

The complete database consists of 8 subdatabases which include 
12193 configurations, equivalent approximately to $1.5 \times 10^5$ atomic environments. The details of each subdatbase are described below (see also Ref.~\onlinecite{dragoni2016energetics}) and also summarized in Tab.~\ref{tab:database_summary} for simplicity. \\
%\item
\textbf{DB1} aims at training around the bcc equilibrium geometry and the elastic response of the bulk. It consists of energies and stresses computed for one-atom cells whose vectors are distorted with respect to the equilibrium bcc primitive cell geometry. The distortions are randomly obtained using a slice-sampling MC algorithm and performed with respect to various reference volumes which are compressed or expanded with respect to the 0~K DFT equilibrium value as reported in Tab.~\ref{tab:database_summary}. \\
%\item 
\textbf{DB2} is used to teach bulk vibrational properties and consists of total energies and forces computed from $3\times3\times3$ and $4\times4\times4$ conventional cubic supercells containing 54 and 128 atoms respectively. The configurations are extracted from MD runs equilibrated at the volumes and temperatures shown in Tab.~\ref{tab:database_summary}.  \\
%\item 
\textbf{DB3} Similarly to DB2 it consists of total energies and forces computed from $3\times3\times3$ cubic supercells generated from MD runs also equilibrated at various volumes and temperatures reported in Tab.~\ref{tab:database_summary}. This subdatabase is used to teach bulk mono-vacancy energetics. As such, the unit cells contain 53 atoms.\\
%\item 
\textbf{DB4} provides information on the di-vacancy energetics. Di-vacancy environments up to third-nearest neighbor are explicitly included. This subdatabase consists of total energies and forces of $4\times4\times4$ conventional cubic supercells containing 126 atoms and obtained from MD equilibrated at 800~K and at the equilibrium volume. \\
%\item 
\textbf{DB5} embodies selected tri-vacancies and small vacancy clusters such as tetra-vacancies and penta-vacancies (see Fig.~\ref{fig:tetrapentaNPCschematics}) that should provide a starting point for describing nano-voids.
We choose those tri-vacancy configurations which lie in low Miller index crystallographic planes \{100\},  \{110\}, and  \{111\} and that, in those planes, are most localized~\footnote{The simple measure 
of localization that we use here is the length of the perimeter connecting the vacancy sites of the perfect bi-dimensional lattice associated to a crystallographic plane.}. 
Total energies and forces from $4\times4\times4$ cubic supercell configurations obtained from MD are used as training quantities.\\
%\item 
\textbf{DB6} embraces relevant self-interstitials environments, including the $\langle 100\rangle$/$\langle 110\rangle$ dumbbell, $\langle 111\rangle$ crowdion, and the tetrahedral and octahedral configurations. 
The configurational space of these point defects is sampled by means of MD performed on cubic $4\times4\times4$ supercells containing 129 atoms at the theoretical equilibrium bulk volume at 0~K. Training is from total energies and forces.
A type of non-parallel di-interstitial configuration (see Fig.~\ref{fig:tetrapentaNPCschematics}) is also considered to cover further defective environments beyond simple self-interstitials. 
All defects of DB6 are sampled with MD performed on $4\times4\times4$ supercells and 140 atoms. Training is from total energies and forces. \\
%\item 

\textbf{DB7} consists of total energies and forces of bulk-terminated surface configurations with \{100\}, \{110\}, \{111\}, and \{211\} crystallographic orientations. 
For this subdatabase we choose supercells which are elongated along $c$ and primitive in the surface plane (at the equilibrium lattice parameter). 
We simulate slabs which are 12 atomic layers thick to minimize interactions between the two surfaces of the slabs. A vacuum separation region of 16~\AA\ is also used to avoid replica interactions in the 
$c$ direction. Molecular dynamics is performed on these cells with the atoms allowed to move only along $z$ in order to gain insight mainly on the out-of-plane surface relaxation of the atoms at the vacuum-slab interface. \\
%\item 
\textbf{DB8} is generated to train on $\gamma$-surfaces, to be able to ensure coverage of local environments found typically around dislocation cores. In particular we consider \{110\} and \{211\} crystallographic orientations which are the most important slip planes for bcc metals. As with the bulk-terminated surfaces, we use supercells elongated along $c$ which contain 12 atomic layers. Configurations are created in a 10$\times$10 grid of slips in directions in the glide plane of the gamma surface. Total energies and forces are used as training quantities. 
%\end{itemize}

\subsection{Computational details}
\label{sec:DBdetails}

MD simulations are all performed in a NVT ensemble with time steps ranging from 2 to 4~fs and a Berendsen thermostat~\cite{berendsen}. Temperatures and volumes are varied as specified in the section above (details are reported in Table~\ref{tab:database_summary}). This is done to ensure a good coverage of the physical properties of interest across the theoretical thermodynamic range of stability of the $\alpha$-phase of iron~\cite{Dragoni}. 
Sampling of self-interstitial defects requires some attention since most of them are metastable states which tend to rapidly relax to more stable configurations during the MD\@. In those cases, we perform very short MD runs at low temperatures trying to capture the transition pathway to lower energy states.

Monte Carlo sampling was originally performed in Ref.~\onlinecite{Wojciech} for bcc tungsten at 300~K and exploiting a slice-sampling technique. Here we simply take the same periodic cells and rescale them to account for the differences in the lattice parameter and elastic constants of ferromagnetic iron and tungsten.

All quantum mechanical calculations of this work are performed in a collinear spin-polarized plane-waves DFT framework as implemented in the \textsf{Quantum ESPRESSO} distribution~\cite{QE}, employing an ultrasoft GGA PBE~\cite{PBE} pseudopotential from the \textsl{0.2.1~pslibrary}~\footnote{\protect\url{http://www.qe-forge.org/gf/project/pslibrary}} with semicore electrons in valence. This pseudopotential has been carefully tested and has proved to be able to reproduce the correct all-electrons behavior for a number of ground sate properties. The exchange-correlation functional used  provides a relatively good description of thermomechanical properties for the $\alpha$-phase\cite{Dragoni}. 
At the same time, it is a reliable choice for reproducing point defects properties~\cite{Dudarev_Derlet}. 

For the accurate calculations mentioned at step (4) of the generation protocol, all input parameters are chosen to ensure convergence to 1~meV/at, 0.01~eV/\AA\, and 0.01~GPa for the energy difference, forces and stresses respectively. 
In particular, a value of 90~Ry on the wavefunction (dual of 12~\footnote{ratio between cutoff on the density and cutoff on the wavefunction for the ultrasoft pseudopotential}) is required for the convergence of energy differences and forces. 
The convergence of stresses instead requires a cutoff value of 144~Ry (dual of 12). It is important to stress that none of these values is however sufficient to ensure proper convergence of the total energy. As a consequence,  
in order to avoid inconsistencies (that would affect the training procedure) between total energies of DB1 and those of DB$x$ with $x > 1$,
we have built DB1 as a combination of  stresses computed at 144~Ry and of total energies computed at 90~Ry.
The BZ is integrated with a Monkhorst-Pack grid and a Marzari-Vanderbilt smearing scheme~\cite{marzarismearing} at an effective temperature of 0.01~Ry. In order to ensure the level of convergence mentioned above, we found critical to choose a sampling density so that for all the subdatabases cells the largest $k$-spacing along any reciprocal cell vector is below 0.03~\AA$^{-1}$. Exact $k$-spacing values for each DB are reported in Tab.~\ref{tab:database_summary} for completeness. Note in the table that slightly different $k$-points densities are used for some of the data. This is due to computational costs (for primitive unit cells one can afford higher densities than for large supercells) but also due to the incommensurate nature of the simulation boxes used in the different DBs. The $k$-spacing has a typical value of 0.025~\AA$^{-1}$ with a standard deviation of 0.005~\AA$^{-1}$.
The \textsf{Aiida materials informatics} infrastructure~\cite{Aiida} has been partially used as a tool to automate submission of accurate calculations of the generation protocol, and to provide easy access to provenance information of all the data of the training database.

For the calculations at step (2) of the generation protocol, i.e.\ those related to the sampling of the quantum mechanical PES, we do not require such level of accuracy. In fact, in this case we use lower cutoff values of 60~Ry (dual 8) with a reduced $k$-sampling of the Brillouin zone.

\onecolumngrid
%\begin{turnpage}
\begin{table*}[h!]
\centering
\setlength{\tabcolsep}{5pt}
\begin{tabular}{@{}l c c c c c c c c c @{}}
\toprule
            &  Target      &  Total \#  & V [\% $V^{el}$] & T [K] & \# atoms in& Simulation  & $k$-spacing & notes \\
            &  property   &  of   atoms  &                       &       &    unit cell           & box              & [\AA$^{-1}$] &         \\
\colrule
%%%%%%
 \multirow{2}{*}{\textbf{DB1}} &  bulk elastic       & \multirow{2}{*}{6001} & \multirow{2}{*}{-0.81/1.08/3.55} & \multirow{2}{*}{300} & \multirow{2}{*}{1} & primitive bcc & \multirow{2}{*}{0.015} &  \multirow{2}{*}{-}   \\
                               &  constants                    &                      &                                  &                      &                    & distorted     &                                              &                       \\
 \colrule
 %%%%%%
 \multirow{2}{*}{\textbf{DB2}} &\multirow{2}{*}{bulk phonons}  &  12474      & \multirow{2}{*}{0.0/$\pm$2.08/3.55} & 400--1400 & 54  & $3\times3\times3$  &\multirow{2}{*}{0.03} &   \multirow{2}{*}{-} \\
                               &                                     & 11520      &                                     & 800               & 128 & $4\times4\times4$      &     			                 &                        \\
 \colrule
 %%%%%%
 \multirow{2}{*}{\textbf{DB3}} &  bulk               &  \multirow{2}{*}{20193}  &  \multirow{2}{*}{0.0/$\pm$2.08}  & \multirow{2}{*}{400--1000} & \multirow{2}{*}{53}  & \multirow{2}{*}{$3\times3\times3$} & \multirow{2}{*}{0.03}  &  \multirow{2}{*}{-}  \\
                               &  mono-vacancies                          &        &                    &                              &  &  &                                               &                      \\
 \colrule
 %%%%%%
 \multirow{2}{*}{\textbf{DB4}} &  bulk               &  \multirow{2}{*}{10836}& \multirow{2}{*}{0.0}  &\multirow{2}{*}{800}             & \multirow{2}{*}{126} & \multirow{2}{*}{$4\times4\times4$} & \multirow{2}{*}{0.03} & \multirow{2}{*}{1-, 2-, 3-$nn$}  \\
                               &  di-vacancies       &                                &            &                       &                                 &     &                   &        &                       \\
 \colrule
 %%%%%%
\multirow{4}{*}{\textbf{DB5}}  & \multirow{2}{*}{tri-vacancies}      &   \multirow{2}{*}{9375}     & \multirow{2}{*}{0.0} & \multirow{2}{*}{800}         & \multirow{2}{*}{125} & \multirow{2}{*}{$4\times4\times4$} &  \multirow{2}{*}{0.03}     & \tiny{[112],[113]},         \\ 
                               &			 	     &            &     &             &     &                   &           & \tiny{[223],[333],[339]}    \\ \cline{2-10}
                               & \multirow{2}{*}{vacancy clusters}   & 1736       & 0.0 & 800--1000   & 124 & $4\times4\times4$ &  0.03     & 4 vac.               \\ 
                               &                                          & 1476       & 0.0 & 600        & 123 & $4\times4\times4$ &  0.03     & 5 vac.               \\                                
 \colrule
 %%%%%%
                               &                               & 2709       &  0.0                  &  100                    & 129                      & $4\times4\times4$& 0.03& dumbbell$_{\langle 100 \rangle}$   \\
                               &                               & 1548       &  0.0                  &  300                    & 129                      & $4\times4\times4$& 0.03& dumbbell$_{\langle 110 \rangle}$ \\
\multirow{2}{*}{\textbf{DB6}}  & self-interstitials  &       4773       &  0.0                  &  100--300                & 129                      & $4\times4\times4$& 0.03& crowdion$_{\langle 111 \rangle}$ \\
                               &                              & 3225       &  0.0                  &  100--300                & 129                      & $4\times4\times4$& 0.03& tetrahedral   \\
                               &                           & 2064       &  0.0                  &  100                    & 129                      & $4\times4\times4$& 0.03& octahedral    \\ \cline{2-10}
                               &  di-interstitials         & 2340       &0.0                    & 300                     & 130                      & $4\times4\times4$& 0.03& non-parallel    \\
 \colrule
 %%%%%%
                               &  bulk                          & 660        & 0.0   & 300  & 12 &  $1\times1\times 6 $ &0.03   & (100)   \\
 \multirow{2}{*}{\textbf{DB7}} &  terminated         &  588        & 0.0   & 300  & 12 &   $1\times1\times 6 $ &0.025     & (110)   \\
                               &  surfaces           & 516        & 0.0   & 300  & 12 &   $1\times1\times 6 $ & 0.04    & (111)   \\
                               &                          & 648        & 0.0   & 300  & 12 &   $1\times1\times 12 $& 0.025     & (211)   \\
 \colrule
 %%%%%%
 \multirow{2}{*}{\textbf{DB8}} & \multirow{2}{*}{$\gamma$ surfaces} &  30000 & 0.0 & 300  & 12  & primitive $xy$ & 0.03   & (110)    \\
                               &                                       & 29388 & 0.0 & 300  & 12  & primitive $xy$ & 0.025   & (211)    \\
 \colrule        
 %%%%%%                      
 \textbf{DB}                   &                                     & \textbf{152070} &                  &                         &                          &        &     &       \\ 
\botrule
\end{tabular}
\caption{Database details used for training the $\alpha$-Fe GAP\@. For each sub database (DB) we report the name of the physical properties focus of the training, the physical quantities explicitly used for training, the number of training local environments, the volume (expressed in percentage variation with respect to the electronic equilibrium value), the temperature, the number of atoms, the simulation box used for the generation of the configurations, the $k$-spacing used for the accurate calculations and other details concerning the type of environments within the DB\@. The 
notation for tri-vacancy identification is taken from Ref.~\onlinecite{beelerVAC}.}
\label{tab:database_summary}
\end{table*}
%\end{turnpage}
\twocolumngrid

\section{Results}
\label{sec:results}
In this section we present the GAP model for bcc iron which has been trained on the database of Table~\ref{tab:database_summary} with generation details reported in Table~\ref{tab:hyperparameters}. Validation is performed in the following sections through an analysis of the energetics and of the  thermomechanic properties of the $\alpha$-phase by comparing with DFT data; comparisons with experiments (when possible) are also reported. DFT calculations which are used for comparison are either taken from the literature or computed in this work with input parameters consistent with those described in Sec.~\ref{sec:DBdetails}. 
The latter are considered part of a testing set and are not used for training.

\subsection{Fundamentals}
\begin{figure}[H]
\centering
  \includegraphics[trim=0mm 0mm 0mm 0mm, clip, width=0.45\textwidth]{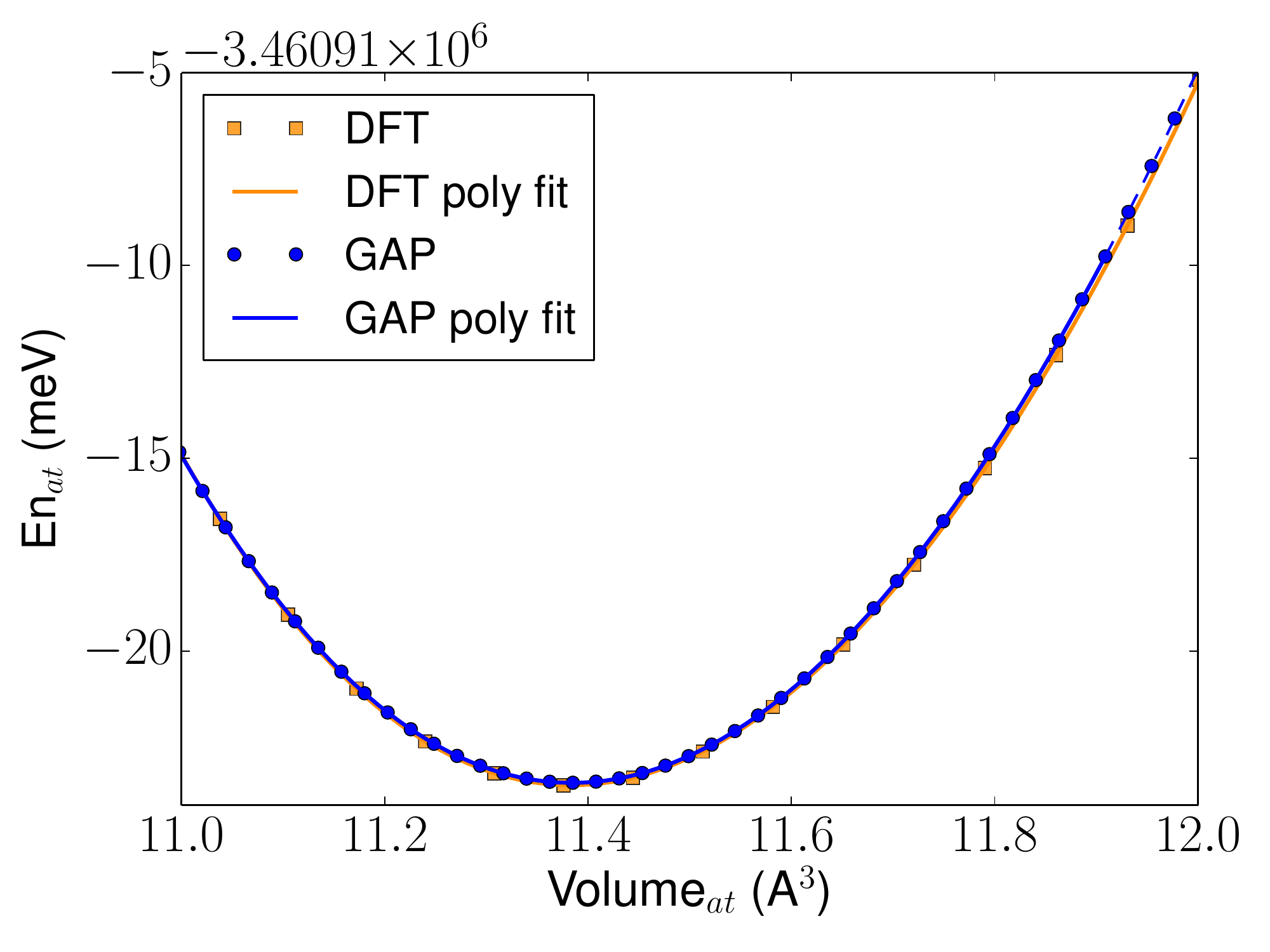} 
  \begin{picture}(0,0)
  \put(-169,49.8){\includegraphics[trim=0mm 0mm 0mm 0mm,height=2.0cm]{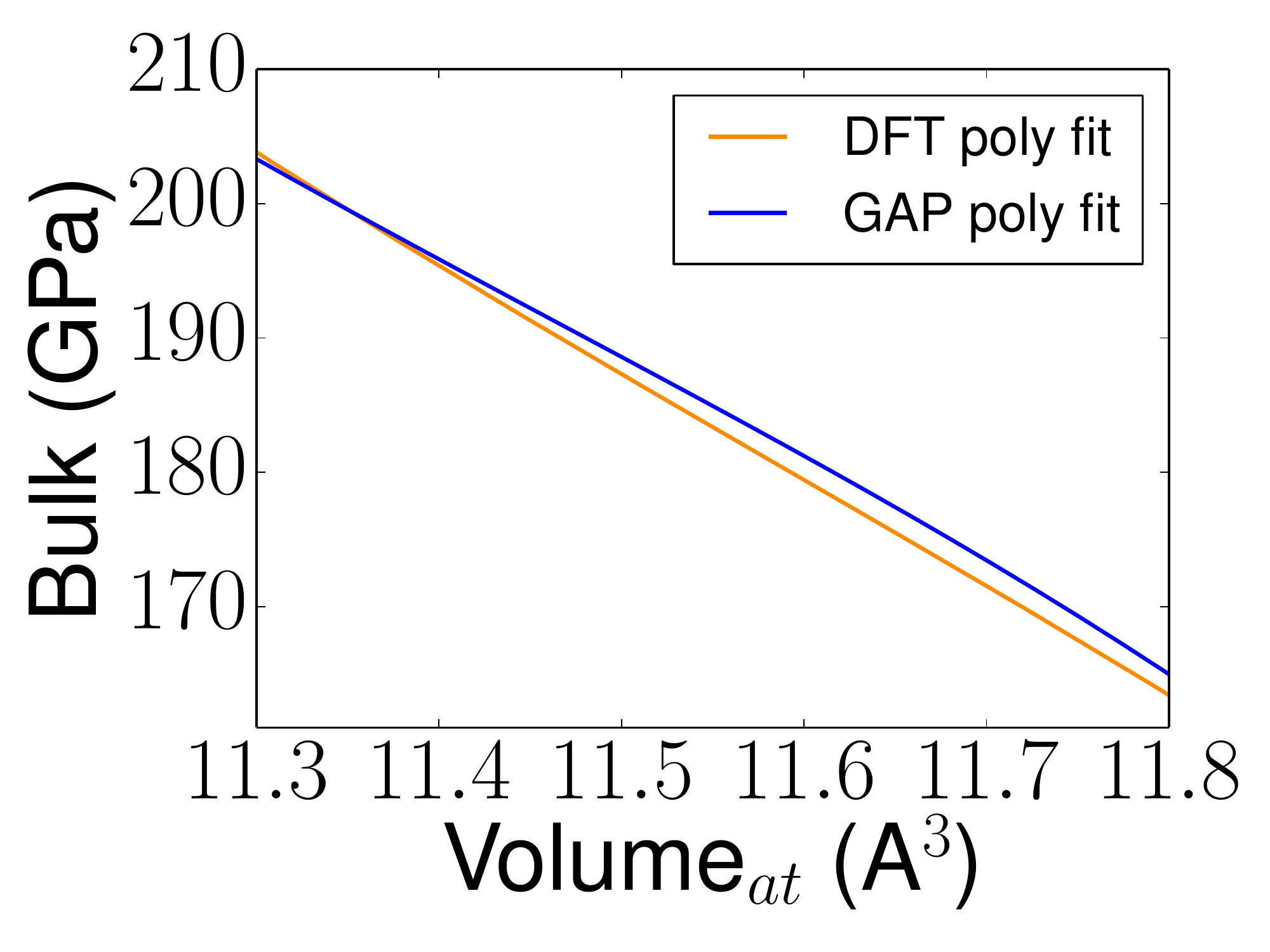}}
  \end{picture}
  \caption{Equation of state of the GAP potential (blue, circles) compared to DFT data (orange, squares) whose calculation details are consistent with those used for the database generation.  In the inset are reported the GAP and DFT bulk moduli obtained analytically from a polynomial fitting of the corresponding total energy curves.}
  \label{fig:EOS_gp}
\end{figure}

We start our analysis showing in Table~\ref{tab:C0K_gp} the lattice parameter $a_0$, the bulk modulus $B_0$, and the elastic constants $C_{11}$, $C_{12}$, $C_{44}$ calculated with GAP at zero temperature (with and without zero-point contributions). Results are in excellent agreement with the quantum mechanical data and, as previously discussed in Ref.~\onlinecite{Dragoni}, reflect the inherent limitations of standard DFT approaches to deal with magnetism~\footnote{A recent work based on a DFT+Gutzwiller approach (see Ref.~\onlinecite{ironGutzwiller}) has shown interesting improvements over standard DFT with respect to the agreement between experiments and theory for the description of the mechanical properties of $\alpha$-iron. One might therefore try to use such approach to generate a database for GAP models with predictions closer to experiments. At the moment, 
however, its computational cost remains an important limitation.}. The equation of state (EOS) reported in Fig.~\ref{fig:EOS_gp} shows how close GAP is with respect to the DFT curve even relatively far from the equilibrium volume. The maximum energy difference between the two EOS curves in the volume interval [11.0:12.0]~\AA$^3$ \ around the electronic equilibrium is $\approx$~0.3~meV~\footnote{This value is well within the DFT uncertainty and highlights the capabilities of our training procedure.}, with a measure $\Delta$ of the distance between the two curves calculated \textit{\`{a} la} Cottenier~\cite{Cottenier} of $0.112$~meV/atom. In the inset of Fig.~\ref{fig:EOS_gp} we also report for reference the GAP and DFT electronic bulk moduli $B(V)=V \frac{\partial^2 E(V)}{\partial V^2}$.

%\begin{adjustbox}{angle=90}
\begin{table}[h]
\centering
\begin{tabular}{@{}l  c c c c c @{}}
\toprule
                 &\multicolumn{2}{c}{GAP}&\multicolumn{2}{c}{DFT}                                    & Expt.                           \\
                 & no ZPE  & ZPE      & no ZPE                         & ZPE                         &                                 \\
\colrule
$a_0$ [\AA]      &   2.834 &  2.839   &  2.834~\cite{Dragoni}          &  2.839~\cite{Dragoni}       &  2.855~\cite{Basinski}    \\
$B_0$ [GPa]      &   198.2 &  191.7   &  199.8$\pm$0.1~\cite{Dragoni}  &194.6$\pm$0.3~\cite{Dragoni} &  170.3$\pm$1~\cite{Adams}       \\
                 &         &          &  196.9~\footnote{This work}    &                             &                                 \\
$C_{11}$         &   285.9 &     -    &  296.7$\pm$0.3~\cite{Dragoni}  &287.9$\pm$0.4~\cite{Dragoni} &  239.5$\pm$1~\cite{Adams}       \\
$C_{12}$         &   154.3 &     -    &  151.4$\pm$0.2~\cite{Dragoni}  &148.0$\pm$0.5~\cite{Dragoni} &  135.7~\cite{Adams}             \\
$C_{44}$         &   103.8 &     -    &  104.7$\pm$0.1~\cite{Dragoni}  &102.2$\pm$0.5~\cite{Dragoni} &  120.7$\pm$0.1~\cite{Adams}     \\
\botrule
\end{tabular}
\caption{Lattice parameter, bulk modulus, and and elastic constants for $\alpha$-iron at zero temperature. GAP results are compared to DFT (with and without quasi-harmonic zero-point energy contributions) and to experimental data  at 0~K. }
\label{tab:C0K_gp}
\end{table}
%\end{adjustbox}

\subsection{Bain path}

\begin{figure}
\centering
  \includegraphics[trim=0mm 0mm 0mm 0mm, clip, width=0.45\textwidth]{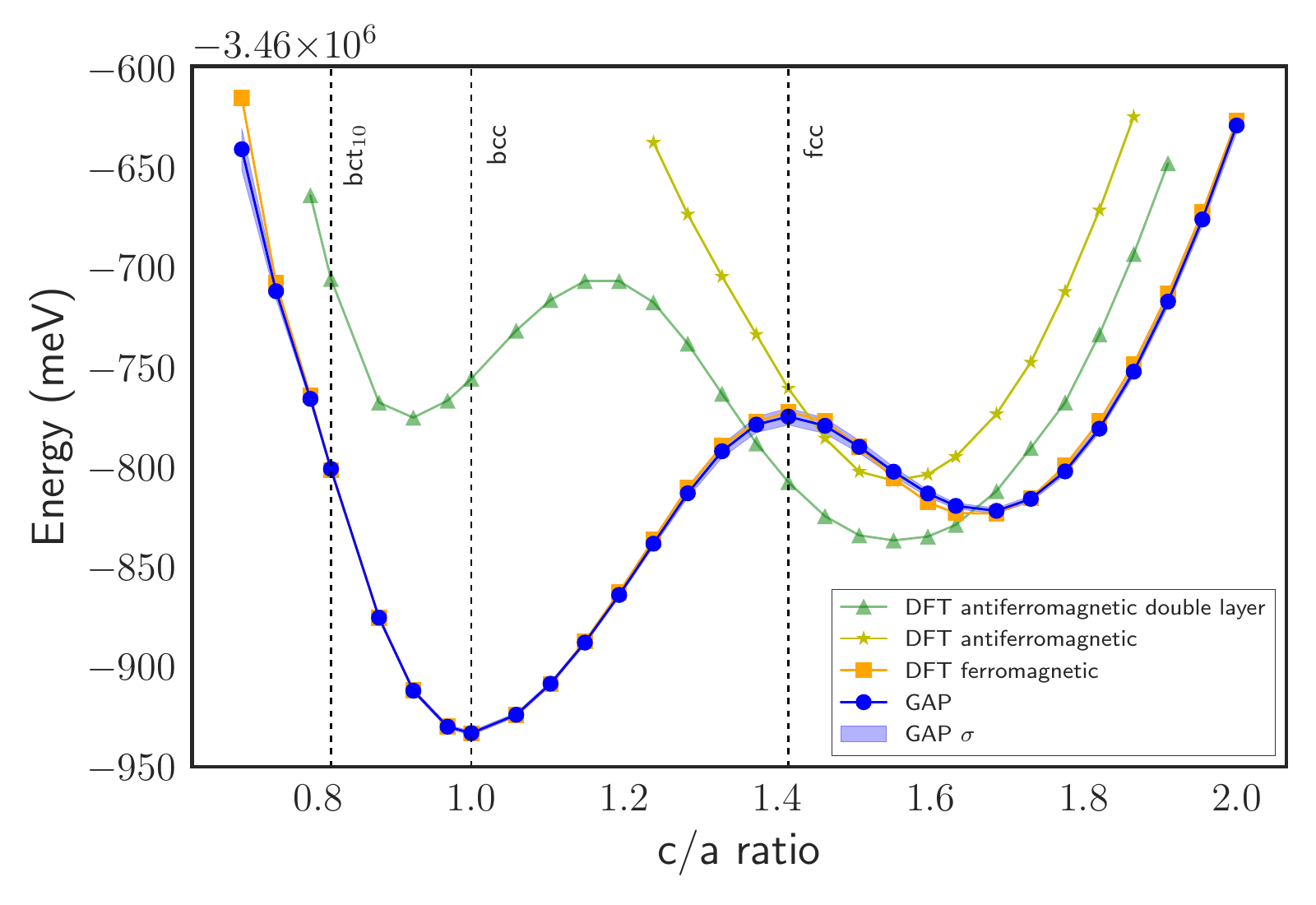}
    \caption{Epitaxial Bain path calculated using GAP and optimizing the volume for each value of $\frac{c}{a}$. 
    DFT energies are calculated using the same volume obtained in the GAP pathway.
    The GAP $\sigma$ is the variance estimated by GAP due to extrapolation outside of the training data.
  }
  \label{fig:bainpath}
\end{figure}

The Bain path traverses the diffusionless transformations between bcc, body centered tetragonal (bct) and fcc crystal structures, by varying $\frac{c}{a}$ for the cell. It shows the relative stabilities of the bcc and fcc phases and the energy barrier for the transformation. Since the GAP training data only includes ferromagnetic bcc data, the Bain path is an interesting test of the performance well outside of the training data. We compute the Bain path using a 2-atom bcc cell, so that $\frac{c}{a} = 1.0$ is the bcc configuration and $\frac{c}{a} = \sqrt[]{2}$ is fcc. At each point along the pathway, the structure is set to a fixed value of $\frac{c}{a}$ and the cell volume and the position of the central atom are relaxed to the minimum energy structure for that value of $\frac{c}{a}$. The GAP calculated Bain path is shown in~\ref{fig:bainpath}. GAP is able to estimate the error in it's prediction, and this variance is also plotted, showing the greatest uncertainty in the prediction for the fcc structure. Each structure has also been re-calculated with DFT (using the GAP optimized volume), and GAP shows notably, an excellent agreement with ferromagnetic DFT across the entire Bain path. Changes in the magnetic ground state complicate the path for iron; we find that an antiferromagnetic double layer magnetic state would stabilize the fcc structure at the GAP optimized volumes, and the true Bain path involves a number of complex magnetic states~\cite{Reith2014,Schonecker2011,Friak2001}. Including magnetic behavior in subsequent development of GAP for iron would be a fascinating challenge.

\subsection{Phonons}
\label{sec:PH_results}
\begin{figure}[H]
\centering
  \includegraphics[trim=0mm 14mm 0mm 0mm, clip, width=0.4\textwidth]{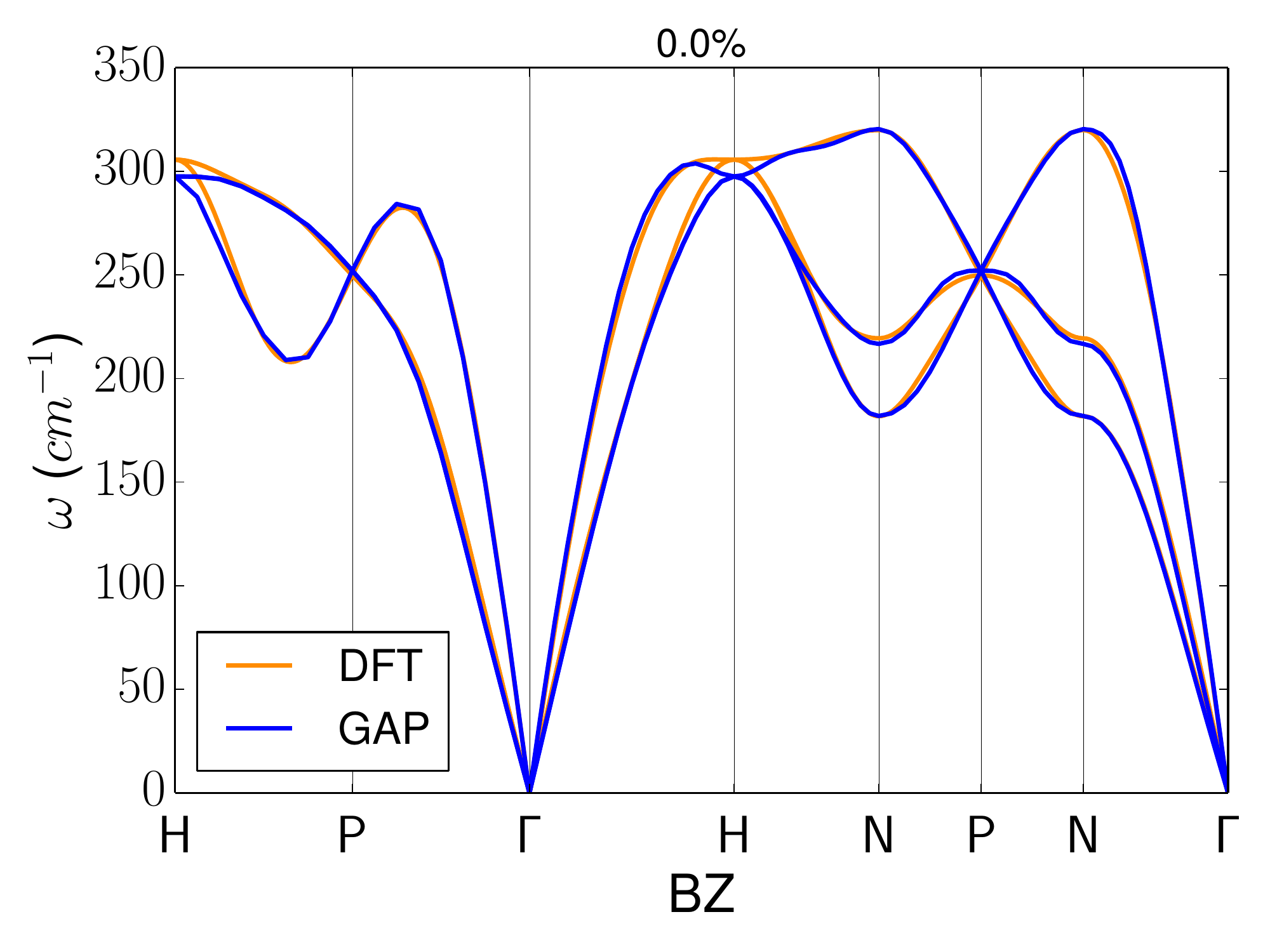} 
  \includegraphics[trim=0mm 14mm 0mm 0mm, clip, width=0.4\textwidth]{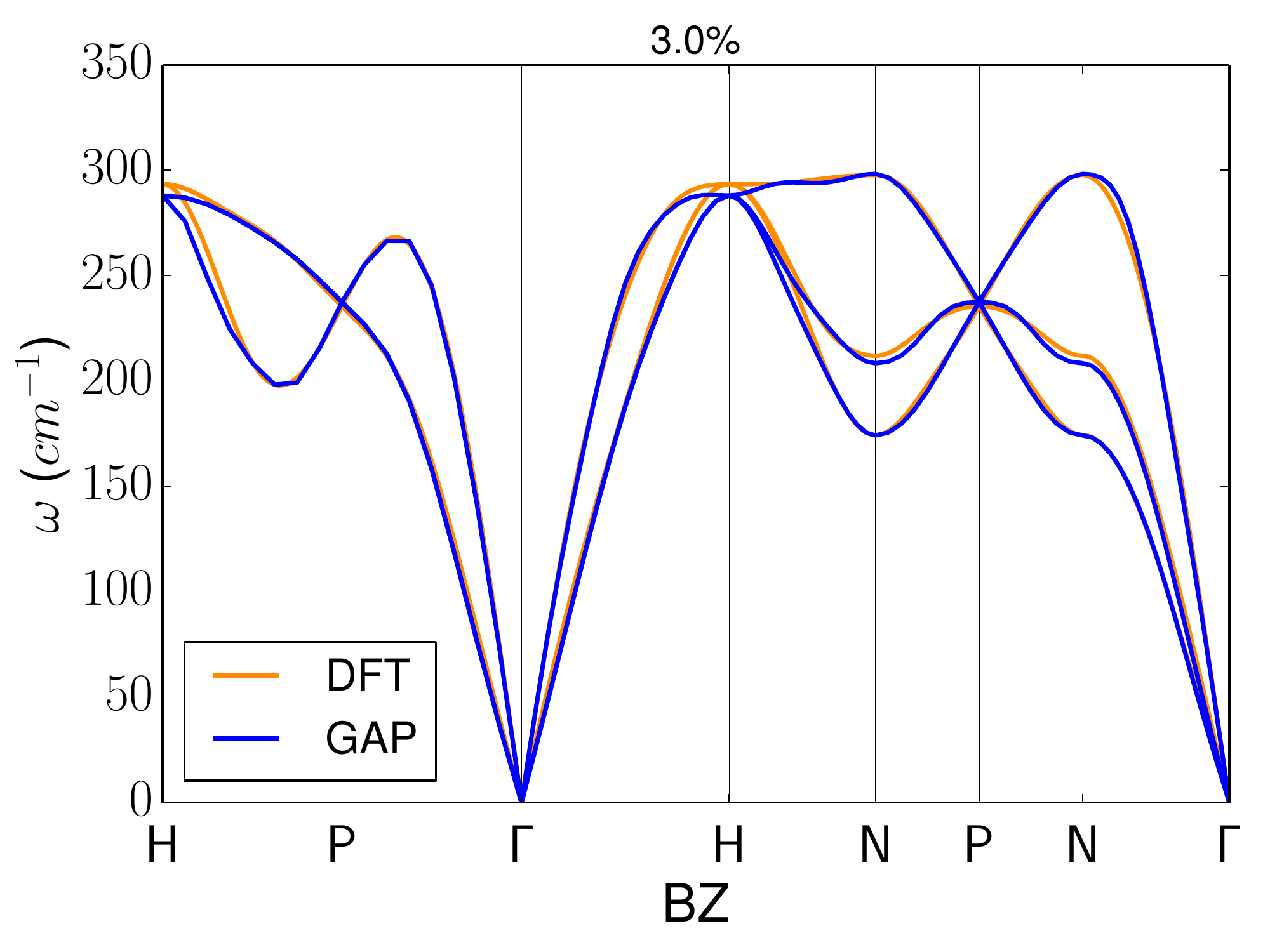} 
  \includegraphics[trim=0mm 0mm 0mm 0mm, clip, width=0.4\textwidth]{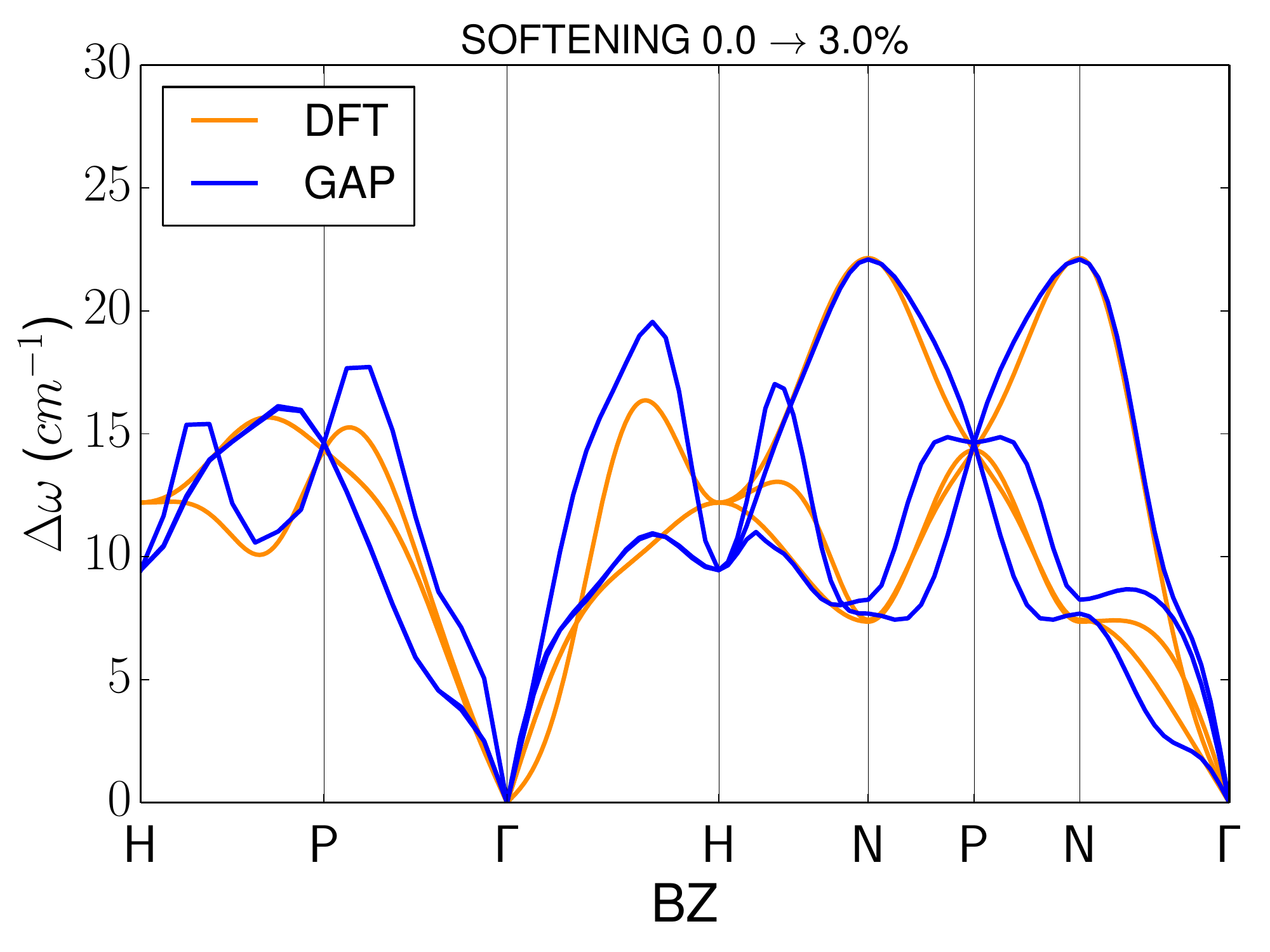} 
  \caption{Top panels: from top to bottom are reported the phonon dispersions computed at (electronic DFT) equilibrium volume and at an equilibrium volume for 1000~K (+3\%~$V_0$). GAP frequencies (orange solid lines) are obtained from frozen-phonon calculations on a supercell and are compared to DFT values (blue solid line) obtained from density-functional perturbation theory on a  $4\times4\times4$ mesh. Bottom panel: softening of the phonon frequencies along the dispersion path due to the change in volume from 0.0\% $\rightarrow$ 3.0\%~$V^{el}_0$. As above, the three blue
  lines are the three GAP modes while the three orange lines are the three DFT modes.  }
  \label{fig:phonon_gp}
\end{figure}

The GAP phonon dispersions are shown along high symmetry paths in the first BZ at the (electronic DFT) equilibrium volume $V_0$ and at an expanded value corresponding approximately to the equilibrium volume at 1000~K predicted by DFT quasi-harmonic theory~\cite{Dragoni} (namely +3.0\%~$V_0$). For each volume, the GAP dynamical matrix is obtained with a frozen-phonon method using a supercell corresponding to a $8\times8\times8$ primitive cell and finite displacements of 0.01~\AA; it is then Fourier interpolated on a denser $32\times32\times32$ mesh to give smoother frequency dispersions. Calculations are performed with the \textsf{QUIP+GAP} code~\cite{github-quip}. Results are compared in Fig.~\ref{fig:phonon_gp} (top panels) to the DFT data from Ref.~\onlinecite{Dragoni}. In the bottom panel of Fig.~\ref{fig:phonon_gp} we also check the phonon softening between the two volumes since phonon softening as a function of volume is critical to the thermal expansion and ultimately for the thermodynamic properties of a material.

\subsection{Bulk thermodynamics}
An important aspect, not always taken under consideration during the validation process of an interatomic potential, is the ability to reproduce finite-temperature properties. Iron has the bcc $\alpha$-phase displaying a ferromagnetic behavior and a Curie temperature of 1043~K. In fact, it has been shown that magnetic excitations come into play for the description of many thermodynamic quantities only above a large fraction of the Curie point~\cite{Koermann,Dragoni}. As such, we can neglect them in first approximation and assume the thermal properties of the $\alpha$-phase as dominated by atomic vibrations. Given that our GAP model provides an excellent description of bulk vibrations, we then expect good finite-temperature performance.

\begin{figure}
\centering
  \includegraphics[trim=0mm 0mm 0mm 0mm, clip, width=0.45\textwidth]{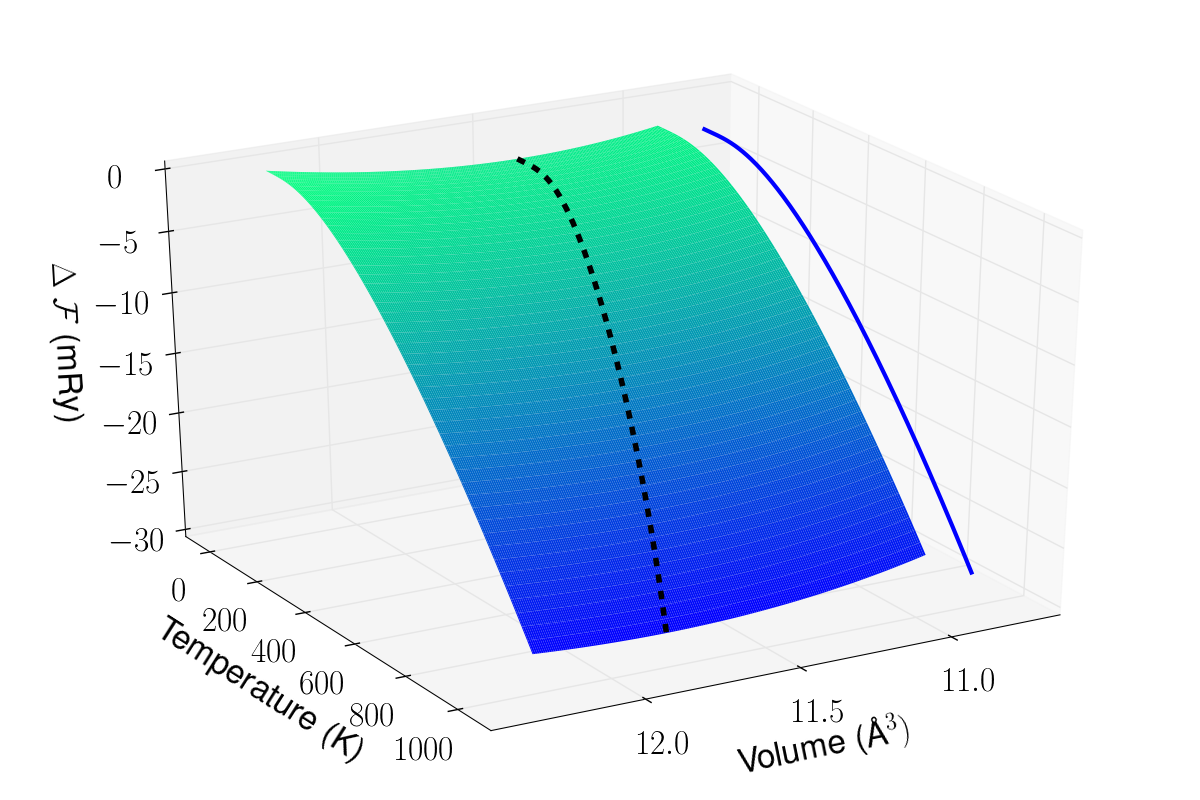} 
  \caption{Helmholtz free energy computed with GAP in the quasi-harmonic approximation. The black dashed curve is the locus of the points where the free energy is minimized at each temperature with   respect to the volume. Its projection in the free energy-temperature plane is also reported.} 
  \label{fig:Helmholtz_gp}
\end{figure}

We start our analysis of the bulk thermal properties making use of the quasi-harmonic approach which provides an accurate tool to access the low-temperature regime taking into account quantum statistical effects. By computing and integrating the phonon dispersions at 57 different volumes, from -3.6\% up to 7.6\%~$V^{el}_0$  in steps of 0.2\% of the electronic equilibrium volume, and using the same calculation details described in paragraph~\ref{sec:PH_results}, we obtain to the Helmholtz free energy (see Fig.~\ref{fig:Helmholtz_gp}). From that, we calculate all the relevant thermodynamic quantities of interest in the quasi-harmonic approximation. In parallel, in order to study the same quantities obtained from QHA in the high-temperature regime, where stronger anharmonicity comes into play and quantum statistical effects lose importance, we use an MD approach. We perform NPT runs at vanishing external pressure to find the equilibrium density at different temperatures from 200~K to 1800~K in steps of 200~K. We use a $8\times8\times8$ supercell with 1024 atoms, a time step of 1~fs with temperature and pressure controlled by a Nose-Hoover chain thermostat~\cite{Nose-Hoover} and and a Parrinello-Rahman barostat~\cite{Parrinello-Rahman} as implemented in the \textsf{LAMMPS}~\cite{lammps} package.

\begin{figure}[H]
\centering
  \includegraphics[trim=0mm 0mm 0mm 0mm, clip, width=0.45\textwidth]{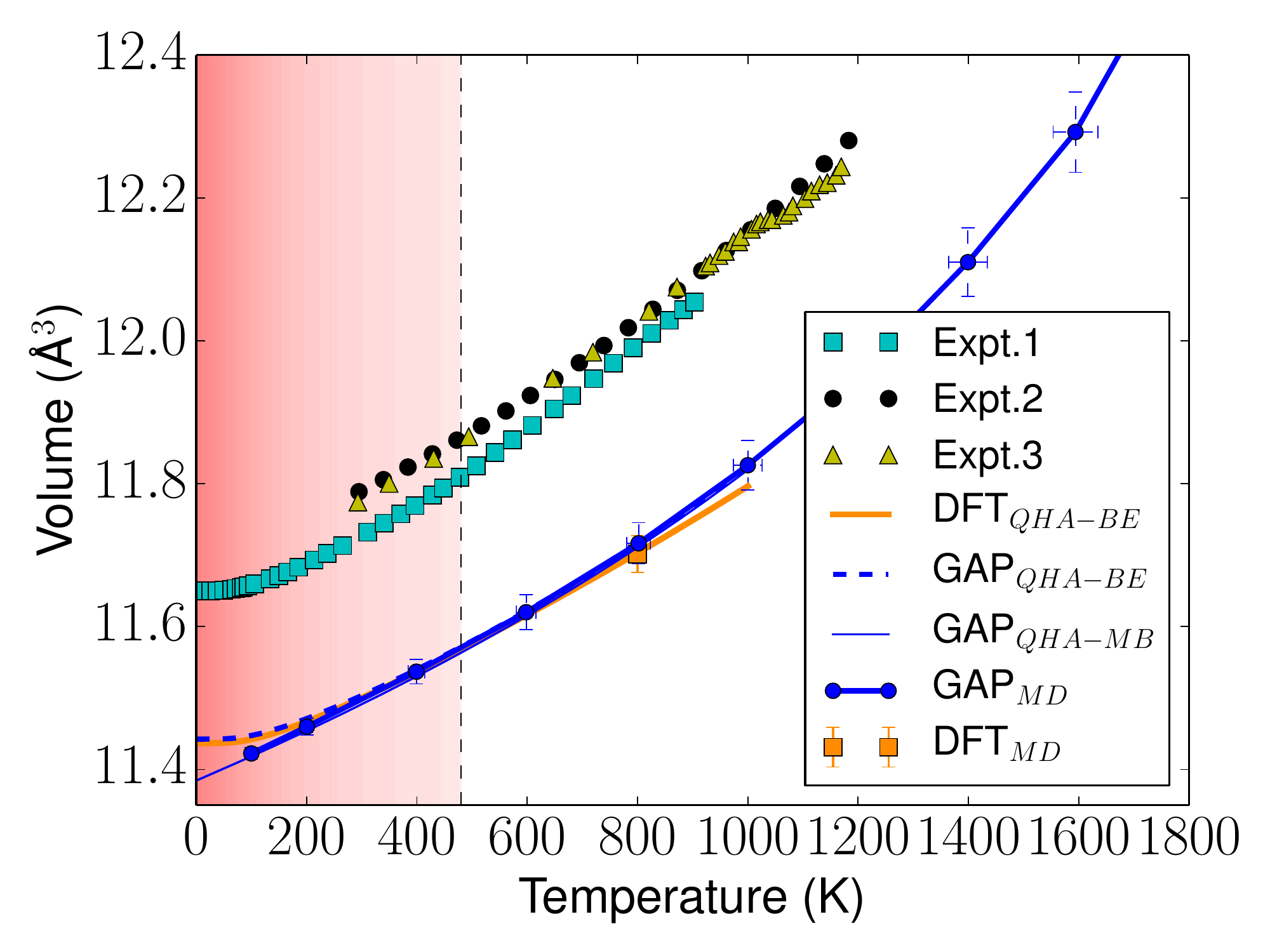} 
  \includegraphics[trim=0mm 0mm 0mm 0mm, clip, width=0.45\textwidth]{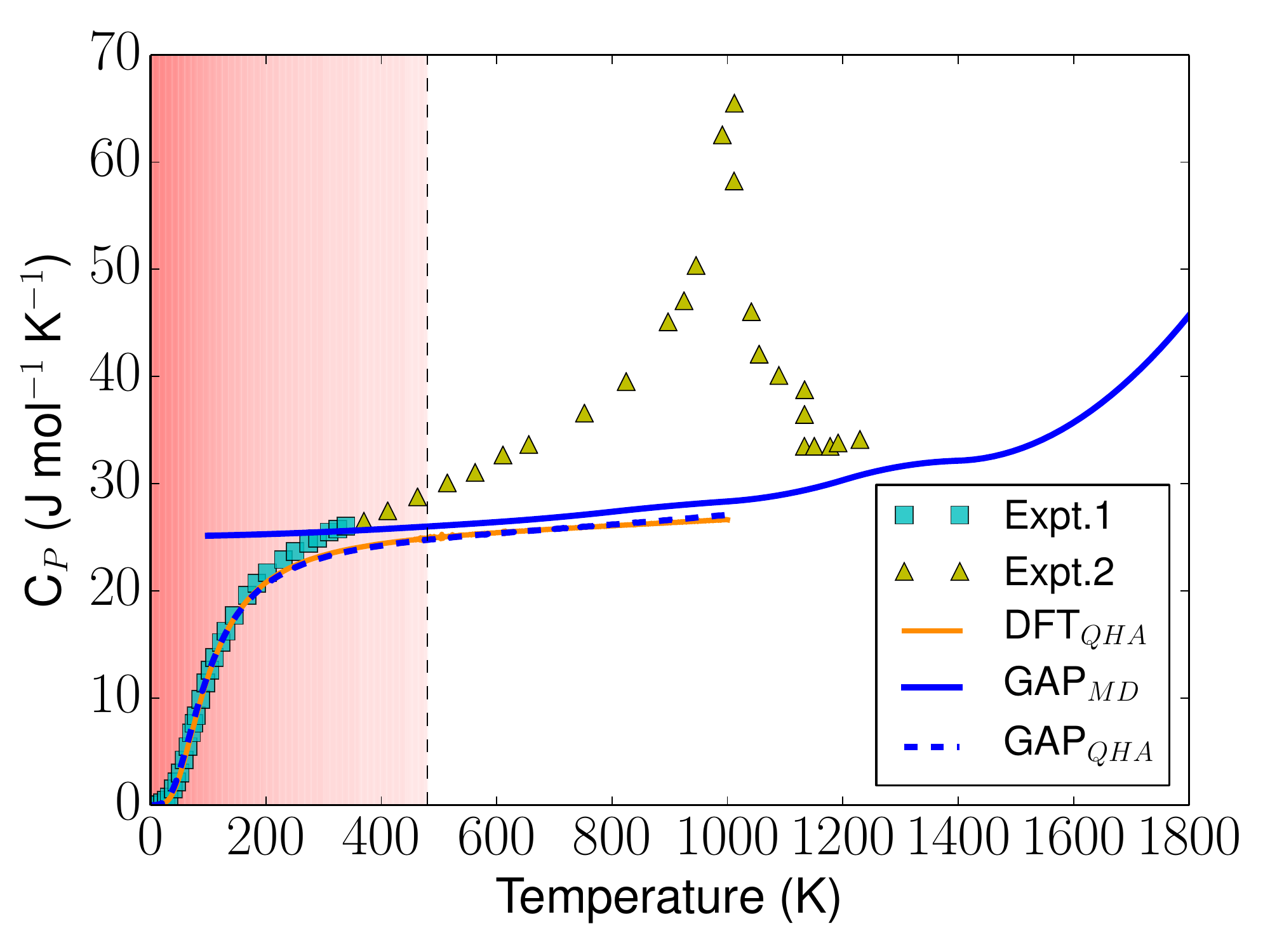}  
  \includegraphics[trim=0mm 0mm 0mm 0mm, clip, width=0.45\textwidth]{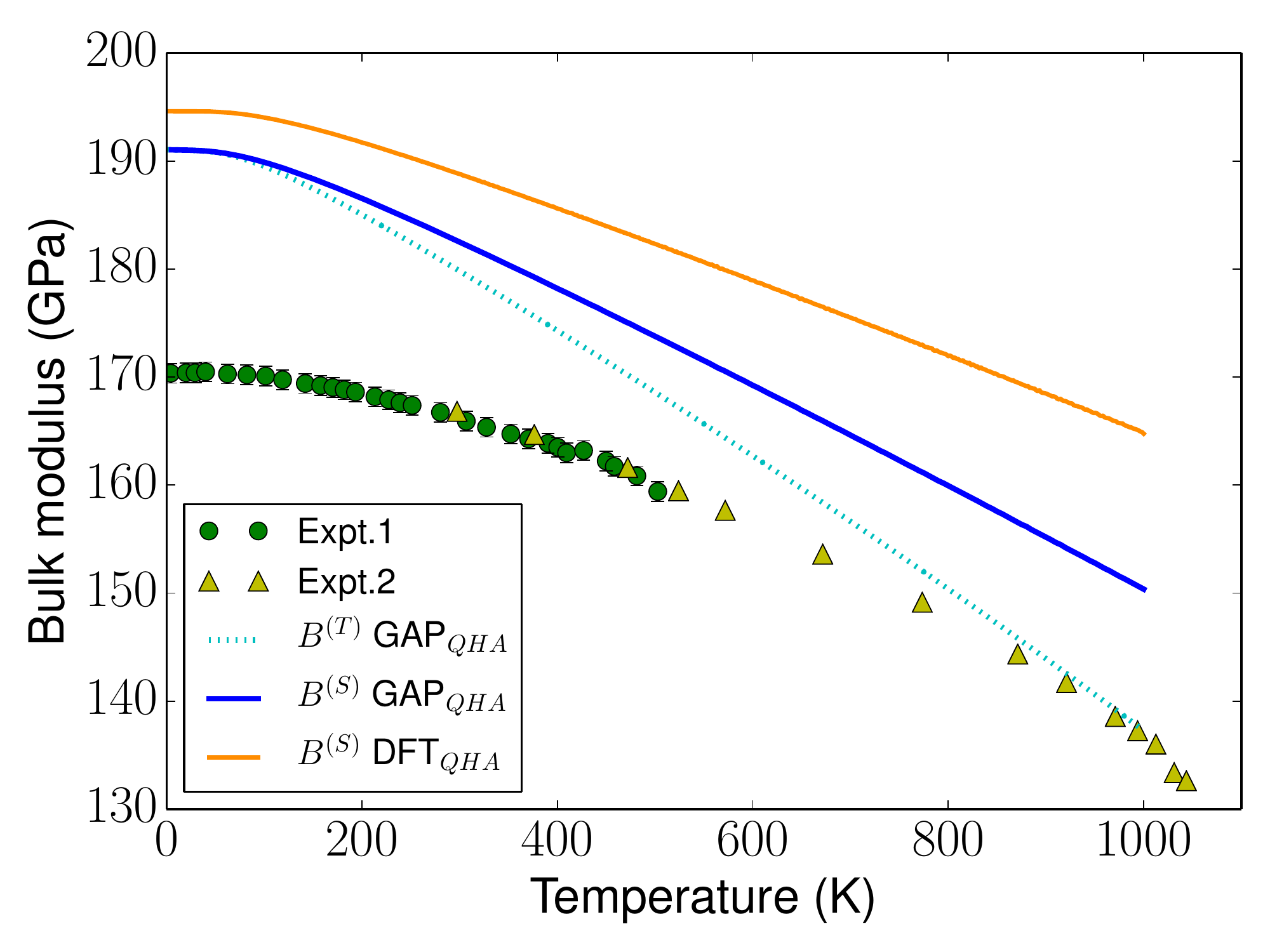} 
  \caption{Top left panel: thermal expansion of GAP obtained from MD (blue solid line) and QHA (blue dashed line is QHA with Bose-Einstein statistics, blue thin line is QHA with Maxwell-Boltzmann   statistics). Results are compared to DFT QHA data (orange solid line), to a single DFT MD point at 800~K, and to experiments (squares, triangles and circles data from   Refs.~\onlinecite{Basinski,Ridley,Seki}).  
  Top right panel: heat capacity at constant pressure as a function of temperature. GAP QHA (dashed blue line) and GAP MD (solid blue line) are compared to DFT QHA (solid orange line) and to   experimental data from Refs.~\onlinecite{Desai,Wallace2} (squares and triangles respectively). Bottom panel: adiabatic (blue solid line) and isothermal (azure dotted line) bulk   modulus as a function of temperature from GAP\@. Comparison with DFT adiabatic data (orange solid line) from Ref.~\onlinecite{Dragoni} and experiments from Refs.~\onlinecite{Adams,Dever} (circles and   triangles).}
  \label{fig:thermal_gp}
\end{figure}

%%%%%%%%%%%%% 
The first quantity that we analyze is the thermal expansion. In Fig.~\ref{fig:thermal_gp} we show the GAP QHA curve which follows the proper quantum Bose-Einstein (BE) statistics, the  GAP QHA  modified to follow the classical Maxwell-Boltzmann (MB) statistics (zero point energy contribution is not included), and the GAP curve resulting from MD calculations. For comparison, we show the DFT QHA (BE) curve~\cite{Dragoni} plus an estimate of the equilibrium volume at 800~K from DFT molecular dynamics. As a reference we also report three sets of experimental data. It is immediately possible to note that the GAP QHA (BE) curve agrees remarkably well with the DFT QHA one up to 1000~K. 
The DFT and GAP results instead underestimate experiments.  As exhaustively discussed in Ref.~\onlinecite{Dragoni} this can be attributed to the DFT PBE functional which has been adopted for the database generation and for the DFT data used for comparison. 
Nonetheless, the experimental thermal trend 
is overall well reproduced. The GAP MD and the GAP QHA curves agree well up to 800~K while they start to deviate above this temperature. The GAP MD curve also matches the DFT MD equilibrium volume at 800~K. This analysis seems to suggest that beyond quasiharmonic effects start to play a role only above 800~K. Interestingly, the MD results overlap with quasi-harmonic results modified to artificially reproduce a classical Maxwell-Boltzmann behavior at temperatures below 200~K. We finally notice that the bcc phase appears mechanically stable up to approximately the experimental melting point~\footnote{Note however that in our MD NPT runs we have allowed only for isotropic fluctuations of the starting cubic box. New runs where the initial box is allowed to fluctuate freely should be performed to confirm this observation.}.
From the knowledge of the temperature-volume relation at equilibrium we then calculate the temperature dependence of other relevant bulk thermodynamic quantities. Heat capacity at constant pressure results are reported in Fig.~\ref{fig:thermal_gp}, including QHA and MD data. As for the thermal expansion, the heat capacities obtained with QHA and MD nicely converge at intermediate temperature. The experimental divergence at the Curie point is related to magnetic entropy~\cite{koerman_specificheat,dudarev_specificheat}; as such it is not captured by our DFT calculations and, consequently, by our model. Within the quasi-harmonic framework, heat capacity is used also to compute the adiabatic bulk modulus thermal behavior starting from the isothermal one as discussed in Ref.~\onlinecite{Dragoni}. 
In Fig.~\ref{fig:thermal_gp} we show that GAP is capable to reproduce well the overall DFT thermal behavior, although slightly underestimating (in the direction of the experimental data) the absolute values.
Since the bulk moduli are second partial derivatives of the Helmholtz free energy, these results reflect the ability of the model to accurately reproduce the details of the bulk quantum mechanical PES.

\subsection{Bulk point defects}
\subsubsection{Atomic vacancies}
\begin{figure}[H]
\centering
  \includegraphics[trim=0mm 0mm 0mm 0mm, clip, width=0.45\textwidth]{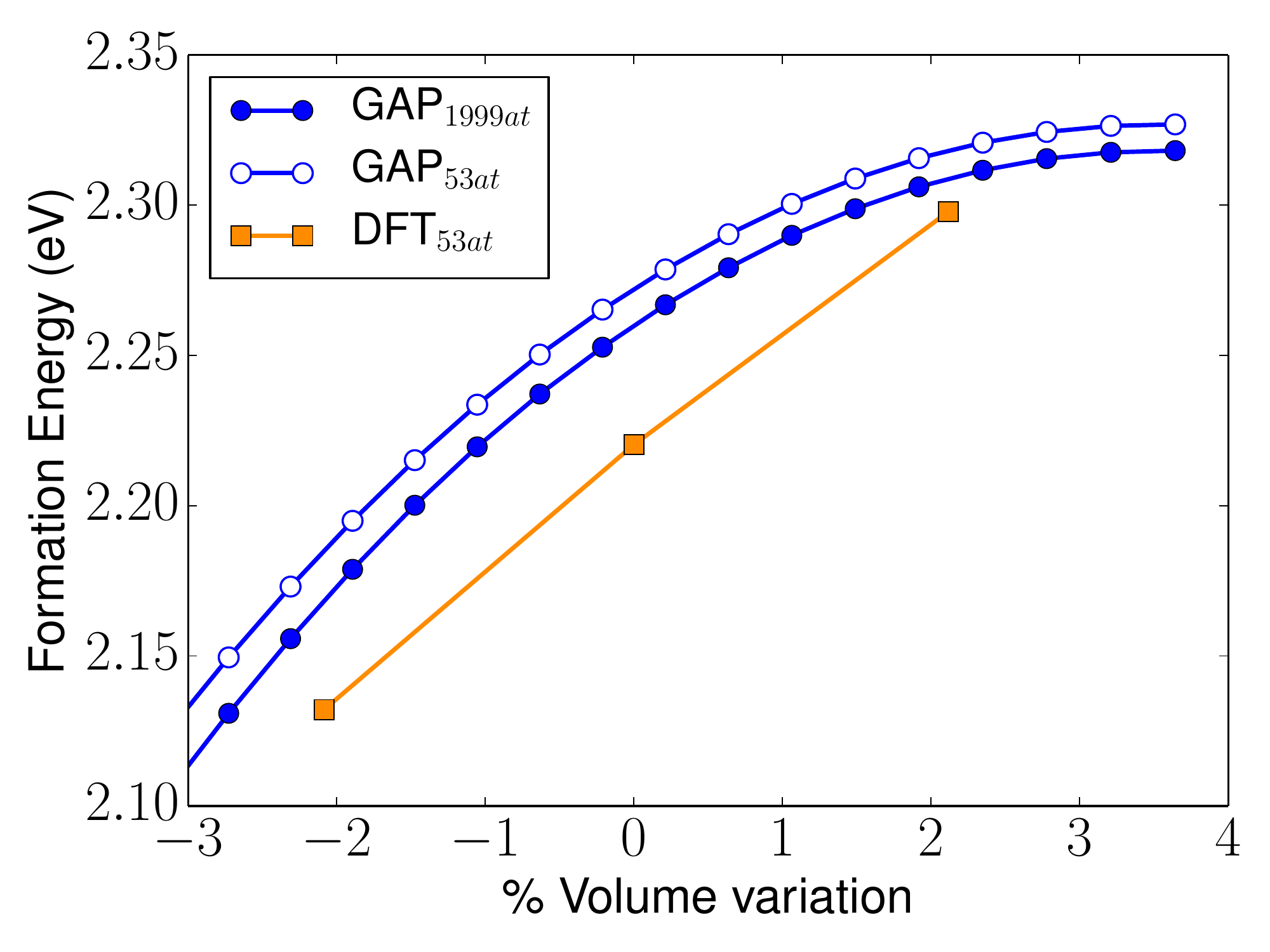} 
  \caption{Mono-vacancy formation energy as a function of percentage volume variation with respect to the (electronic) equilibrium value. GAP results (blue full and empty circles) are obtained relaxing atomic positions in a $10\times10\times10$ and in a $3\times3\times3$ conventional cubic supercells, while DFT data are obtained relaxing only atomic positions of a $3\times3\times3$ cubic supercell and are reported as orange circles.}
  \label{fig:monovac_gp}
\end{figure}

\begin{figure}[H]
\centering
  \includegraphics[trim=0mm 0mm 0mm 0mm, clip, width=0.45\textwidth]{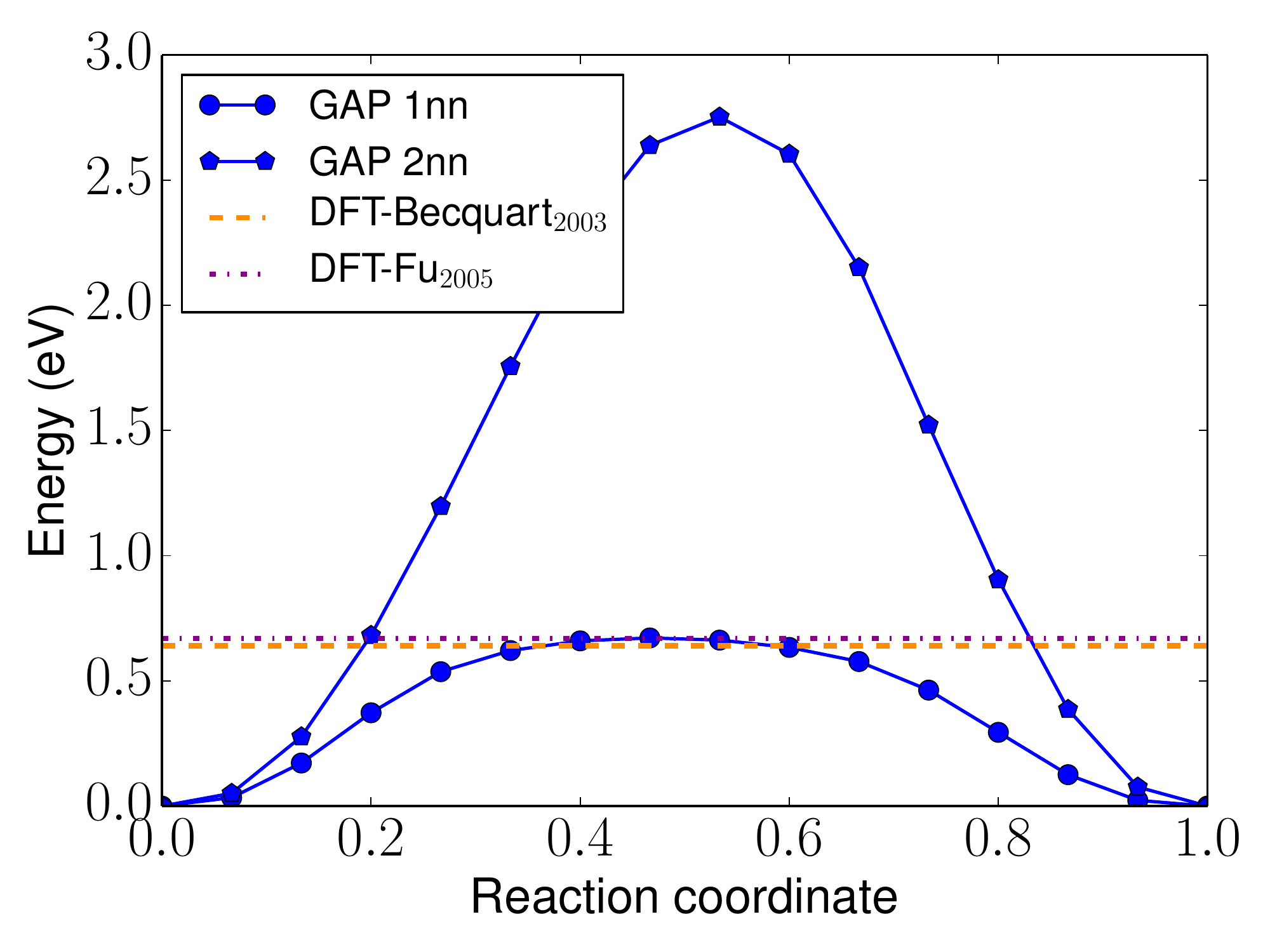} 
  \caption{Migration energy profiles of a mono-vacancy diffusing to a first- and second-nearest neighbor site (blue circles and pentagons respectively) obtained from nudged-elastic-band calculations. 
  The calculations are performed using 249 atoms in a $5\times5\times5$ cubic supercell. DFT mono-vacancy first-nearest neighbor migration energy barriers are taken from Refs.~\onlinecite{nature_chuchun,divacancy_DFT1}
  (dashed lines).}
  \label{fig:monovac_migration_gp}
\end{figure}

\begin{figure}[H]
\centering
  \includegraphics[trim=0mm 0mm 0mm 0mm, clip, width=0.45\textwidth]{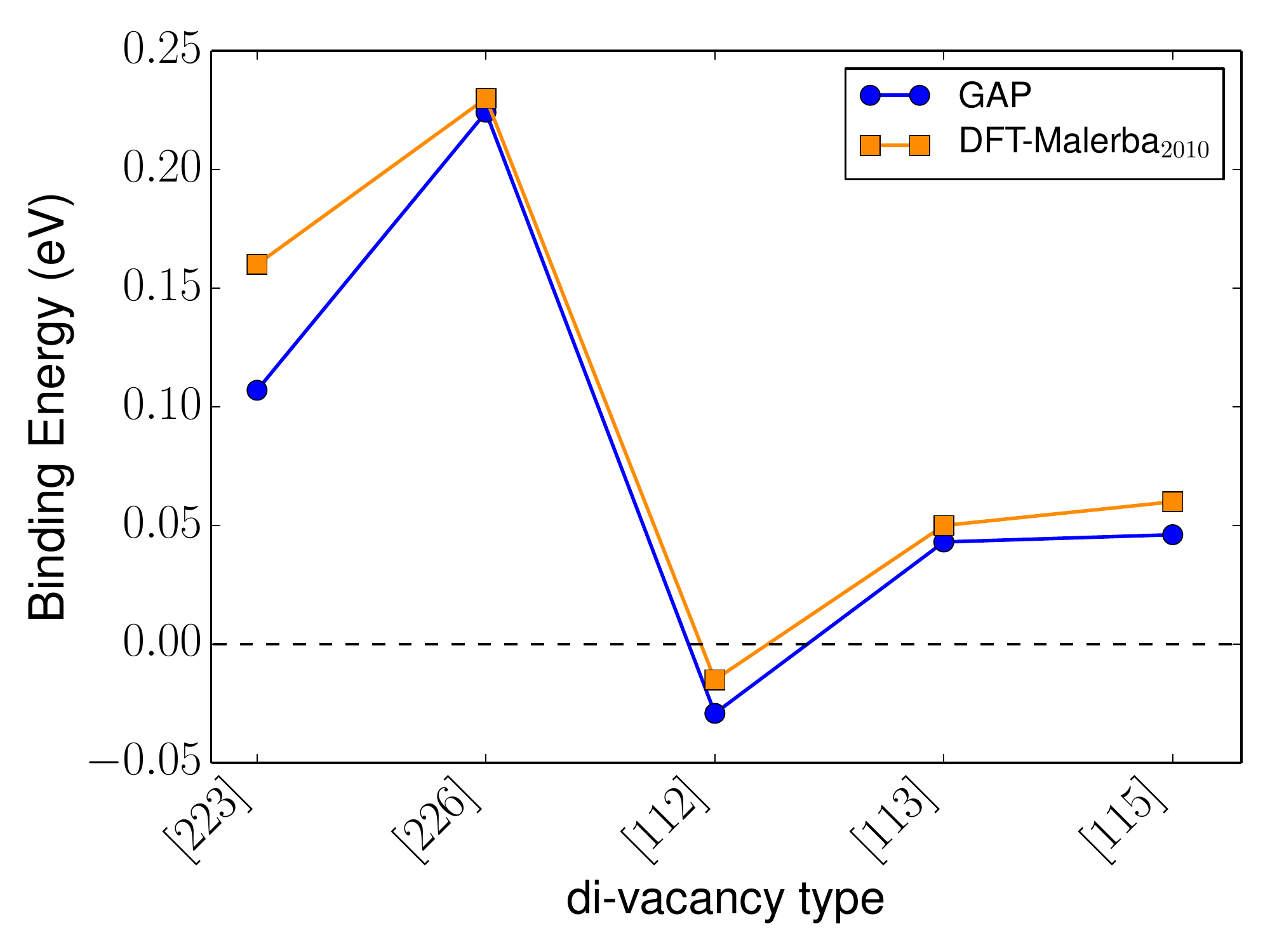} 
  \caption{Di-vacancy binding energy landscape. GAP results (blue circles connected by solid lines) are compared to DFT data 
  from Ref.~\onlinecite{malerba2010} (orange squares). Formation energy values are also reported in Tab.~\ref{tab:energetics_gp}. 
  }
  \label{fig:di_vacancies}
\end{figure}

\begin{figure}[H]
\centering
  \includegraphics[trim=0mm 0mm 0mm 0mm, clip, width=0.45\textwidth]{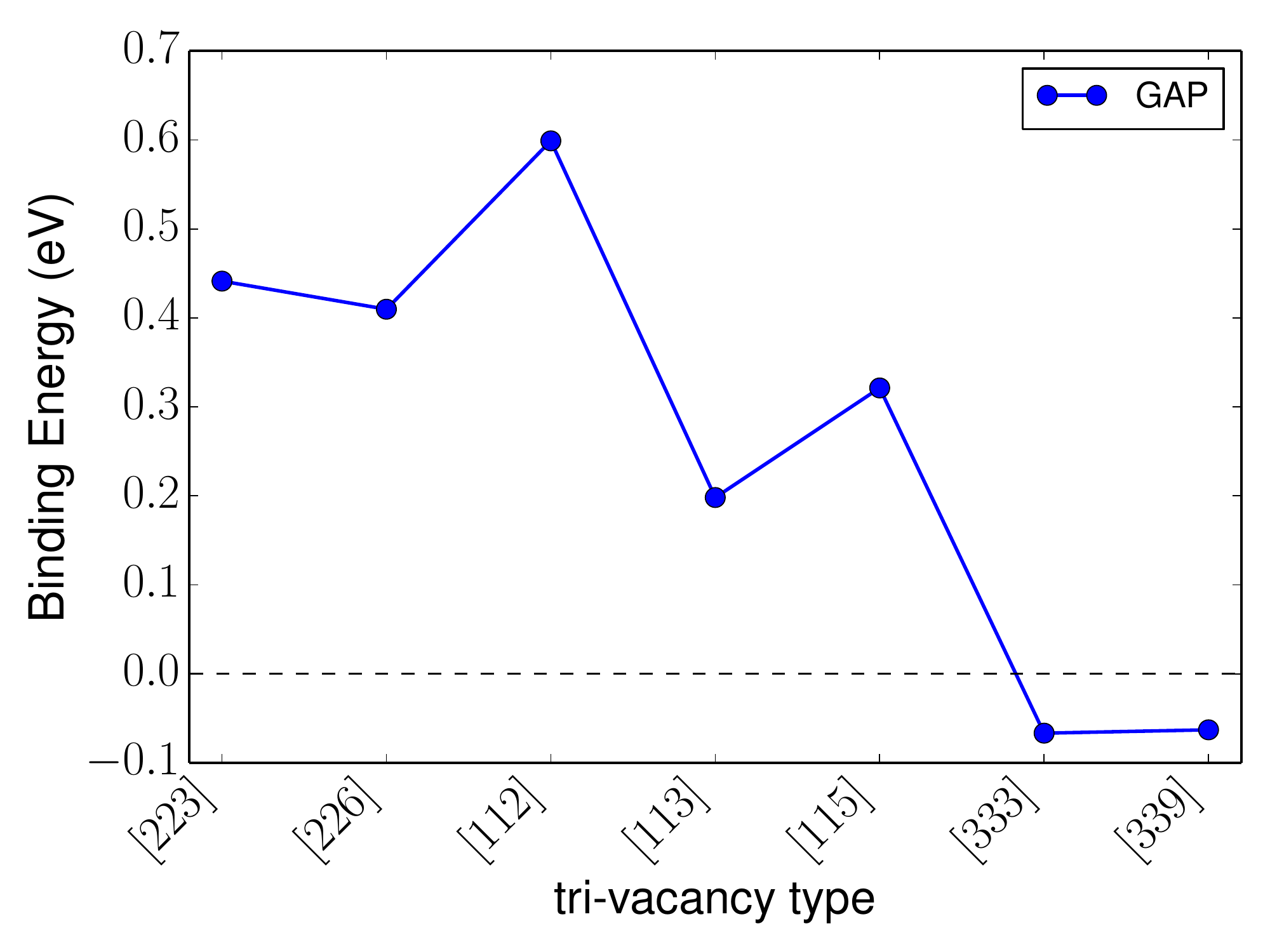} 
  \caption{Tri-vacancy binding energy landscape. Formation energy values are also reported in Tab.~\ref{tab:energetics_gp}.
  }
  \label{fig:tri_vacancies}
\end{figure}

Real crystals are far from being perfect, and contain defects that can be e.g.\ point-like or extended in space. Their study is fundamental for understanding the microscopic processes that govern the actual response of a macroscopic system under different external conditions. It is therefore important to test the capabilities of GAP in describing the energetics of some simple defects. We start from the mono-vacancy, which consists of a missing atom in an infinite lattice. This missing atom is assumed to be isolated, i.e.\ not interacting with any other defect in the surroundings. The energy of formation of a mono-vacancy, i.e.\ the cost of removing an atom from the perfect bulk, at the equilibrium volume is reported in Tab.~\ref{tab:energetics_gp}. 
In addition, the dependence of the formation energy upon volume~\footnote{The calculation of this quantity is performed at constant volume with relaxed atomic positions in a cubic cell containing 53/1999 atoms.} is shown in Fig.~\ref{fig:monovac_gp}.  We also compute the energy profile or minimum energy path for a mono-vacancy migration to a first-, second- and third-nearest neighbor site through nudge-elastic band~\cite{NEB} (NEB) calculations. The energy profiles are reported in Fig.~\ref{fig:monovac_migration_gp} and the corresponding migration energy barriers are summarized in Tab.~\ref{tab:energetics_gp}. Results are closely consistent with DFT calculations. Next, we consider di-vacancy defects, where two missing atoms are simultaneously present and interact with each other in the crystal. The formation energy and binding energy of first-, second-, third-, fourth-, and fifth-nearest neighbor di-vacancies at the zero pressure condition are reported in Fig.~\ref{fig:di_vacancies} and summarized in Tab.~\ref{tab:energetics_gp}. In agreement with  Refs.~\onlinecite{malerba2010,djurabekova2010kinetics} the binding energy of the third-nearest neighbor is negative, thus suggesting the instability of such configuration compared to the condition of two isolated mono-vacancies. As expected from DFT calculations, but contrarily to most of the semi-empirical models available in the literature~\cite{malerba2010}, the fifth-$nn$ configuration is reported to be positive. Selected tri-vacancy binding energies are reported in Fig.~\ref{fig:tri_vacancies} and summarized in Tab.~\ref{tab:energetics_gp} along with formation energies.  Note that the tri-vacancy types are here identified by means of the Beeler notation~\cite{beelerVAC}. Results suggest the suppression of binding on the $\{111\}$ plane with a ground state [112] configuration as predicted by  DFT~\cite{hayward2013interplay}. At variance with DFT the ordering of formation of the [226] and [223] configurations is swapped while, in accordance with DFT, the cost of formation of the [223] configuration is predicted lower than the [115] one. It is worth to note that neither the [226] nor the [115] are included in the training database. This fact suggests that some caution is needed when the potential is used as an extrapolation. 
Although we have included a few four and five vacancies configurations in the training database, an extensive analysis of tetra- and penta-vacancies will be performed elsewhere.

\subsubsection{Self-interstititals}
\begin{figure}[H]
\centering
  \includegraphics[trim=0mm 0mm 0mm 0mm, clip, width=0.45\textwidth]{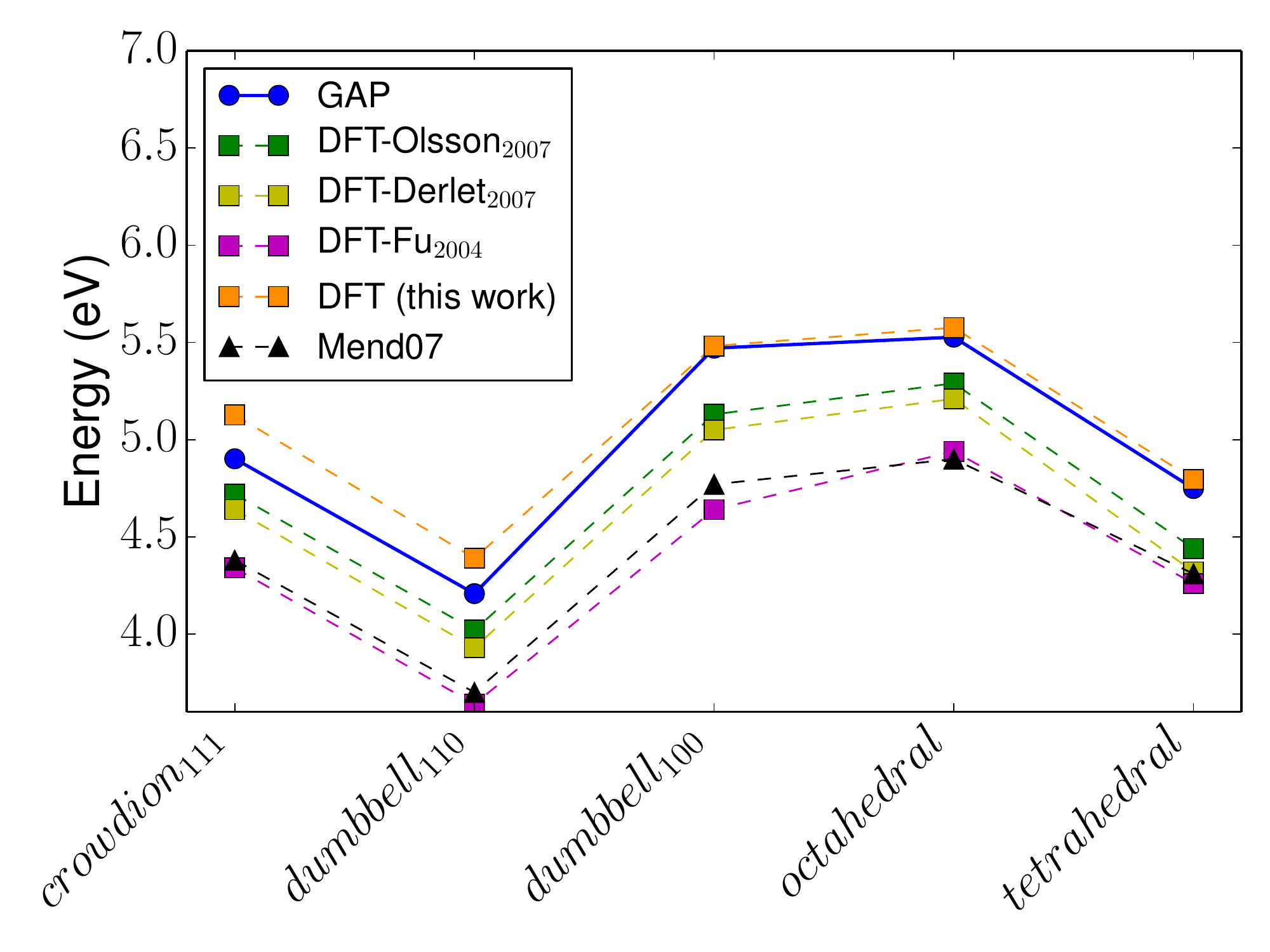} 
  \caption{Selected self-interstitial formation energies from GAP (using $10\times10\times10$ cubic supercell) and compared to DFT data from various works in the literature~\cite{Dudarev_Derlet,Olsson,Fu_SIA}. 
  Results are also compared to   a successful EAM model well known in the literature~\cite{malerba2010}, the so-called Mendelev07.  GAP and DFT results from Refs.~\cite{Dudarev_Derlet,Fu_SIA} are  obtained at zero pressure. The DFT data of this work  have been instead computed using a constant volume plus atomic relaxation. Note however that according to the authors of Ref.~\cite{Olsson} there is no significant difference between constant pressure  and constant volume with atomic relaxation conditions for a cubic supercell of 129 atoms. }
  \label{fig:SIA_gp}
\end{figure}

\begin{figure}[H]
\centering
  \includegraphics[trim=0mm 0mm 0mm 0mm, clip, width=0.4\textwidth]{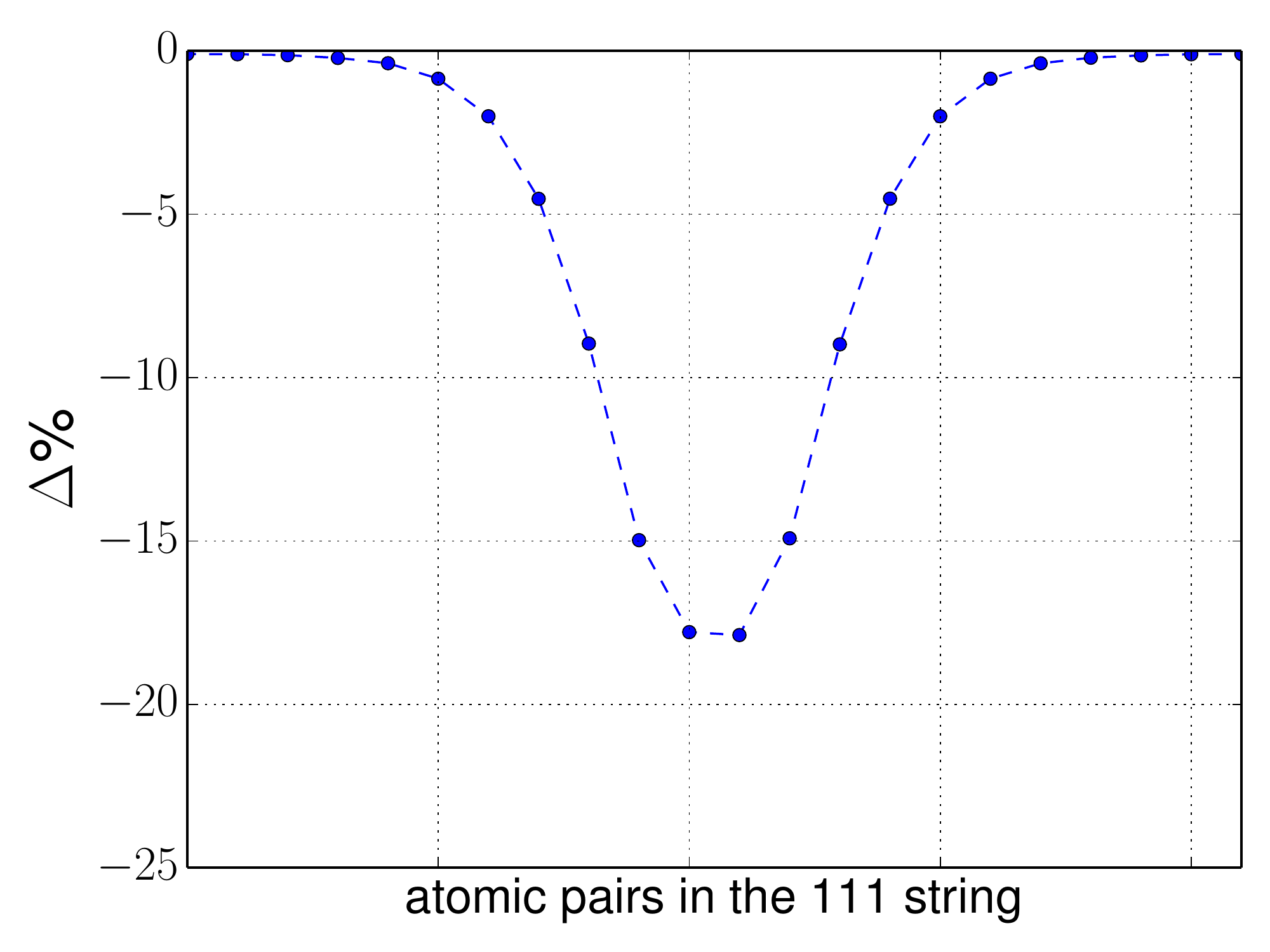} 
  \caption{Crowdion displacement field along the 111 string as predicted by GAP\@. $\Delta\%=(d-a^{el}_{111})/a^{el}_{111} \times 100$, with $d$ being the distance between two consecutive atoms in the 111   string of the relaxed defective configuration, and $a^{el}_{111}=\sqrt{3/2} \, a^{el}_0$ the same quantity in a perfect lattice at the (electronic) equilibrium. The minimum of the curve indicates where  the   atomic pairs are closest. Here we use a $10\times10\times10$  cubic supercell containing 2001 atoms in total so that we have 21 atomic distances to consider along the $\langle111\rangle$ direction. }
  \label{fig:crowdion_renorm_gp}
\end{figure}

Next we consider self-interstitial atoms (SIAs), with particular interest for the crowdion$_{111}$, dumbbell$_{110}$, dumbbell$_{100}$, tetrahedral and octahedral configurations. These are in fact the simple self-interstitial defects in bcc iron~\cite{malerba2010} which are considered the most relevant in the study of damage and aging of steel reactor vessels under strong irradiation. 
GAP formation and binding energies are evaluated at zero pressure. Results are reported in Tab.~\ref{tab:energetics_gp} and summarized in Fig.~\ref{fig:SIA_gp}. As expected from our DFT results~\footnote{Performed at the constant equilibrium volume with atomic relaxation in a $3\times3\times3$ cubic cell} and other DFT studies~\cite{Dudarev_Derlet}, we find that the most stable GAP interstitial is the dumbbell$_{110}$ configuration, followed by the tetrahedral and crowdion$_{111}$ ones. The renormalization of the atomic distances of the atoms of the 111 string  along the $\langle111\rangle$ direction are reported in Fig.~\ref{fig:crowdion_renorm_gp} for reference.
We have also computed the migration energy barrier for a dumbbell$_{110}$ to jump to a first-nearest neighbor site. The jump mechanism to the first-nearest neighbor is consistent with the one observed within the DFT framework~\cite{malerba2010}. The migration energy barrier $E_{m_{\langle 110\rangle}}^{jump}$ of such mechanism is reported again in Tab.~\ref{tab:energetics_gp}.
At variance with most of the models available in the literature~\cite{malerba2010}, the GAP model presented here is able to reproduce the relative ordering of binding energies of the non-parallel and $\langle110\rangle$ dumbbell di-interstitial configurations.  All GAP calculation on self-interstitials are performed at zero pressure in a $10\times10\times10$ cubic supercell.

\subsection{Free surfaces}
Bulk-terminated surfaces can be regarded as a type of extended defect. The energy cost of creating a bulk terminated surface and its dependence on the crystallographic orientation influences the growth and equilibrium shape of the crystal during crystallization. Here we calculate the surface formation energy of four crystallographic surface orientations which are considered most relevant for bcc structures~\cite{spencer2002,blonski2007}. These are the low index surfaces \{110\}, \{100\}, \{112\} which are lowest in energy compared to other orientations, plus the \{111\} surface orientation for the sake of completeness. The formation energy ordering obtained for GAP is reported in Tab.~\ref{tab:energetics_gp} and agrees well both qualitatively and quantitatively with the DFT results. 

\subsection{Gamma surfaces and Dislocations}
\begin{figure*}[h]
\centering
\begin{subfigure}%{0.5\textwidth}
\centering
  \textbf{GAP \; \; \; \; \; \; \; \; \; \; \; \; \; \; \; \; \; \; \; \; \; \; \; \; \; \; \; \; \; \; DFT  }\par\medskip 
  \includegraphics[trim=6mm 0mm 0mm 2mm, clip, width=0.45\linewidth]{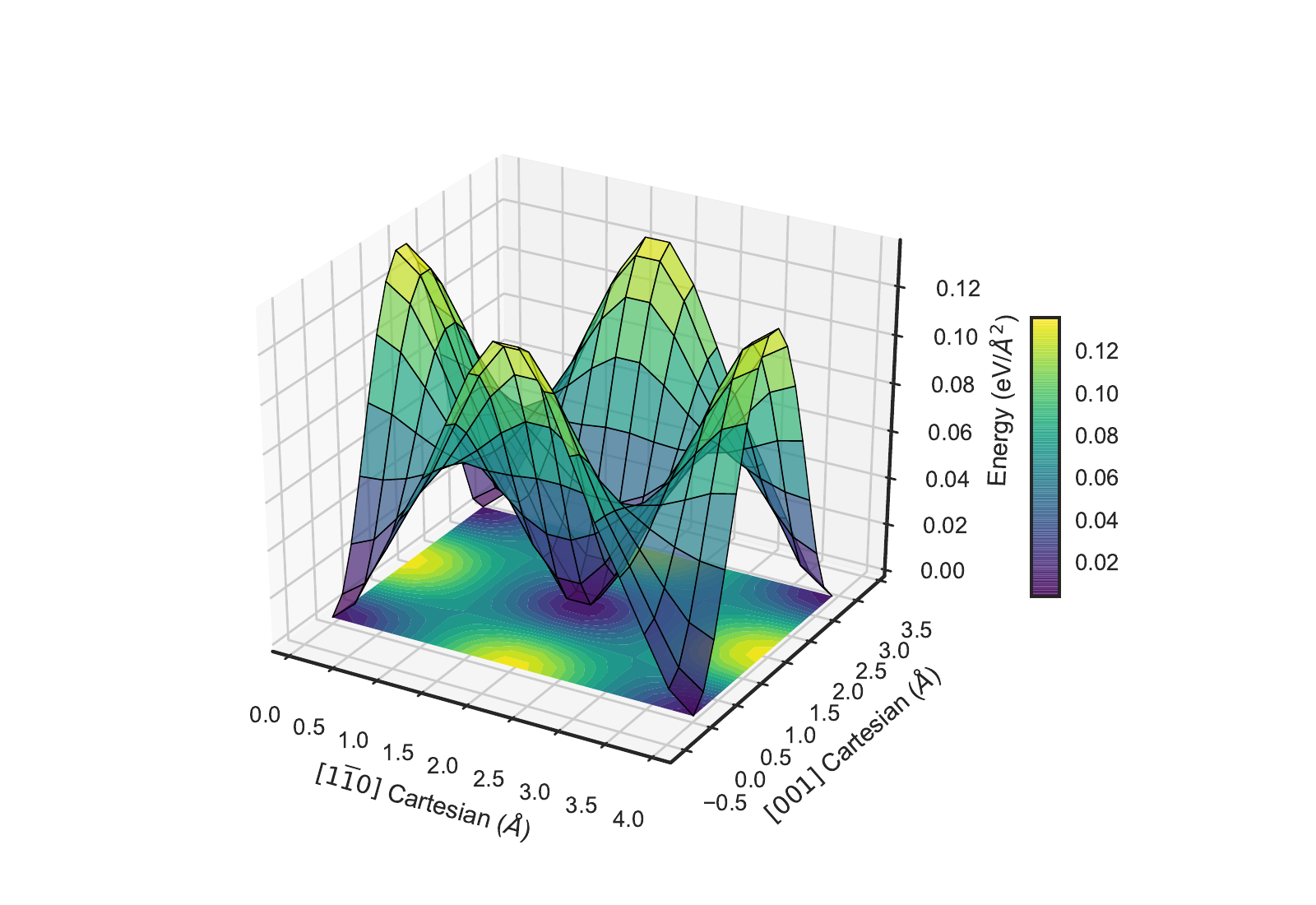}
  \includegraphics[trim=6mm 0mm 0mm 2mm, clip, width=0.45\linewidth]{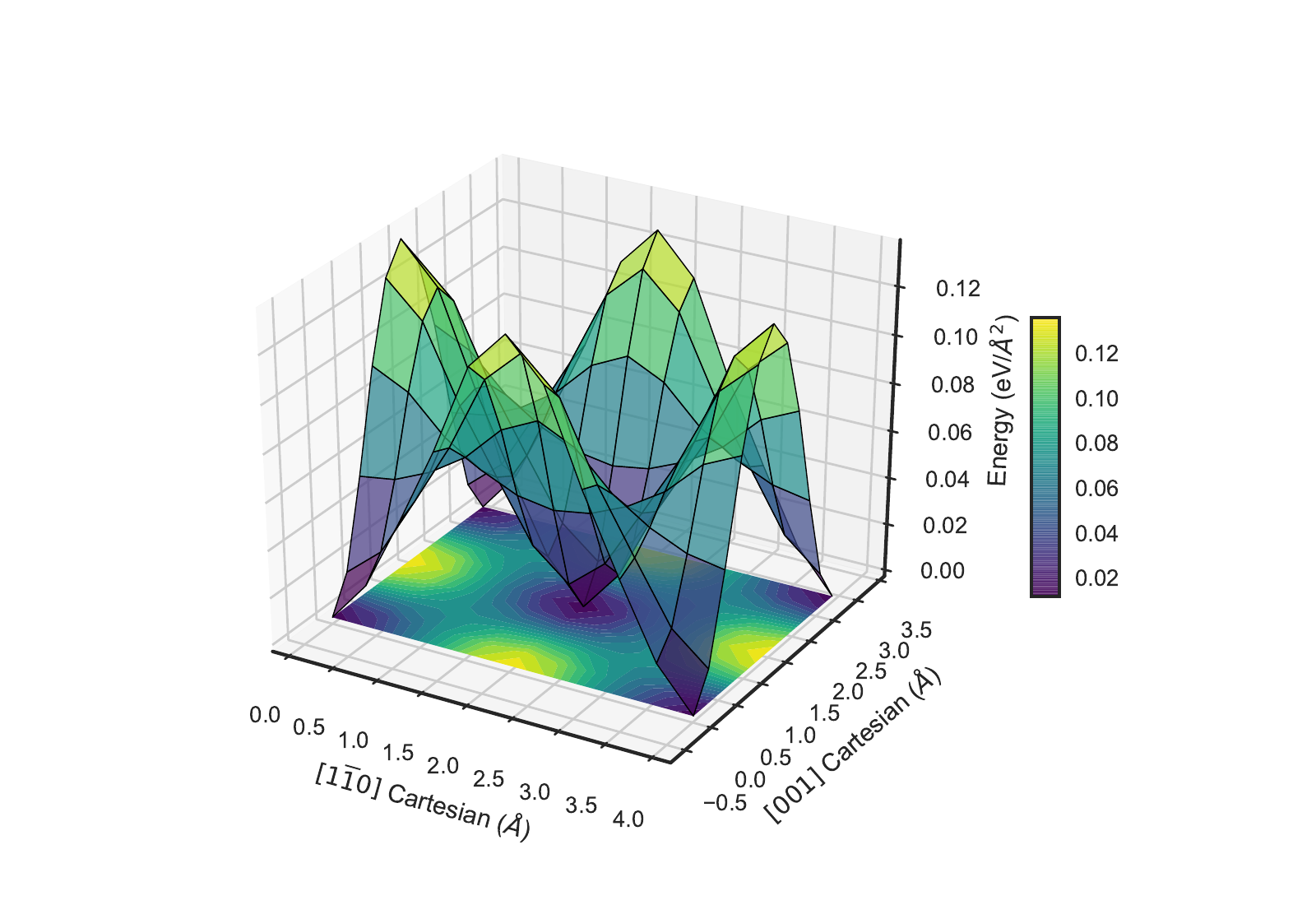}
\end{subfigure}
\begin{subfigure}%{0.5\textwidth}
\centering
  \includegraphics[trim=6mm 0mm 0mm 2mm, clip, width=0.45\textwidth]{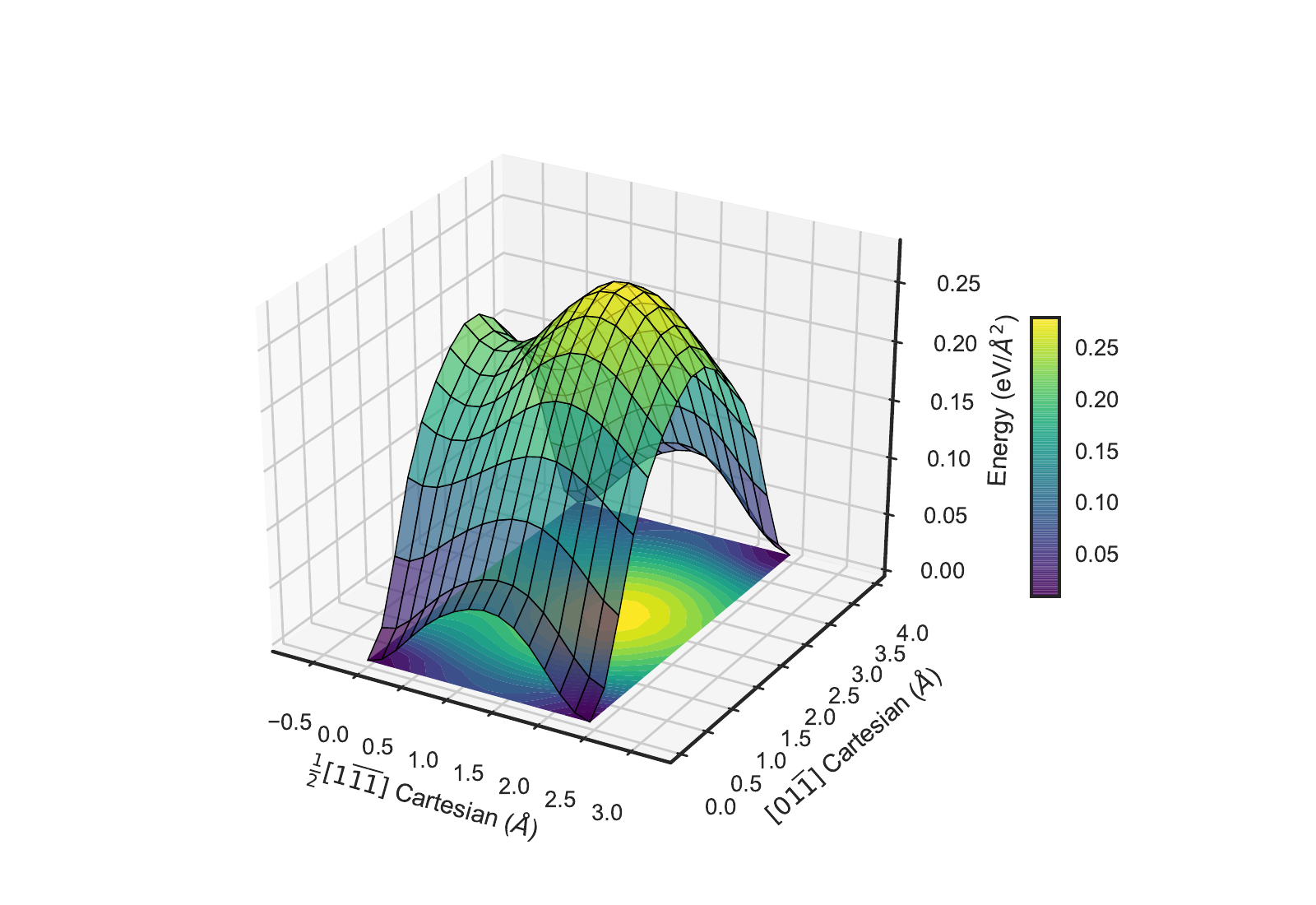}
  \includegraphics[trim=6mm 0mm 0mm 2mm, clip, width=0.45\textwidth]{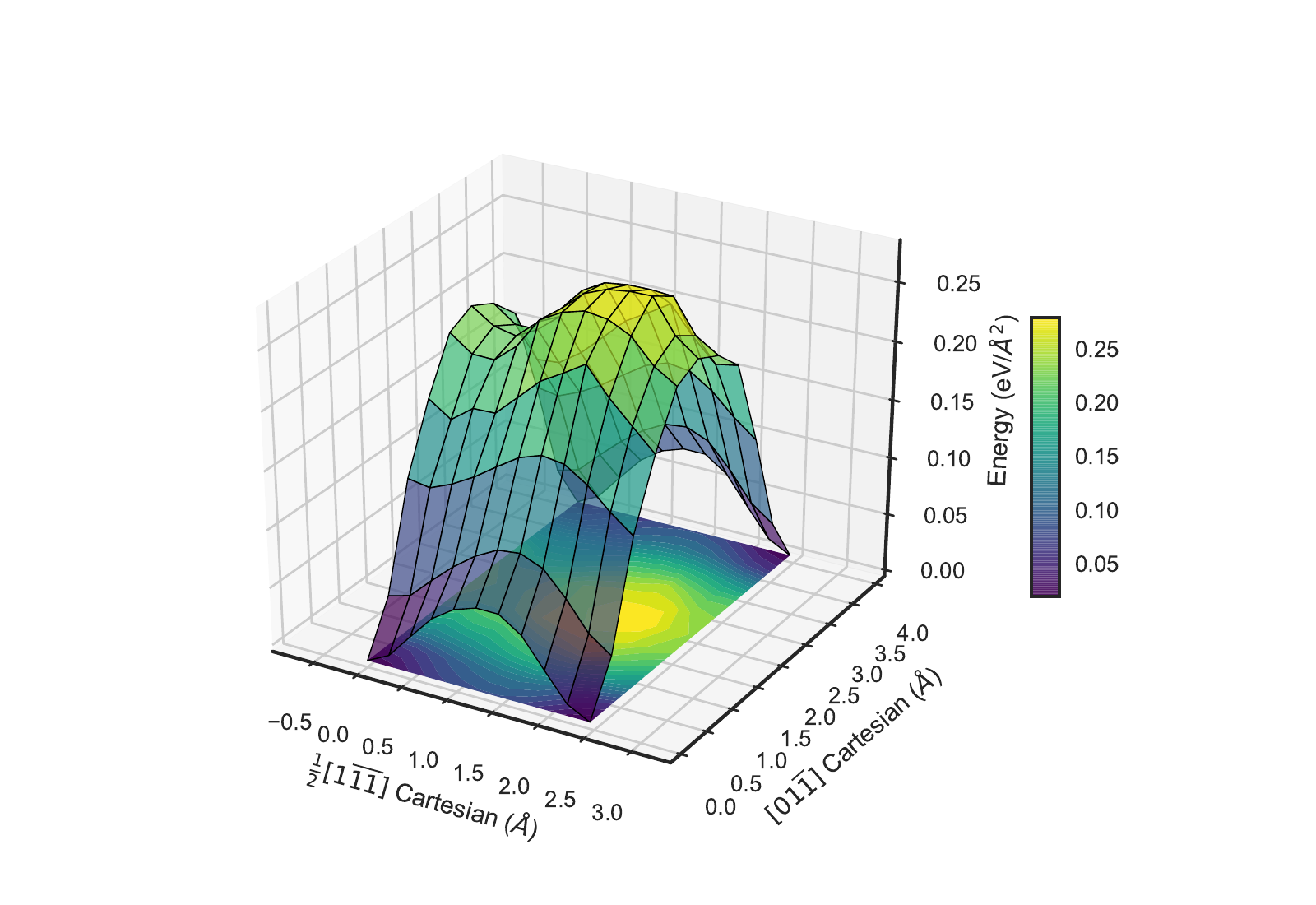}
\end{subfigure}
    \caption{Gamma surfaces computed in GAP (left column) and DFT (right column) for the \{110\} (top row) and \{211\} (bottom row) crystallographic orientation with atomic relaxation in the direction perpendicular to the surface.}
\label{fig:Gamma_surf}
\end{figure*}

We shall now examine the ability of the GAP potential to reproduce gamma surfaces, as introduced by Vitek~\cite{vitek}. Gamma surfaces, or generalized stacking faults, are two dimensional functions describing the energy change due to a relative displacement of two halves of a crystal with respect to each other across a glide plane. Such surfaces are obtained by computing the energy associated to all possible relative shear displacement vectors spanning a given crystallographic plane. However, due to the crystal periodicity, the displacement vectors that need to be considered to fully characterize any gamma surface are bound by the lattice vectors of the crystallographic plane under consideration. Gamma surfaces provide a way for finding potential stacking faults in metals by looking at local minima in the computed energy landscape and their details have direct impact on the structure of screw dislocations core. Here we restrict our analysis to the \{110\} and \{112\} crystallographic planes which, due to their dense packing, are the most important slip planes in bcc metals. Gamma surface calculations are performed using slanted cells of 12 atoms with the long direction oriented perpendicular to the gamma surface. The crystal cell is distorted, without moving the atoms, in a grid of displacements in the $[1\bar{1}0]$ and $[001]$ directions for the \{110\} gamma surface, and $[1\bar{1}\bar{1}]$ and $[0\bar{1}1]$ directions for the \{112\} gamma surface. Atoms are relaxed in the direction normal to the glide plane before evaluation of the total energy. GAP results are reported on the right column of Fig.~\ref{fig:Gamma_surf} and appear in good agreement with DFT data reported on the left column of the same figure.

\begin{figure}
\centering
\begin{subfigure}
\centering
  \textbf{\; \; \; \;  GAP \; \; \; \; \; \; \; \; \; \; \; \; \; \; \; \; \; \; \; \; DFT \; \; \; \;  \; \; \; \; \; \; \; \; \; \; \; \; EAM }(\textit{Mendelev}) \par\medskip
  \includegraphics[trim=0mm 0mm 0mm 0mm, clip, width=0.3\textwidth]{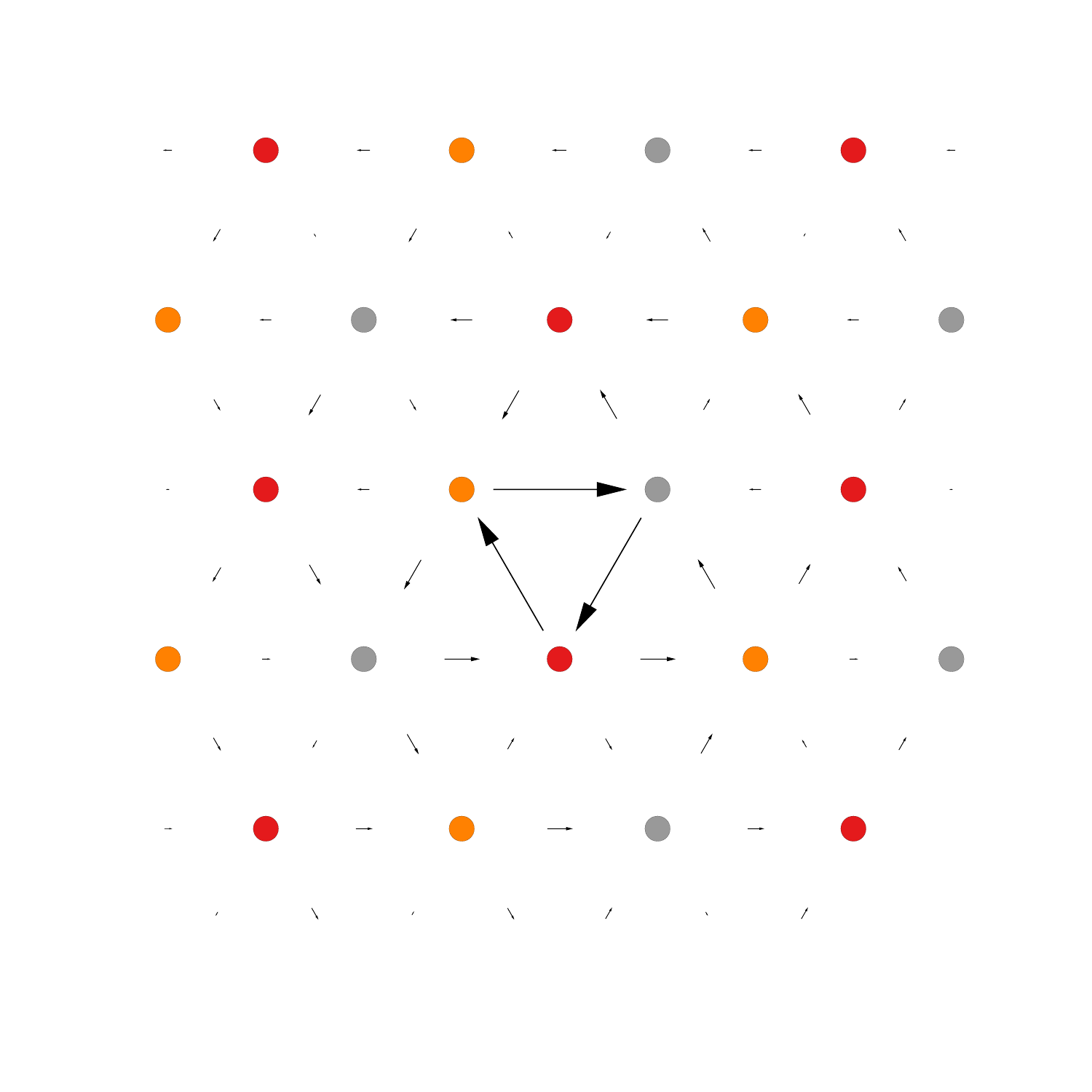} 
  \includegraphics[trim=0mm 0mm 0mm 0mm, clip, width=0.3\textwidth]{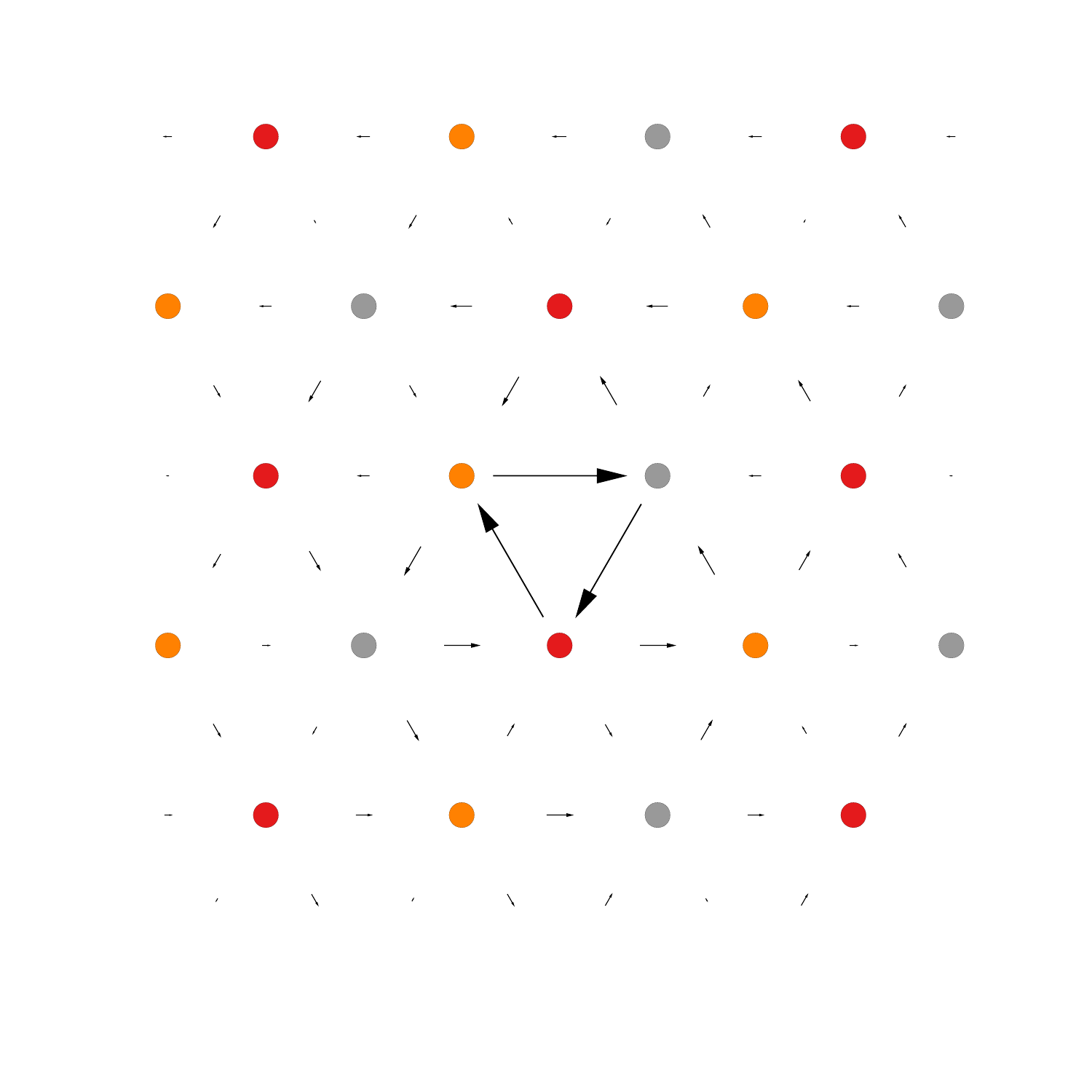}
  \includegraphics[trim=0mm 0mm 0mm 0mm, clip, width=0.3\textwidth]{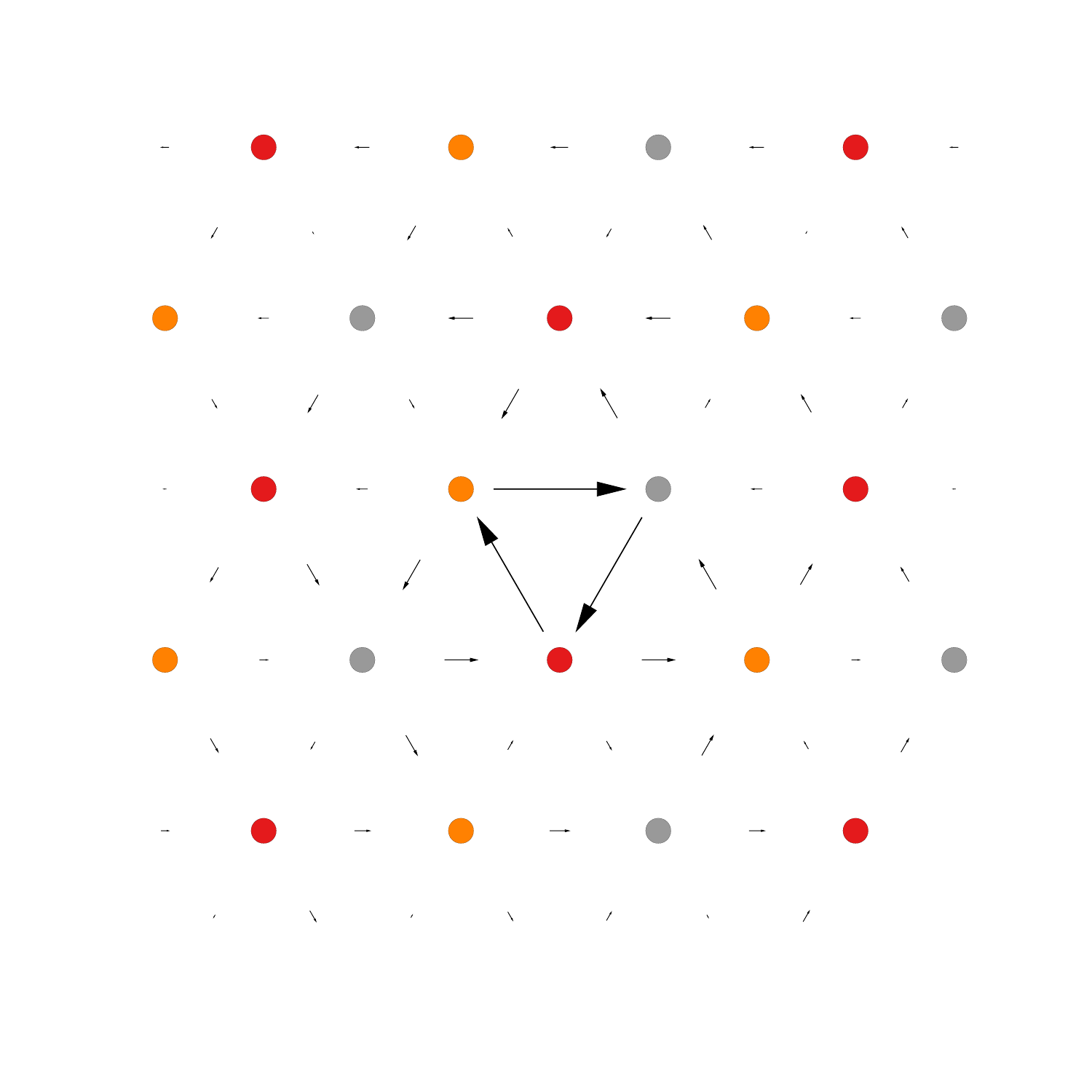}
\end{subfigure}
\begin{subfigure}
\centering
  \includegraphics[trim=0mm 0mm 0mm 0mm, clip, width=0.3\textwidth]{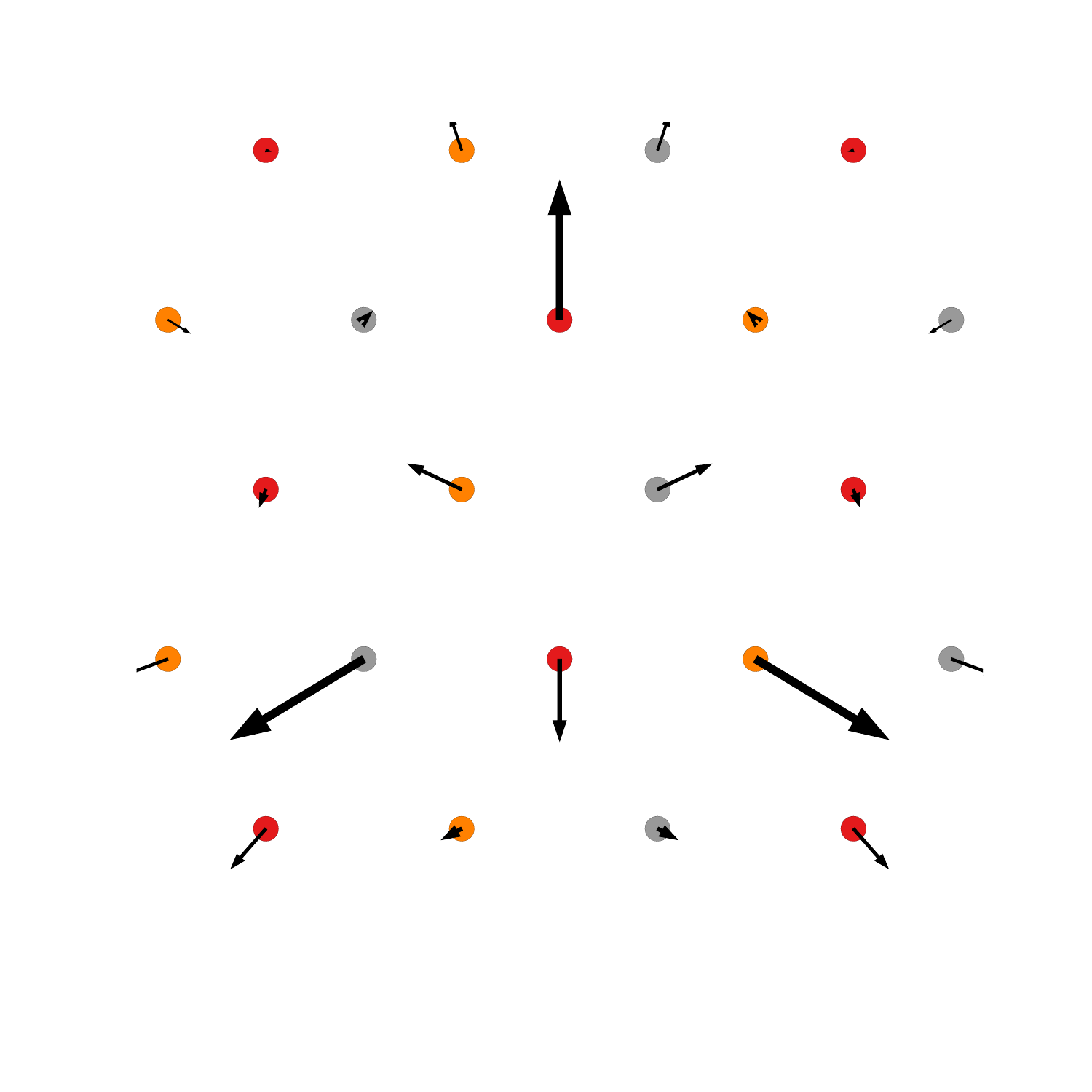} 
  \includegraphics[trim=0mm 0mm 0mm 0mm, clip, width=0.3\textwidth]{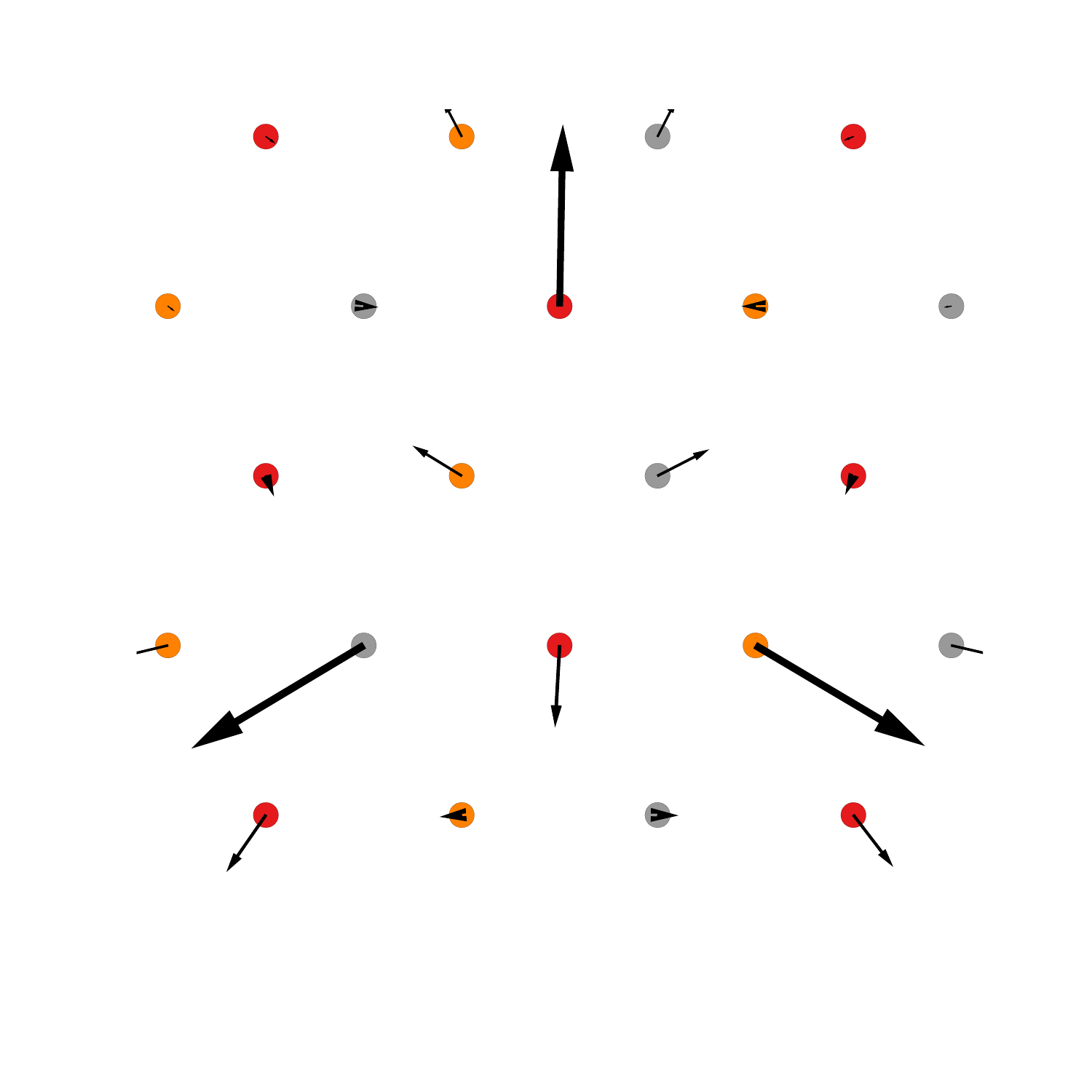}
  \includegraphics[trim=0mm 0mm 0mm 0mm, clip, width=0.3\textwidth]{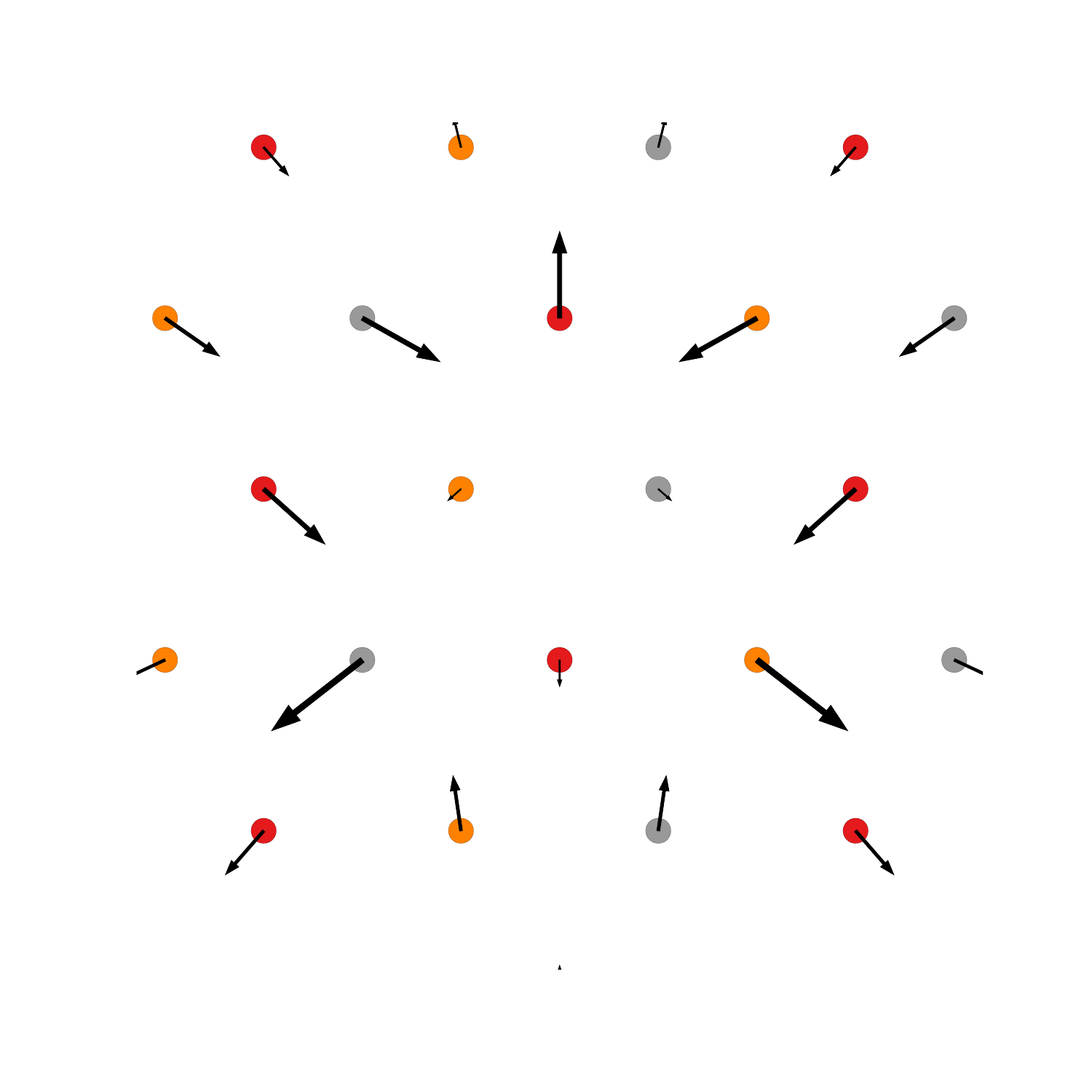}
\end{subfigure}
    \caption{Differential displacement maps of the screw dislocation core structure obtained with GAP (left column), DFT (centre column) and Mendelev potential (right column). The compact, non-degenerate core structure satisfying D3 point-group symmetry is consistent with earlier first-principles findings~\cite{frederiksen2003density,domain2005simulation,ventelon2007core}.
  Circles of different colors represent atoms belonging to different parallel planes with the ABCABC stacking sequence of the $\langle111\rangle$ zone before introduction of the dislocations. The top row shows out-of-plane screw displacements and the bottom row shows the in-plane edge displacements (magnified 20 times).}
  \label{fig:Core_structure_map}
\end{figure}

We proceed further with our validation process by assessing the Peierls energy barriers for a $1/2 \langle 111 \rangle$ screw dislocation gliding along any of the equivalent $\langle 112 \rangle$ directions.  As a first step we determine the stable structure for the dislocation core predicted by the potential. For the simulation of dislocations with PBCs, quadrupolar arrangements of easy-core $\frac{1}{2}\langle 111\rangle$ are created by making a dislocation dipole in a slanted cell, which would be equivalent to a square arrangement of dislocations in a square cell~\cite{itakura2012first,ventelon2007core}. The simulation cell lattice parameters used are:
\begin{align}
 \vec{a} &= N_x \vec{v}_{11\bar{2}} \\
 \vec{b} &= \frac{N_x}{2}\vec{v}_{11\bar{2}} + N_y\vec{v}_{1\bar{1}0} + \frac{1}{2}\vec{v}_{111} \\
 \vec{c} &= \vec{v}_{111} 
\end{align}
which is equivalent to half a cell of $N_x\vec{v}_{11\bar{2}} \times N_y\vec{v}_{1\bar{1}0}$, where $\vec{v}$ are the directions in the bulk lattice, and the integer values of $N_x$ and $N_y$ are chosen to make the arrangement of dislocations as close to square as possible. Two cells are used in this study, one with 135 atoms ($5 \times 9$), for which DFT calculations can also be performed for validation, where dislocations are separated by \SI{\sim 17}{\angstrom}, and a larger cell containing 2330 atoms ($21 \times 37$) where dislocations are separated by \SI{\sim 70}{\angstrom}. Atoms are displaced in the $z$ direction according to linear elastic theory around the dislocation core positions.
All atomic positions and the cell vectors are allowed to relax. The differential displacement map reported in Fig.~\ref{fig:Core_structure_map} shows the screw components of the screw dislocation core structure (out of plane displacements~\cite{vitek2004core}) computed with GAP\@.
Results are in agreement with DFT~\cite{frederiksen2003density,domain2005simulation,ventelon2007core} having a non-degenerate compact core structure with a D3 point-group symmetry. Separate plots of the in-plane edge components (magnified 20 times) show that GAP more closely matches the structure obtained with DFT than with the Mendelev potential.

\begin{figure}
\centering
  \includegraphics[trim=0mm 0mm 0mm 0mm, clip, width=0.45\textwidth]{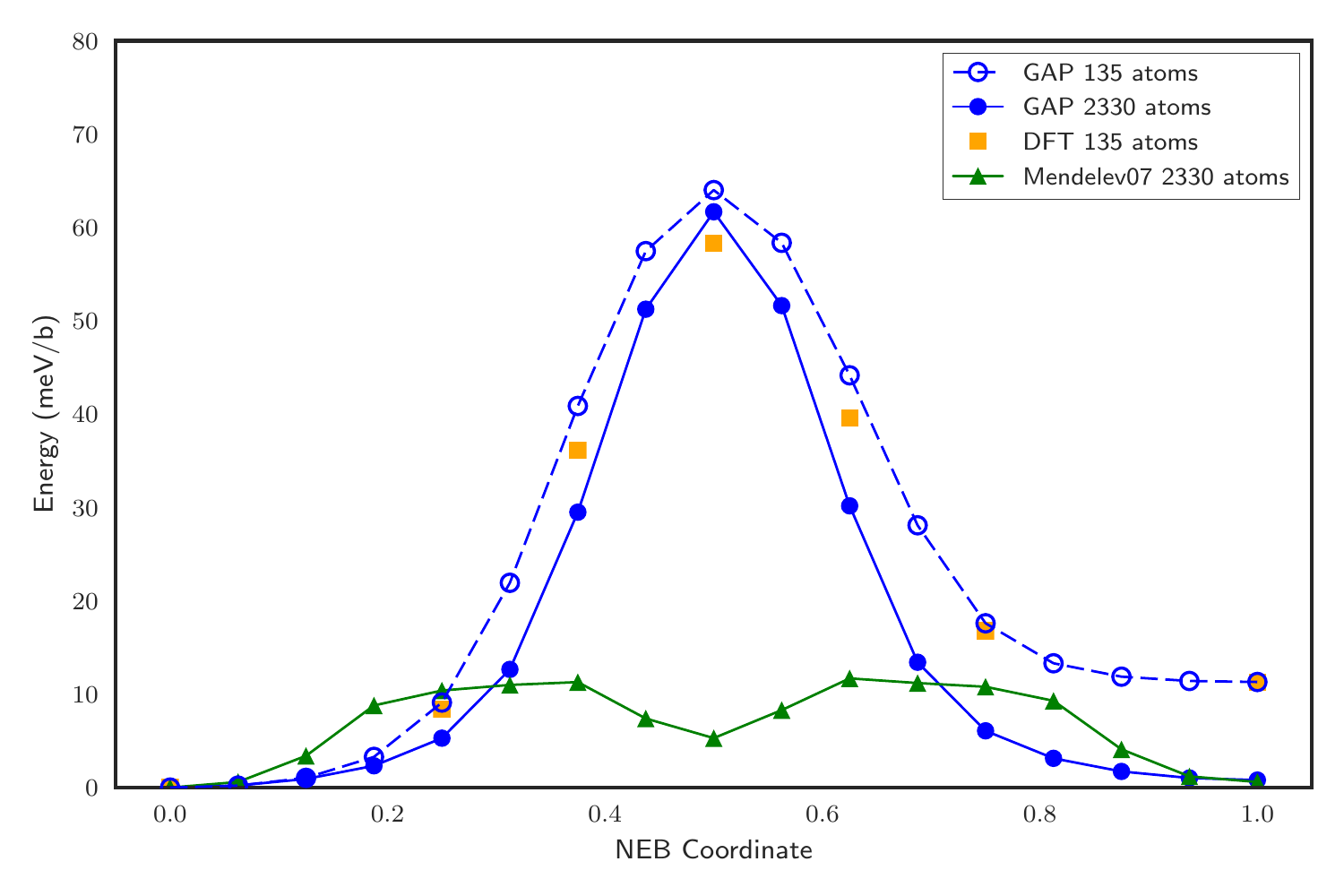} 
  \caption{Peierls energy barrier of a $1/2 \langle 111 \rangle$ screw dislocation gliding in the $\langle 112 \rangle$ direction.
  GAP data are computed using two different cells containing 135 and 2330 atoms with two dislocations in a quadrupolar arrangement. 
  DFT energies are computed in this work for the structures obtained from the 135 atoms GAP NEB pathway.
  The Mendelev semi-empirical EAM potential curve is shown for comparison using a 2330 atom cell.
 }
  \label{fig:Peierls_barrier}
\end{figure}

The Peierls barrier is calculated by performing a NEB calculation with climbing images\cite{NEB,bahn2002object,kolsbjerg2016automated} between the initial configuration and with one dislocation moved by $\vec{v}_{11\bar{2}}$. The Peierls plot for the GAP potential shows a single saddle point in qualitative accordance with earlier DFT findings~\cite{ventelon2007core}, whereas the Mendelev pathway has a double hump due to an incorrectly stabilised split-core structure~\cite{Ventelon2013}. The asymmetry in the barrier plot of Fig.~\ref{fig:Peierls_barrier} is due to moving only one of the dislocations in the cell so the final configuration deviates from an exact square quadrupole in the final arrangement. This finite-size effect vanishes for sufficiently large simulation boxes. The DFT reference energies for the Peierls barrier are generated by recalculating the structures obtained from the 135 atom GAP NEB with DFT (for computational efficiency reasons). The value of the Peierls barrier, \SI{64}{\meV\per\b} (where b is the Burgers vector), is in good agreement with our DFT calculations. The largest deviation from DFT calculated forces is for atoms in the dislocation core at the saddle point, and does not exceed \SI{0.1}{\eV\per\angstrom}. Although the Peierls barrier seems high in comparison to barriers of $\approx$\SI{40}{\meV\per\b} found in the literature~\cite{ventelon2007core,Dezerald2014AbMetals}, the barrier itself is shown to vary by \SIrange{10}{20}{\meV\per\b} for different DFT methods~\cite{Ventelon2013}, and may also be sensitive to the method used to find the transition state, so we only make quantitative comparisons with our own calculations.

\onecolumngrid

\onecolumngrid

\twocolumngrid

%\begin{turnpage}{angle=90}
\begin{table*}[h]
\centering
\setlength{\tabcolsep}{12pt}
\renewcommand{\arraystretch}{1.2}
\begin{tabular}{@{}l c c c c @{}}
\toprule
                                    &GAP   &   DFT     & Other DFT calcs.                                     & Expt.   \\
                                    &      &{\scriptsize (this work)}&   &  \\
\colrule
$E^{v}_f$ [eV]                      & 2.26 &  2.22    & 2.10~\cite{tateyama2003HFe}                          &  1.6, 2.2~\cite{monovac_posannih1,de1983positron}   \\ 
$E^{v}_{m1NN}$                      & 0.67 &   -      & 0.64~\cite{divacancy_DFT1},0.67~\cite{nature_chuchun}&  -      \\
$E^{v}_{m2NN}$                      & 2.75 &   -      &  -                                                   &  -      \\
$E^{v}_{m3NN}$                      & 5.63 &   -      &  -                                                   &  -      \\
$E_f^{1NNv}$                        & 4.41 &   4.24   & 4.01~\cite{tateyama2003HFe}                          &  -      \\
$E_f^{2NNv}$                        & 4.30 &   4.20   & 3.95~\cite{tateyama2003HFe}                          &  -      \\
$E_f^{3NNv}$                        & 4.55 &   4.45   & -                                                    &  -      \\
$E_f^{4NNv}$                        & 4.48 &   -      & -                                                    &  -      \\
$E_f^{5NNv}$                        & 4.47 &   -      & -                                                    &  -      \\
$E_b^{1NNv}$                        & 0.11 &   0.20   & 0.14, 0.08,0.16~\cite{divacancy_DFT1,tateyama2003HFe,malerba2010}     &  -      \\
$E_b^{2NNv}$                        & 0.22 &   0.24   & 0.28, 0.15,$\sim$0.2~\cite{divacancy_DFT1,tateyama2003HFe,christian2002theory,dudarev2013density}  &  -      \\
$E_b^{3NNv}$                        &-0.03 &  -0.01   & -0.02~\cite{divacancy_DFT1},-0.015~\cite{malerba2010}&  -      \\
$E_b^{4NNv}$                        & 0.04 &   -      & 0.05~\cite{malerba2010}                              &  0.09~\cite{malerba2010}      \\
$E_b^{5NNv}$                        & 0.05 &   -      & 0.06~\cite{malerba2010}                              &  0.06~\cite{malerba2010}      \\ 
$E_f^{[112]v}$                      & 6.19 &   -  &$E_f^{[112]v}< E_f^{[226]v}\;$ \cite{hayward2013interplay}  &  -      \\
$E_f^{[226]v}$                      & 6.38 &   -  &$E_f^{[226]v}< E_f^{[223]v}\;$ \cite{hayward2013interplay}  &  -      \\
$E_f^{[223]v}$                      & 6.35 &   -  &$E_f^{[223]v}< E_f^{[115]v}\;$ \cite{hayward2013interplay}  &  -      \\
$E_f^{[115]v}$                      & 6.47 &   -  & -                         &  -      \\
$E_f^{[113]v}$                      & 6.59 &   -  & -                         &  -      \\
$E_f^{[333]v}$                      & 6.85 &   -  & -                         &  -      \\
$E_f^{[339]v}$                      & 6.85 &   -  & -                         &  -      \\
$E_b^{[112]v}$                      & 0.60 &   -  & -                         &  -      \\
$E_b^{[226]v}$                      & 0.41 &   -  & -                         &  -      \\
$E_b^{[223]v}$                      & 0.44 &   -  & -                         &  -      \\
$E_b^{[115]v}$                      & 0.32 &   -  & -                         &  -      \\
$E_b^{[113]v}$                      & 0.20 &   -  & -                         &  -      \\
$E_b^{[333]v}$                      &-0.07 &   -  & -                         &  -      \\
$E_b^{[339]v}$                      &-0.06 &   -  & -                         &  -      \\
$E^{SIA}_{f_{\langle 110\rangle}}$  & 4.21 &   4.37   & 3.93~\cite{Dudarev_Derlet}            &  -      \\
$E^{SIA}_{f_{\langle 111\rangle}}$  & 4.90 &   5.13   & 4.64~\cite{Dudarev_Derlet}            &  -      \\
$E^{SIA}_{f_{\langle 100\rangle}}$  & 5.47 &   5.48   & 5.05~\cite{Dudarev_Derlet}            &  -      \\
$E^{SIA}_{f_{tet}}$                 & 4.75 &   4.79   & 4.32~\cite{Dudarev_Derlet}            &  -      \\
$E^{SIA}_{f_{oct}}$                 & 5.53 &   5.58   & 5.21~\cite{Dudarev_Derlet}            &  -      \\
$E_{m_{\langle 110\rangle}}^{jump}$ & 0.31 &   -      & 0.30, 0.34~\cite{malerba2010,Fu_SIA}  &  -      \\ 
$E^{diSIA}_{NPC}$                   & 7.54 &   7.84   & 7.04~\cite{NPC-diSIA}                 &  -      \\
$E^{diSIA}_{110}$                   & 8.36 &   8.95   & 7.15~\cite{NPC-diSIA}                 &  -      \\
\colrule
$E^{110}$ [J/m$^2$]                 & 2.499&   2.495  & 2.37, 2.25, 2.27~\cite{blonski2007,blonski2004,spencer2002} &  -      \\   
$E^{100}$                           & 2.547&   2.543  & 2.47, 2.25, 2.29~\cite{blonski2007,blonski2004,spencer2002} &  -      \\ 
$E^{211}$                           & 2.612&   2.629  & 2.50~\cite{blonski2007}                                     &  -      \\
$E^{111}$                           & 2.756&   2.752  & 2.58, 2.54, 2.52~\cite{blonski2007,blonski2004}             &  -      \\
\botrule
\end{tabular}
\caption{Formation and binding energies of defected configurations from GAP\@. Results are obtained with fully relaxed cells except for NEB calculations which are performed at the equilibrium (electronic DFT) volume and compared to DFT (from this work and from the literature) and, when possible, to experimental data at 0~K. }
\label{tab:energetics_gp}
\end{table*}

\section{Remarks and Conclusions}
\label{sec:conclusions}
We have generated a Gaussian Approximation Potential for $\alpha$-iron by training on DFT total energies, forces and stresses for approximately $150$k local atomic environments. The GAP model is presented and validated against DFT data not included in the training protocol, either computed in this work or taken from the literature. Results show that the new model is able to reproduce DFT energetics and thermodynamics with great accuracy, including energetics of point defects such as mono-, di-, and tri-vacancies and of self-interstitials and di-interstitials. Notably, the potential is able to reproduce a positive 5-$nn$ di-vacancy binding energy and the correct ordering of binding energies for the non-parallel and parallel $\langle110\rangle$ di-interstitials, rectifying some of the weaknesses displayed by the EAM interatomic potentials available in the literature~\cite{malerba2010}. Selected generalized stacking-faults and formation energy of selected free surfaces are also reproduced from a qualitative and quantitative point of view. The compact, non-degenerate core structure of the 1/2$\langle111\rangle$ screw dislocation and the associated Peierls energy barrier are also consistent with DFT.
In order to achieve such accuracy we found it essential to use first-principles data with a high degree of convergence to the DFT Born-Oppenheimer PES, in particular the $k$-point sampling needs to be high because supercells of different sizes cannot have a congruent Brillouin sampling, and the plane wave cutoff is high enough that energies, forces and virials are all converged.

We stress that, as pointed out in previous works, the model is built to interpolate between known atomic environments but does not extrapolate to completely new configurations.  Caution is therefore always suggested when dealing with such cases. In this work we have tried to ensure transferability of the model by creating an extended  training database that provides a good coverage of environments across the thermodynamic range of stability of the $\alpha$-phase of iron. Such a database can be further extended in a modular way to include new environments which are relevant to a specific line of research. To this end we have pointed out all the details of the data generation protocol needed to preserve the accuracy of the current database.

The DFT data used for the training are always performed in a collinear spin-polarized approximation starting from a ferromagnetic ordering. As such, the model can only reproduce reliable thermomechanical properties up to two-thirds of the Curie point, while the high-temperature paramagnetic behavior governed by magnetic disorder cannot be correctly captured. 
In order to study high-temperature bcc phases of iron, one needs to train the paramagnetic PES\@. In fact, however, accessing the paramagnetic PES with standard DFT calculations is a non-trivial task~\cite{phonon-magnon,neuge2016}. An alternative route to magnetism is to generalize the GAP formalism to treat magnetic degrees of freedom in a semi-classical way. This approach will possibly be a future direction of investigation.

The computational cost of GAP is higher than simple analytical models, at around 60~ms/atom/cpu-core, so at this stage, the method is too expensive to tackle multi-million atom and/or nano-second calculations using moderate computational resources. However, its linear scaling cost with respect to the number of processors, combined with its high accuracy, makes the methodology suitable to access intermediate time and size scales which are not accessible by first-principles. 
This paves the way for example to the use of interatomic potentials for studying thermodynamics of real materials with a reliability never achieved before.

The training database is freely available on the \textsf{Materials Cloud Archive}~\cite{mca} and at \url{www.libatoms.org}.
All software and data necessary for the reproduction of the results are freely available at \url{www.libatoms.org}.

\section{Acknowledgements}
The authors gratefully acknowledge the financial support from the Swiss National Science Foundation (SNSF – Project No. 200021–143636). 
This work was supported by a grant from the Swiss National Supercomputing Centre (CSCS) under project ID \textit{ch3}.
This work was supported by the Engineering and Physical Sciences Research Council (EPSRC) [Programme grant number EP/L014742/1].
This work used the ARCHER UK National Supercomputing Service (http://www.archer.ac.uk).

\bibliography{DRAFT_1}

\begin{thebibliography}{107}%
\makeatletter
\providecommand \@ifxundefined [1]{%
 \@ifx{#1\undefined}
}%
\providecommand \@ifnum [1]{%
 \ifnum #1\expandafter \@firstoftwo
 \else \expandafter \@secondoftwo
 \fi
}%
\providecommand \@ifx [1]{%
 \ifx #1\expandafter \@firstoftwo
 \else \expandafter \@secondoftwo
 \fi
}%
\providecommand \natexlab [1]{#1}%
\providecommand \enquote  [1]{``#1''}%
\providecommand \bibnamefont  [1]{#1}%
\providecommand \bibfnamefont [1]{#1}%
\providecommand \citenamefont [1]{#1}%
\providecommand \href@noop [0]{\@secondoftwo}%
\providecommand \href [0]{\begingroup \@sanitize@url \@href}%
\providecommand \@href[1]{\@@startlink{#1}\@@href}%
\providecommand \@@href[1]{\endgroup#1\@@endlink}%
\providecommand \@sanitize@url [0]{\catcode `\\12\catcode `\$12\catcode
  `\&12\catcode `\#12\catcode `\^12\catcode `\_12\catcode `\%12\relax}%
\providecommand \@@startlink[1]{}%
\providecommand \@@endlink[0]{}%
\providecommand \url  [0]{\begingroup\@sanitize@url \@url }%
\providecommand \@url [1]{\endgroup\@href {#1}{\urlprefix }}%
\providecommand \urlprefix  [0]{URL }%
\providecommand \Eprint [0]{\href }%
\providecommand \doibase [0]{http://dx.doi.org/}%
\providecommand \selectlanguage [0]{\@gobble}%
\providecommand \bibinfo  [0]{\@secondoftwo}%
\providecommand \bibfield  [0]{\@secondoftwo}%
\providecommand \translation [1]{[#1]}%
\providecommand \BibitemOpen [0]{}%
\providecommand \bibitemStop [0]{}%
\providecommand \bibitemNoStop [0]{.\EOS\space}%
\providecommand \EOS [0]{\spacefactor3000\relax}%
\providecommand \BibitemShut  [1]{\csname bibitem#1\endcsname}%
\let\auto@bib@innerbib\@empty
%</preamble>
\bibitem [{\citenamefont {Moroni}\ \emph {et~al.}(1997)\citenamefont {Moroni},
  \citenamefont {Kresse}, \citenamefont {Hafner},\ and\ \citenamefont
  {Furthm{\"u}ller}}]{moroni1997ultrasoft}%
  \BibitemOpen
  \bibfield  {author} {\bibinfo {author} {\bibfnamefont {E.}~\bibnamefont
  {Moroni}}, \bibinfo {author} {\bibfnamefont {G.}~\bibnamefont {Kresse}},
  \bibinfo {author} {\bibfnamefont {J.}~\bibnamefont {Hafner}}, \ and\ \bibinfo
  {author} {\bibfnamefont {J.}~\bibnamefont {Furthm{\"u}ller}},\ }\href@noop {}
  {\bibfield  {journal} {\bibinfo  {journal} {Physical Review B}\ }\textbf
  {\bibinfo {volume} {56}},\ \bibinfo {pages} {15629} (\bibinfo {year}
  {1997})}\BibitemShut {NoStop}%
\bibitem [{\citenamefont {Alf}\ \emph {et~al.}(1999)\citenamefont {Alf},
  \citenamefont {Gillan},\ and\ \citenamefont {Price}}]{alf1999melting}%
  \BibitemOpen
  \bibfield  {author} {\bibinfo {author} {\bibfnamefont {D.}~\bibnamefont
  {Alf}}, \bibinfo {author} {\bibfnamefont {M.}~\bibnamefont {Gillan}}, \ and\
  \bibinfo {author} {\bibfnamefont {G.}~\bibnamefont {Price}},\ }\href@noop {}
  {\bibfield  {journal} {\bibinfo  {journal} {Nature}\ }\textbf {\bibinfo
  {volume} {401}},\ \bibinfo {pages} {462} (\bibinfo {year}
  {1999})}\BibitemShut {NoStop}%
\bibitem [{\citenamefont {Dal~Corso}\ and\ \citenamefont
  {de~Gironcoli}(2000)}]{DalCorsodeGir}%
  \BibitemOpen
  \bibfield  {author} {\bibinfo {author} {\bibfnamefont {A.}~\bibnamefont
  {Dal~Corso}}\ and\ \bibinfo {author} {\bibfnamefont {S.}~\bibnamefont
  {de~Gironcoli}},\ }\href {\doibase 10.1103/PhysRevB.62.273} {\bibfield
  {journal} {\bibinfo  {journal} {Phys. Rev. B}\ }\textbf {\bibinfo {volume}
  {62}},\ \bibinfo {pages} {273} (\bibinfo {year} {2000})}\BibitemShut
  {NoStop}%
\bibitem [{\citenamefont {Laio}\ \emph {et~al.}(2000)\citenamefont {Laio},
  \citenamefont {Bernard}, \citenamefont {Chiarotti}, \citenamefont
  {Scandolo},\ and\ \citenamefont {Tosatti}}]{laio2000physics}%
  \BibitemOpen
  \bibfield  {author} {\bibinfo {author} {\bibfnamefont {A.}~\bibnamefont
  {Laio}}, \bibinfo {author} {\bibfnamefont {S.}~\bibnamefont {Bernard}},
  \bibinfo {author} {\bibfnamefont {G.}~\bibnamefont {Chiarotti}}, \bibinfo
  {author} {\bibfnamefont {S.}~\bibnamefont {Scandolo}}, \ and\ \bibinfo
  {author} {\bibfnamefont {E.}~\bibnamefont {Tosatti}},\ }\href@noop {}
  {\bibfield  {journal} {\bibinfo  {journal} {Science}\ }\textbf {\bibinfo
  {volume} {287}},\ \bibinfo {pages} {1027} (\bibinfo {year}
  {2000})}\BibitemShut {NoStop}%
\bibitem [{\citenamefont {Hatt}\ \emph {et~al.}(2010)\citenamefont {Hatt},
  \citenamefont {Melot},\ and\ \citenamefont {Narasimhan}}]{shobhana}%
  \BibitemOpen
  \bibfield  {author} {\bibinfo {author} {\bibfnamefont {A.~J.}\ \bibnamefont
  {Hatt}}, \bibinfo {author} {\bibfnamefont {B.~C.}\ \bibnamefont {Melot}}, \
  and\ \bibinfo {author} {\bibfnamefont {S.}~\bibnamefont {Narasimhan}},\
  }\href {\doibase 10.1103/PhysRevB.82.134418} {\bibfield  {journal} {\bibinfo
  {journal} {Phys. Rev. B}\ }\textbf {\bibinfo {volume} {82}},\ \bibinfo
  {pages} {134418} (\bibinfo {year} {2010})}\BibitemShut {NoStop}%
\bibitem [{\citenamefont {Pozzo}\ \emph {et~al.}(2012)\citenamefont {Pozzo},
  \citenamefont {Davies}, \citenamefont {Gubbins},\ and\ \citenamefont
  {Alfe}}]{pozzo2012thermal}%
  \BibitemOpen
  \bibfield  {author} {\bibinfo {author} {\bibfnamefont {M.}~\bibnamefont
  {Pozzo}}, \bibinfo {author} {\bibfnamefont {C.}~\bibnamefont {Davies}},
  \bibinfo {author} {\bibfnamefont {D.}~\bibnamefont {Gubbins}}, \ and\
  \bibinfo {author} {\bibfnamefont {D.}~\bibnamefont {Alfe}},\ }\href@noop {}
  {\bibfield  {journal} {\bibinfo  {journal} {Nature}\ }\textbf {\bibinfo
  {volume} {485}},\ \bibinfo {pages} {355} (\bibinfo {year}
  {2012})}\BibitemShut {NoStop}%
\bibitem [{\citenamefont {Dragoni}\ \emph {et~al.}(2015)\citenamefont
  {Dragoni}, \citenamefont {Ceresoli},\ and\ \citenamefont
  {Marzari}}]{Dragoni}%
  \BibitemOpen
  \bibfield  {author} {\bibinfo {author} {\bibfnamefont {D.}~\bibnamefont
  {Dragoni}}, \bibinfo {author} {\bibfnamefont {D.}~\bibnamefont {Ceresoli}}, \
  and\ \bibinfo {author} {\bibfnamefont {N.}~\bibnamefont {Marzari}},\ }\href
  {\doibase 10.1103/PhysRevB.91.104105} {\bibfield  {journal} {\bibinfo
  {journal} {Phys. Rev. B}\ }\textbf {\bibinfo {volume} {91}},\ \bibinfo
  {pages} {104105} (\bibinfo {year} {2015})}\BibitemShut {NoStop}%
\bibitem [{\citenamefont {K\"ormann}\ \emph {et~al.}(2008)\citenamefont
  {K\"ormann}, \citenamefont {Dick}, \citenamefont {Grabowski}, \citenamefont
  {Hallstedt}, \citenamefont {Hickel},\ and\ \citenamefont
  {Neugebauer}}]{Koermann}%
  \BibitemOpen
  \bibfield  {author} {\bibinfo {author} {\bibfnamefont {F.}~\bibnamefont
  {K\"ormann}}, \bibinfo {author} {\bibfnamefont {A.}~\bibnamefont {Dick}},
  \bibinfo {author} {\bibfnamefont {B.}~\bibnamefont {Grabowski}}, \bibinfo
  {author} {\bibfnamefont {B.}~\bibnamefont {Hallstedt}}, \bibinfo {author}
  {\bibfnamefont {T.}~\bibnamefont {Hickel}}, \ and\ \bibinfo {author}
  {\bibfnamefont {J.}~\bibnamefont {Neugebauer}},\ }\href {\doibase
  10.1103/PhysRevB.78.033102} {\bibfield  {journal} {\bibinfo  {journal} {Phys.
  Rev. B}\ }\textbf {\bibinfo {volume} {78}},\ \bibinfo {pages} {033102}
  (\bibinfo {year} {2008})}\BibitemShut {NoStop}%
\bibitem [{\citenamefont {K\"ormann}\ \emph {et~al.}(2012)\citenamefont
  {K\"ormann}, \citenamefont {Dick}, \citenamefont {Grabowski}, \citenamefont
  {Hickel},\ and\ \citenamefont {Neugebauer}}]{Neugebauer-phonon}%
  \BibitemOpen
  \bibfield  {author} {\bibinfo {author} {\bibfnamefont {F.}~\bibnamefont
  {K\"ormann}}, \bibinfo {author} {\bibfnamefont {A.}~\bibnamefont {Dick}},
  \bibinfo {author} {\bibfnamefont {B.}~\bibnamefont {Grabowski}}, \bibinfo
  {author} {\bibfnamefont {T.}~\bibnamefont {Hickel}}, \ and\ \bibinfo {author}
  {\bibfnamefont {J.}~\bibnamefont {Neugebauer}},\ }\href {\doibase
  10.1103/PhysRevB.85.125104} {\bibfield  {journal} {\bibinfo  {journal} {Phys.
  Rev. B}\ }\textbf {\bibinfo {volume} {85}},\ \bibinfo {pages} {125104}
  (\bibinfo {year} {2012})}\BibitemShut {NoStop}%
\bibitem [{\citenamefont {K\"ormann}\ \emph {et~al.}(2014)\citenamefont
  {K\"ormann}, \citenamefont {Grabowski}, \citenamefont {Dutta}, \citenamefont
  {Hickel}, \citenamefont {Mauger}, \citenamefont {Fultz},\ and\ \citenamefont
  {Neugebauer}}]{phonon-magnon}%
  \BibitemOpen
  \bibfield  {author} {\bibinfo {author} {\bibfnamefont {F.}~\bibnamefont
  {K\"ormann}}, \bibinfo {author} {\bibfnamefont {B.}~\bibnamefont
  {Grabowski}}, \bibinfo {author} {\bibfnamefont {B.}~\bibnamefont {Dutta}},
  \bibinfo {author} {\bibfnamefont {T.}~\bibnamefont {Hickel}}, \bibinfo
  {author} {\bibfnamefont {L.}~\bibnamefont {Mauger}}, \bibinfo {author}
  {\bibfnamefont {B.}~\bibnamefont {Fultz}}, \ and\ \bibinfo {author}
  {\bibfnamefont {J.}~\bibnamefont {Neugebauer}},\ }\href {\doibase
  10.1103/PhysRevLett.113.165503} {\bibfield  {journal} {\bibinfo  {journal}
  {Phys. Rev. Lett.}\ }\textbf {\bibinfo {volume} {113}},\ \bibinfo {pages}
  {165503} (\bibinfo {year} {2014})}\BibitemShut {NoStop}%
\bibitem [{\citenamefont {Sha}\ and\ \citenamefont {Cohen}(2006)}]{sha}%
  \BibitemOpen
  \bibfield  {author} {\bibinfo {author} {\bibfnamefont {X.}~\bibnamefont
  {Sha}}\ and\ \bibinfo {author} {\bibfnamefont {R.~E.}\ \bibnamefont
  {Cohen}},\ }\href {\doibase 10.1103/PhysRevB.73.104303} {\bibfield  {journal}
  {\bibinfo  {journal} {Phys. Rev. B}\ }\textbf {\bibinfo {volume} {73}},\
  \bibinfo {pages} {104303} (\bibinfo {year} {2006})}\BibitemShut {NoStop}%
\bibitem [{\citenamefont {Daw}\ \emph {et~al.}(1993)\citenamefont {Daw},
  \citenamefont {Foiles},\ and\ \citenamefont {Baskes}}]{EAM}%
  \BibitemOpen
  \bibfield  {author} {\bibinfo {author} {\bibfnamefont {M.~S.}\ \bibnamefont
  {Daw}}, \bibinfo {author} {\bibfnamefont {S.~M.}\ \bibnamefont {Foiles}}, \
  and\ \bibinfo {author} {\bibfnamefont {M.~I.}\ \bibnamefont {Baskes}},\
  }\href@noop {} {\bibfield  {journal} {\bibinfo  {journal} {Materials Science
  Reports}\ }\textbf {\bibinfo {volume} {9}},\ \bibinfo {pages} {251} (\bibinfo
  {year} {1993})}\BibitemShut {NoStop}%
\bibitem [{\citenamefont {Finnis}\ and\ \citenamefont {Sinclair}(1984)}]{FS}%
  \BibitemOpen
  \bibfield  {author} {\bibinfo {author} {\bibfnamefont {M.}~\bibnamefont
  {Finnis}}\ and\ \bibinfo {author} {\bibfnamefont {J.}~\bibnamefont
  {Sinclair}},\ }\href@noop {} {\bibfield  {journal} {\bibinfo  {journal}
  {Philosophical Magazine A}\ }\textbf {\bibinfo {volume} {50}},\ \bibinfo
  {pages} {45} (\bibinfo {year} {1984})}\BibitemShut {NoStop}%
\bibitem [{\citenamefont {Chen}\ \emph {et~al.}(1986)\citenamefont {Chen},
  \citenamefont {Voter},\ and\ \citenamefont {Srolovitz}}]{Srolovitz}%
  \BibitemOpen
  \bibfield  {author} {\bibinfo {author} {\bibfnamefont {S.~P.}\ \bibnamefont
  {Chen}}, \bibinfo {author} {\bibfnamefont {A.~F.}\ \bibnamefont {Voter}}, \
  and\ \bibinfo {author} {\bibfnamefont {D.~J.}\ \bibnamefont {Srolovitz}},\
  }\href {\doibase 10.1103/PhysRevLett.57.1308} {\bibfield  {journal} {\bibinfo
   {journal} {Phys. Rev. Lett.}\ }\textbf {\bibinfo {volume} {57}},\ \bibinfo
  {pages} {1308} (\bibinfo {year} {1986})}\BibitemShut {NoStop}%
\bibitem [{\citenamefont {Ercolessi}\ \emph {et~al.}(1986)\citenamefont
  {Ercolessi}, \citenamefont {Parrinello},\ and\ \citenamefont
  {Tosatti}}]{Glue}%
  \BibitemOpen
  \bibfield  {author} {\bibinfo {author} {\bibfnamefont {F.}~\bibnamefont
  {Ercolessi}}, \bibinfo {author} {\bibfnamefont {M.}~\bibnamefont
  {Parrinello}}, \ and\ \bibinfo {author} {\bibfnamefont {E.}~\bibnamefont
  {Tosatti}},\ }\href@noop {} {\bibfield  {journal} {\bibinfo  {journal}
  {Surface Science}\ }\textbf {\bibinfo {volume} {177}},\ \bibinfo {pages}
  {314} (\bibinfo {year} {1986})}\BibitemShut {NoStop}%
\bibitem [{\citenamefont {Mendelev}\ \emph {et~al.}(2003)\citenamefont
  {Mendelev}, \citenamefont {Han}, \citenamefont {Srolovitz}, \citenamefont
  {Ackland}, \citenamefont {Sun},\ and\ \citenamefont {Asta}}]{mendelev}%
  \BibitemOpen
  \bibfield  {author} {\bibinfo {author} {\bibfnamefont {M.~I.}\ \bibnamefont
  {Mendelev}}, \bibinfo {author} {\bibfnamefont {S.}~\bibnamefont {Han}},
  \bibinfo {author} {\bibfnamefont {D.~J.}\ \bibnamefont {Srolovitz}}, \bibinfo
  {author} {\bibfnamefont {G.~J.}\ \bibnamefont {Ackland}}, \bibinfo {author}
  {\bibfnamefont {D.~Y.}\ \bibnamefont {Sun}}, \ and\ \bibinfo {author}
  {\bibfnamefont {M.}~\bibnamefont {Asta}},\ }\href {\doibase
  10.1080/14786430310001613264} {\bibfield  {journal} {\bibinfo  {journal}
  {Philosophical Magazine}\ }\textbf {\bibinfo {volume} {83}},\ \bibinfo
  {pages} {3977} (\bibinfo {year} {2003})}\BibitemShut {NoStop}%
\bibitem [{\citenamefont {Ackland}\ \emph {et~al.}(2004)\citenamefont
  {Ackland}, \citenamefont {Mendelev}, \citenamefont {Srolovitz}, \citenamefont
  {Han},\ and\ \citenamefont {Barashev}}]{ackland2004development}%
  \BibitemOpen
  \bibfield  {author} {\bibinfo {author} {\bibfnamefont {G.}~\bibnamefont
  {Ackland}}, \bibinfo {author} {\bibfnamefont {M.}~\bibnamefont {Mendelev}},
  \bibinfo {author} {\bibfnamefont {D.}~\bibnamefont {Srolovitz}}, \bibinfo
  {author} {\bibfnamefont {S.}~\bibnamefont {Han}}, \ and\ \bibinfo {author}
  {\bibfnamefont {A.}~\bibnamefont {Barashev}},\ }\href@noop {} {\bibfield
  {journal} {\bibinfo  {journal} {Journal of Physics: Condensed Matter}\
  }\textbf {\bibinfo {volume} {16}},\ \bibinfo {pages} {S2629} (\bibinfo {year}
  {2004})}\BibitemShut {NoStop}%
\bibitem [{\citenamefont {Malerba}\ \emph {et~al.}(2010)\citenamefont
  {Malerba}, \citenamefont {Marinica}, \citenamefont {Anento}, \citenamefont
  {Bj{\"o}rkas}, \citenamefont {Nguyen}, \citenamefont {Domain}, \citenamefont
  {Djurabekova}, \citenamefont {Olsson}, \citenamefont {Nordlund},
  \citenamefont {Serra} \emph {et~al.}}]{malerba2010}%
  \BibitemOpen
  \bibfield  {author} {\bibinfo {author} {\bibfnamefont {L.}~\bibnamefont
  {Malerba}}, \bibinfo {author} {\bibfnamefont {M.-C.}\ \bibnamefont
  {Marinica}}, \bibinfo {author} {\bibfnamefont {N.}~\bibnamefont {Anento}},
  \bibinfo {author} {\bibfnamefont {C.}~\bibnamefont {Bj{\"o}rkas}}, \bibinfo
  {author} {\bibfnamefont {H.}~\bibnamefont {Nguyen}}, \bibinfo {author}
  {\bibfnamefont {C.}~\bibnamefont {Domain}}, \bibinfo {author} {\bibfnamefont
  {F.}~\bibnamefont {Djurabekova}}, \bibinfo {author} {\bibfnamefont
  {P.}~\bibnamefont {Olsson}}, \bibinfo {author} {\bibfnamefont
  {K.}~\bibnamefont {Nordlund}}, \bibinfo {author} {\bibfnamefont
  {A.}~\bibnamefont {Serra}},  \emph {et~al.},\ }\href@noop {} {\bibfield
  {journal} {\bibinfo  {journal} {Journal of nuclear materials}\ }\textbf
  {\bibinfo {volume} {406}},\ \bibinfo {pages} {19} (\bibinfo {year}
  {2010})}\BibitemShut {NoStop}%
\bibitem [{\citenamefont {Terentyev}\ \emph {et~al.}(2008)\citenamefont
  {Terentyev}, \citenamefont {Klaver}, \citenamefont {Olsson}, \citenamefont
  {Marinica}, \citenamefont {Willaime}, \citenamefont {Domain},\ and\
  \citenamefont {Malerba}}]{NPC-diSIA}%
  \BibitemOpen
  \bibfield  {author} {\bibinfo {author} {\bibfnamefont {D.~A.}\ \bibnamefont
  {Terentyev}}, \bibinfo {author} {\bibfnamefont {T.~P.~C.}\ \bibnamefont
  {Klaver}}, \bibinfo {author} {\bibfnamefont {P.}~\bibnamefont {Olsson}},
  \bibinfo {author} {\bibfnamefont {M.-C.}\ \bibnamefont {Marinica}}, \bibinfo
  {author} {\bibfnamefont {F.}~\bibnamefont {Willaime}}, \bibinfo {author}
  {\bibfnamefont {C.}~\bibnamefont {Domain}}, \ and\ \bibinfo {author}
  {\bibfnamefont {L.}~\bibnamefont {Malerba}},\ }\href {\doibase
  10.1103/PhysRevLett.100.145503} {\bibfield  {journal} {\bibinfo  {journal}
  {Phys. Rev. Lett.}\ }\textbf {\bibinfo {volume} {100}},\ \bibinfo {pages}
  {145503} (\bibinfo {year} {2008})}\BibitemShut {NoStop}%
\bibitem [{\citenamefont {Ventelon}\ and\ \citenamefont
  {Willaime}(2007)}]{ventelon2007core}%
  \BibitemOpen
  \bibfield  {author} {\bibinfo {author} {\bibfnamefont {L.}~\bibnamefont
  {Ventelon}}\ and\ \bibinfo {author} {\bibfnamefont {F.}~\bibnamefont
  {Willaime}},\ }\href@noop {} {\bibfield  {journal} {\bibinfo  {journal}
  {Journal of Computer-Aided Materials Design}\ }\textbf {\bibinfo {volume}
  {14}},\ \bibinfo {pages} {85} (\bibinfo {year} {2007})}\BibitemShut {NoStop}%
\bibitem [{\citenamefont {Dragoni}\ \emph {et~al.}(2016)\citenamefont
  {Dragoni}, \citenamefont {Ceresoli},\ and\ \citenamefont
  {Marzari}}]{dragoni2016vibrational}%
  \BibitemOpen
  \bibfield  {author} {\bibinfo {author} {\bibfnamefont {D.}~\bibnamefont
  {Dragoni}}, \bibinfo {author} {\bibfnamefont {D.}~\bibnamefont {Ceresoli}}, \
  and\ \bibinfo {author} {\bibfnamefont {N.}~\bibnamefont {Marzari}},\
  }\href@noop {} {\bibfield  {journal} {\bibinfo  {journal} {arXiv preprint
  arXiv:1605.03334}\ } (\bibinfo {year} {2016})}\BibitemShut {NoStop}%
\bibitem [{\citenamefont {Baskes}\ \emph {et~al.}(1989)\citenamefont {Baskes},
  \citenamefont {Nelson},\ and\ \citenamefont {Wright}}]{MEAM}%
  \BibitemOpen
  \bibfield  {author} {\bibinfo {author} {\bibfnamefont {M.~I.}\ \bibnamefont
  {Baskes}}, \bibinfo {author} {\bibfnamefont {J.~S.}\ \bibnamefont {Nelson}},
  \ and\ \bibinfo {author} {\bibfnamefont {A.~F.}\ \bibnamefont {Wright}},\
  }\href {\doibase 10.1103/PhysRevB.40.6085} {\bibfield  {journal} {\bibinfo
  {journal} {Phys. Rev. B}\ }\textbf {\bibinfo {volume} {40}},\ \bibinfo
  {pages} {6085} (\bibinfo {year} {1989})}\BibitemShut {NoStop}%
\bibitem [{\citenamefont {Mrovec}\ \emph {et~al.}(2011)\citenamefont {Mrovec},
  \citenamefont {Nguyen-Manh}, \citenamefont {Els\"asser},\ and\ \citenamefont
  {Gumbsch}}]{BOP_iron}%
  \BibitemOpen
  \bibfield  {author} {\bibinfo {author} {\bibfnamefont {M.}~\bibnamefont
  {Mrovec}}, \bibinfo {author} {\bibfnamefont {D.}~\bibnamefont {Nguyen-Manh}},
  \bibinfo {author} {\bibfnamefont {C.}~\bibnamefont {Els\"asser}}, \ and\
  \bibinfo {author} {\bibfnamefont {P.}~\bibnamefont {Gumbsch}},\ }\href
  {\doibase 10.1103/PhysRevLett.106.246402} {\bibfield  {journal} {\bibinfo
  {journal} {Phys. Rev. Lett.}\ }\textbf {\bibinfo {volume} {106}},\ \bibinfo
  {pages} {246402} (\bibinfo {year} {2011})}\BibitemShut {NoStop}%
\bibitem [{\citenamefont {Drautz}\ and\ \citenamefont
  {Pettifor}(2011)}]{BOP_magnetic}%
  \BibitemOpen
  \bibfield  {author} {\bibinfo {author} {\bibfnamefont {R.}~\bibnamefont
  {Drautz}}\ and\ \bibinfo {author} {\bibfnamefont {D.~G.}\ \bibnamefont
  {Pettifor}},\ }\href {\doibase 10.1103/PhysRevB.84.214114} {\bibfield
  {journal} {\bibinfo  {journal} {Phys. Rev. B}\ }\textbf {\bibinfo {volume}
  {84}},\ \bibinfo {pages} {214114} (\bibinfo {year} {2011})}\BibitemShut
  {NoStop}%
\bibitem [{\citenamefont {Ford}\ \emph {et~al.}(2014)\citenamefont {Ford},
  \citenamefont {Drautz}, \citenamefont {Hammerschmidt},\ and\ \citenamefont
  {Pettifor}}]{BOP_iron_drautz}%
  \BibitemOpen
  \bibfield  {author} {\bibinfo {author} {\bibfnamefont {M.~E.}\ \bibnamefont
  {Ford}}, \bibinfo {author} {\bibfnamefont {R.}~\bibnamefont {Drautz}},
  \bibinfo {author} {\bibfnamefont {T.}~\bibnamefont {Hammerschmidt}}, \ and\
  \bibinfo {author} {\bibfnamefont {D.~G.}\ \bibnamefont {Pettifor}},\ }\href
  {http://stacks.iop.org/0965-0393/22/i=3/a=034005} {\bibfield  {journal}
  {\bibinfo  {journal} {Modelling and Simulation in Materials Science and
  Engineering}\ }\textbf {\bibinfo {volume} {22}},\ \bibinfo {pages} {034005}
  (\bibinfo {year} {2014})}\BibitemShut {NoStop}%
\bibitem [{\citenamefont {Chiesa}\ \emph {et~al.}(2011)\citenamefont {Chiesa},
  \citenamefont {Derlet}, \citenamefont {Dudarev},\ and\ \citenamefont
  {Van~Swygenhoven}}]{magneticEAM}%
  \BibitemOpen
  \bibfield  {author} {\bibinfo {author} {\bibfnamefont {S.}~\bibnamefont
  {Chiesa}}, \bibinfo {author} {\bibfnamefont {P.}~\bibnamefont {Derlet}},
  \bibinfo {author} {\bibfnamefont {S.}~\bibnamefont {Dudarev}}, \ and\
  \bibinfo {author} {\bibfnamefont {H.}~\bibnamefont {Van~Swygenhoven}},\
  }\href@noop {} {\bibfield  {journal} {\bibinfo  {journal} {Journal of
  Physics: Condensed Matter}\ }\textbf {\bibinfo {volume} {23}},\ \bibinfo
  {pages} {206001} (\bibinfo {year} {2011})}\BibitemShut {NoStop}%
\bibitem [{\citenamefont {Hepburn}\ and\ \citenamefont
  {Ackland}(2008)}]{metallic-covalent-EAM}%
  \BibitemOpen
  \bibfield  {author} {\bibinfo {author} {\bibfnamefont {D.~J.}\ \bibnamefont
  {Hepburn}}\ and\ \bibinfo {author} {\bibfnamefont {G.~J.}\ \bibnamefont
  {Ackland}},\ }\href {\doibase 10.1103/PhysRevB.78.165115} {\bibfield
  {journal} {\bibinfo  {journal} {Phys. Rev. B}\ }\textbf {\bibinfo {volume}
  {78}},\ \bibinfo {pages} {165115} (\bibinfo {year} {2008})}\BibitemShut
  {NoStop}%
\bibitem [{\citenamefont {Bart\'ok}\ \emph {et~al.}(2010)\citenamefont
  {Bart\'ok}, \citenamefont {Payne}, \citenamefont {Kondor},\ and\
  \citenamefont {Cs\'anyi}}]{Bartok_GAP}%
  \BibitemOpen
  \bibfield  {author} {\bibinfo {author} {\bibfnamefont {A.~P.}\ \bibnamefont
  {Bart\'ok}}, \bibinfo {author} {\bibfnamefont {M.~C.}\ \bibnamefont {Payne}},
  \bibinfo {author} {\bibfnamefont {R.}~\bibnamefont {Kondor}}, \ and\ \bibinfo
  {author} {\bibfnamefont {G.}~\bibnamefont {Cs\'anyi}},\ }\href {\doibase
  10.1103/PhysRevLett.104.136403} {\bibfield  {journal} {\bibinfo  {journal}
  {Phys. Rev. Lett.}\ }\textbf {\bibinfo {volume} {104}},\ \bibinfo {pages}
  {136403} (\bibinfo {year} {2010})}\BibitemShut {NoStop}%
\bibitem [{\citenamefont {Behler}\ and\ \citenamefont
  {Parrinello}(2007)}]{BehlerParr}%
  \BibitemOpen
  \bibfield  {author} {\bibinfo {author} {\bibfnamefont {J.}~\bibnamefont
  {Behler}}\ and\ \bibinfo {author} {\bibfnamefont {M.}~\bibnamefont
  {Parrinello}},\ }\href {\doibase 10.1103/PhysRevLett.98.146401} {\bibfield
  {journal} {\bibinfo  {journal} {Phys. Rev. Lett.}\ }\textbf {\bibinfo
  {volume} {98}},\ \bibinfo {pages} {146401} (\bibinfo {year}
  {2007})}\BibitemShut {NoStop}%
\bibitem [{\citenamefont {Seko}\ \emph {et~al.}(2014)\citenamefont {Seko},
  \citenamefont {Takahashi},\ and\ \citenamefont {Tanaka}}]{seko2014sparse}%
  \BibitemOpen
  \bibfield  {author} {\bibinfo {author} {\bibfnamefont {A.}~\bibnamefont
  {Seko}}, \bibinfo {author} {\bibfnamefont {A.}~\bibnamefont {Takahashi}}, \
  and\ \bibinfo {author} {\bibfnamefont {I.}~\bibnamefont {Tanaka}},\
  }\href@noop {} {\bibfield  {journal} {\bibinfo  {journal} {Physical Review
  B}\ }\textbf {\bibinfo {volume} {90}},\ \bibinfo {pages} {024101} (\bibinfo
  {year} {2014})}\BibitemShut {NoStop}%
\bibitem [{\citenamefont {Artrith}\ and\ \citenamefont
  {Urban}(2016)}]{artrith2016}%
  \BibitemOpen
  \bibfield  {author} {\bibinfo {author} {\bibfnamefont {N.}~\bibnamefont
  {Artrith}}\ and\ \bibinfo {author} {\bibfnamefont {A.}~\bibnamefont
  {Urban}},\ }\href@noop {} {\bibfield  {journal} {\bibinfo  {journal}
  {Computational Materials Science}\ }\textbf {\bibinfo {volume} {114}},\
  \bibinfo {pages} {135} (\bibinfo {year} {2016})}\BibitemShut {NoStop}%
\bibitem [{\citenamefont {Botu}\ \emph {et~al.}(2016)\citenamefont {Botu},
  \citenamefont {Batra}, \citenamefont {Chapman},\ and\ \citenamefont
  {Ramprasad}}]{Ramprasad2016}%
  \BibitemOpen
  \bibfield  {author} {\bibinfo {author} {\bibfnamefont {V.}~\bibnamefont
  {Botu}}, \bibinfo {author} {\bibfnamefont {R.}~\bibnamefont {Batra}},
  \bibinfo {author} {\bibfnamefont {J.}~\bibnamefont {Chapman}}, \ and\
  \bibinfo {author} {\bibfnamefont {R.}~\bibnamefont {Ramprasad}},\ }\href@noop
  {} {\bibfield  {journal} {\bibinfo  {journal} {The Journal of Physical
  Chemistry C}\ } (\bibinfo {year} {2016})}\BibitemShut {NoStop}%
\bibitem [{\citenamefont {Miwa}\ and\ \citenamefont {Ohno}(2016)}]{Kazutoshi}%
  \BibitemOpen
  \bibfield  {author} {\bibinfo {author} {\bibfnamefont {K.}~\bibnamefont
  {Miwa}}\ and\ \bibinfo {author} {\bibfnamefont {H.}~\bibnamefont {Ohno}},\
  }\href {\doibase 10.1103/PhysRevB.94.184109} {\bibfield  {journal} {\bibinfo
  {journal} {Phys. Rev. B}\ }\textbf {\bibinfo {volume} {94}},\ \bibinfo
  {pages} {184109} (\bibinfo {year} {2016})}\BibitemShut {NoStop}%
\bibitem [{\citenamefont {Dolgirev}\ \emph {et~al.}(2016)\citenamefont
  {Dolgirev}, \citenamefont {Kruglov},\ and\ \citenamefont
  {Oganov}}]{Oganov2016}%
  \BibitemOpen
  \bibfield  {author} {\bibinfo {author} {\bibfnamefont {P.~E.}\ \bibnamefont
  {Dolgirev}}, \bibinfo {author} {\bibfnamefont {I.~A.}\ \bibnamefont
  {Kruglov}}, \ and\ \bibinfo {author} {\bibfnamefont {A.~R.}\ \bibnamefont
  {Oganov}},\ }\href@noop {} {\bibfield  {journal} {\bibinfo  {journal} {AIP
  Advances}\ }\textbf {\bibinfo {volume} {6}},\ \bibinfo {pages} {085318}
  (\bibinfo {year} {2016})}\BibitemShut {NoStop}%
\bibitem [{\citenamefont {Shapeev}(2016)}]{shapeev}%
  \BibitemOpen
  \bibfield  {author} {\bibinfo {author} {\bibfnamefont {A.~V.}\ \bibnamefont
  {Shapeev}},\ }\href@noop {} {\bibfield  {journal} {\bibinfo  {journal}
  {Multiscale Modeling \& Simulation}\ }\textbf {\bibinfo {volume} {14}},\
  \bibinfo {pages} {1153} (\bibinfo {year} {2016})}\BibitemShut {NoStop}%
\bibitem [{\citenamefont {Faraji}\ \emph {et~al.}(2017)\citenamefont {Faraji},
  \citenamefont {Ghasemi}, \citenamefont {Rostami}, \citenamefont
  {Rasoulkhani}, \citenamefont {Schaefer}, \citenamefont {Goedecker},\ and\
  \citenamefont {Amsler}}]{Goedecker}%
  \BibitemOpen
  \bibfield  {author} {\bibinfo {author} {\bibfnamefont {S.}~\bibnamefont
  {Faraji}}, \bibinfo {author} {\bibfnamefont {S.~A.}\ \bibnamefont {Ghasemi}},
  \bibinfo {author} {\bibfnamefont {S.}~\bibnamefont {Rostami}}, \bibinfo
  {author} {\bibfnamefont {R.}~\bibnamefont {Rasoulkhani}}, \bibinfo {author}
  {\bibfnamefont {B.}~\bibnamefont {Schaefer}}, \bibinfo {author}
  {\bibfnamefont {S.}~\bibnamefont {Goedecker}}, \ and\ \bibinfo {author}
  {\bibfnamefont {M.}~\bibnamefont {Amsler}},\ }\href@noop {} {\bibfield
  {journal} {\bibinfo  {journal} {Physical Review B}\ }\textbf {\bibinfo
  {volume} {95}},\ \bibinfo {pages} {104105} (\bibinfo {year}
  {2017})}\BibitemShut {NoStop}%
\bibitem [{\citenamefont {Glielmo}\ \emph {et~al.}(2017)\citenamefont
  {Glielmo}, \citenamefont {Sollich},\ and\ \citenamefont {De~Vita}}]{DeVita}%
  \BibitemOpen
  \bibfield  {author} {\bibinfo {author} {\bibfnamefont {A.}~\bibnamefont
  {Glielmo}}, \bibinfo {author} {\bibfnamefont {P.}~\bibnamefont {Sollich}}, \
  and\ \bibinfo {author} {\bibfnamefont {A.}~\bibnamefont {De~Vita}},\ }\href
  {\doibase 10.1103/PhysRevB.95.214302} {\bibfield  {journal} {\bibinfo
  {journal} {Phys. Rev. B}\ }\textbf {\bibinfo {volume} {95}},\ \bibinfo
  {pages} {214302} (\bibinfo {year} {2017})}\BibitemShut {NoStop}%
\bibitem [{\citenamefont {MacKay}(2005)}]{mackay}%
  \BibitemOpen
  \bibfield  {author} {\bibinfo {author} {\bibfnamefont {D.~J.~C.}\
  \bibnamefont {MacKay}},\ }\href@noop {} {\emph {\bibinfo {title} {Information
  Theory, Inference, and Learning Algorithms}}}\ (\bibinfo  {publisher}
  {Cambridge University Press},\ \bibinfo {year} {2005})\BibitemShut {NoStop}%
\bibitem [{\citenamefont {Rasmussen}\ and\ \citenamefont
  {Williams}(2006)}]{GPbook}%
  \BibitemOpen
  \bibfield  {author} {\bibinfo {author} {\bibfnamefont {C.~E.}\ \bibnamefont
  {Rasmussen}}\ and\ \bibinfo {author} {\bibfnamefont {C.~K.~I.}\ \bibnamefont
  {Williams}},\ }\href@noop {} {\emph {\bibinfo {title} {Gaussian Processes for
  Machine Learning}}}\ (\bibinfo  {publisher} {MIT Press},\ \bibinfo {year}
  {2006})\BibitemShut {NoStop}%
\bibitem [{\citenamefont {Bart{\'o}k}\ \emph {et~al.}(2013)\citenamefont
  {Bart{\'o}k}, \citenamefont {Kondor},\ and\ \citenamefont
  {Cs{\'a}nyi}}]{Bartok_envir}%
  \BibitemOpen
  \bibfield  {author} {\bibinfo {author} {\bibfnamefont {A.~P.}\ \bibnamefont
  {Bart{\'o}k}}, \bibinfo {author} {\bibfnamefont {R.}~\bibnamefont {Kondor}},
  \ and\ \bibinfo {author} {\bibfnamefont {G.}~\bibnamefont {Cs{\'a}nyi}},\
  }\href@noop {} {\bibfield  {journal} {\bibinfo  {journal} {Physical Review
  B}\ }\textbf {\bibinfo {volume} {87}},\ \bibinfo {pages} {184115} (\bibinfo
  {year} {2013})}\BibitemShut {NoStop}%
\bibitem [{\citenamefont {Szlachta}\ \emph {et~al.}(2014)\citenamefont
  {Szlachta}, \citenamefont {Bart\'ok},\ and\ \citenamefont
  {Cs\'anyi}}]{Wojciech}%
  \BibitemOpen
  \bibfield  {author} {\bibinfo {author} {\bibfnamefont {W.~J.}\ \bibnamefont
  {Szlachta}}, \bibinfo {author} {\bibfnamefont {A.~P.}\ \bibnamefont
  {Bart\'ok}}, \ and\ \bibinfo {author} {\bibfnamefont {G.}~\bibnamefont
  {Cs\'anyi}},\ }\href {\doibase 10.1103/PhysRevB.90.104108} {\bibfield
  {journal} {\bibinfo  {journal} {Phys. Rev. B}\ }\textbf {\bibinfo {volume}
  {90}},\ \bibinfo {pages} {104108} (\bibinfo {year} {2014})}\BibitemShut
  {NoStop}%
\bibitem [{\citenamefont {Deringer}\ and\ \citenamefont
  {Csanyi}(2017)}]{Deringer_aC}%
  \BibitemOpen
  \bibfield  {author} {\bibinfo {author} {\bibfnamefont {V.~L.}\ \bibnamefont
  {Deringer}}\ and\ \bibinfo {author} {\bibfnamefont {G.}~\bibnamefont
  {Csanyi}},\ }\href@noop {} {\bibfield  {journal} {\bibinfo  {journal}
  {Physical Review B}\ }\textbf {\bibinfo {volume} {95}},\ \bibinfo {pages}
  {094203} (\bibinfo {year} {2017})}\BibitemShut {NoStop}%
\bibitem [{\citenamefont {De}\ \emph {et~al.}(2016)\citenamefont {De},
  \citenamefont {Bartok}, \citenamefont {Csanyi},\ and\ \citenamefont
  {Ceriotti}}]{PCCP-SOAP}%
  \BibitemOpen
  \bibfield  {author} {\bibinfo {author} {\bibfnamefont {S.}~\bibnamefont
  {De}}, \bibinfo {author} {\bibfnamefont {A.~P.}\ \bibnamefont {Bartok}},
  \bibinfo {author} {\bibfnamefont {G.}~\bibnamefont {Csanyi}}, \ and\ \bibinfo
  {author} {\bibfnamefont {M.}~\bibnamefont {Ceriotti}},\ }\href@noop {}
  {\bibfield  {journal} {\bibinfo  {journal} {Phys. Chem. Chem. Phys.}\
  }\textbf {\bibinfo {volume} {18}},\ \bibinfo {pages} {13754} (\bibinfo {year}
  {2016})}\BibitemShut {NoStop}%
\bibitem [{\citenamefont {Bernstein}\ \emph {et~al.}(2017)\citenamefont
  {Bernstein}, \citenamefont {Kermode} \emph {et~al.}}]{Kermode}%
  \BibitemOpen
  \bibfield  {author} {\bibinfo {author} {\bibfnamefont {N.}~\bibnamefont
  {Bernstein}}, \bibinfo {author} {\bibfnamefont {J.}~\bibnamefont {Kermode}},
  \emph {et~al.},\ }\href@noop {} {\bibfield  {journal} {\bibinfo  {journal}
  {Bulletin of the American Physical Society}\ }\textbf {\bibinfo {volume}
  {62}} (\bibinfo {year} {2017})}\BibitemShut {NoStop}%
\bibitem [{\citenamefont {Bartok}\ and\ \citenamefont
  {Csanyi}(2015)}]{gap-tutorial}%
  \BibitemOpen
  \bibfield  {author} {\bibinfo {author} {\bibfnamefont {A.~P.}\ \bibnamefont
  {Bartok}}\ and\ \bibinfo {author} {\bibfnamefont {G.}~\bibnamefont
  {Csanyi}},\ }\href@noop {} {\bibfield  {journal} {\bibinfo  {journal} {Int.
  J. Quant. Chem.}\ }\textbf {\bibinfo {volume} {115}},\ \bibinfo {pages}
  {1051} (\bibinfo {year} {2015})}\BibitemShut {NoStop}%
\bibitem [{\citenamefont {Qui\~nonero Candela}\ and\ \citenamefont
  {Rasmussen}(2005)}]{sparseGP}%
  \BibitemOpen
  \bibfield  {author} {\bibinfo {author} {\bibfnamefont {J.}~\bibnamefont
  {Qui\~nonero Candela}}\ and\ \bibinfo {author} {\bibfnamefont {C.~E.}\
  \bibnamefont {Rasmussen}},\ }\href@noop {} {\bibfield  {journal} {\bibinfo
  {journal} {J. of Machine Learning Research}\ }\textbf {\bibinfo {volume}
  {6}},\ \bibinfo {pages} {1939} (\bibinfo {year} {2005})}\BibitemShut
  {NoStop}%
\bibitem [{\citenamefont {Mahoney}\ and\ \citenamefont {Drineas}(2009)}]{CUR}%
  \BibitemOpen
  \bibfield  {author} {\bibinfo {author} {\bibfnamefont {M.~W.}\ \bibnamefont
  {Mahoney}}\ and\ \bibinfo {author} {\bibfnamefont {P.}~\bibnamefont
  {Drineas}},\ }\href@noop {} {\bibfield  {journal} {\bibinfo  {journal} {Proc.
  Nat. Acad. Sci.}\ }\textbf {\bibinfo {volume} {106}},\ \bibinfo {pages} {697}
  (\bibinfo {year} {2009})}\BibitemShut {NoStop}%
\bibitem [{git()}]{github-quip}%
  \BibitemOpen
  \href@noop {} {}\bibinfo {howpublished}
  {\url{http://www.github.com/libAtoms/QUIP}}\BibitemShut {NoStop}%
\bibitem [{\citenamefont {Dragoni}(2016)}]{dragoni2016energetics}%
  \BibitemOpen
  \bibfield  {author} {\bibinfo {author} {\bibfnamefont {D.}~\bibnamefont
  {Dragoni}},\ }\emph {\bibinfo {title} {Energetics and thermodynamics of
  $\alpha$-iron from first-principles and machine-learning potentials}},\
  \href@noop {} {Ph.D. thesis},\ \bibinfo  {school} {{\'E}COLE POLYTECHNIQUE
  F{\'E}D{\'E}RALE DE LAUSANNE} (\bibinfo {year} {2016})\BibitemShut {NoStop}%
\bibitem [{Note1()}]{Note1}%
  \BibitemOpen
  \bibinfo {note} {The simple measure of localization that we use here is the
  length of the perimeter connecting the vacancy sites of the perfect
  bi-dimensional lattice associated to a crystallographic plane.}\BibitemShut
  {Stop}%
\bibitem [{\citenamefont {Berendsen}\ \emph {et~al.}(1984)\citenamefont
  {Berendsen}, \citenamefont {Postma}, \citenamefont {van Gunsteren},
  \citenamefont {DiNola},\ and\ \citenamefont {Haak}}]{berendsen}%
  \BibitemOpen
  \bibfield  {author} {\bibinfo {author} {\bibfnamefont {H.~J.}\ \bibnamefont
  {Berendsen}}, \bibinfo {author} {\bibfnamefont {J.~v.}\ \bibnamefont
  {Postma}}, \bibinfo {author} {\bibfnamefont {W.~F.}\ \bibnamefont {van
  Gunsteren}}, \bibinfo {author} {\bibfnamefont {A.}~\bibnamefont {DiNola}}, \
  and\ \bibinfo {author} {\bibfnamefont {J.}~\bibnamefont {Haak}},\ }\href@noop
  {} {\bibfield  {journal} {\bibinfo  {journal} {The Journal of chemical
  physics}\ }\textbf {\bibinfo {volume} {81}},\ \bibinfo {pages} {3684}
  (\bibinfo {year} {1984})}\BibitemShut {NoStop}%
\bibitem [{\citenamefont {Giannozzi}\ \emph {et~al.}(2009)\citenamefont
  {Giannozzi}, \citenamefont {Baroni}, \citenamefont {Bonini}, \citenamefont
  {Calandra}, \citenamefont {Car}, \citenamefont {Cavazzoni}, \citenamefont
  {Ceresoli}, \citenamefont {Chiarotti}, \citenamefont {Cococcioni},
  \citenamefont {Dabo}, \citenamefont {{Dal Corso}}, \citenamefont
  {de~Gironcoli}, \citenamefont {Fabris}, \citenamefont {Fratesi},
  \citenamefont {Gebauer}, \citenamefont {Gerstmann}, \citenamefont
  {Gougoussis}, \citenamefont {Kokalj}, \citenamefont {Lazzeri}, \citenamefont
  {Martin-Samos}, \citenamefont {Marzari}, \citenamefont {Mauri}, \citenamefont
  {Mazzarello}, \citenamefont {Paolini}, \citenamefont {Pasquarello},
  \citenamefont {Paulatto}, \citenamefont {Sbraccia}, \citenamefont {Scandolo},
  \citenamefont {Sclauzero}, \citenamefont {Seitsonen}, \citenamefont
  {Smogunov}, \citenamefont {Umari},\ and\ \citenamefont {Wentzcovitch}}]{QE}%
  \BibitemOpen
  \bibfield  {author} {\bibinfo {author} {\bibfnamefont {P.}~\bibnamefont
  {Giannozzi}}, \bibinfo {author} {\bibfnamefont {S.}~\bibnamefont {Baroni}},
  \bibinfo {author} {\bibfnamefont {N.}~\bibnamefont {Bonini}}, \bibinfo
  {author} {\bibfnamefont {M.}~\bibnamefont {Calandra}}, \bibinfo {author}
  {\bibfnamefont {R.}~\bibnamefont {Car}}, \bibinfo {author} {\bibfnamefont
  {C.}~\bibnamefont {Cavazzoni}}, \bibinfo {author} {\bibfnamefont
  {D.}~\bibnamefont {Ceresoli}}, \bibinfo {author} {\bibfnamefont {G.~L.}\
  \bibnamefont {Chiarotti}}, \bibinfo {author} {\bibfnamefont {M.}~\bibnamefont
  {Cococcioni}}, \bibinfo {author} {\bibfnamefont {I.}~\bibnamefont {Dabo}},
  \bibinfo {author} {\bibfnamefont {A.}~\bibnamefont {{Dal Corso}}}, \bibinfo
  {author} {\bibfnamefont {S.}~\bibnamefont {de~Gironcoli}}, \bibinfo {author}
  {\bibfnamefont {S.}~\bibnamefont {Fabris}}, \bibinfo {author} {\bibfnamefont
  {G.}~\bibnamefont {Fratesi}}, \bibinfo {author} {\bibfnamefont
  {R.}~\bibnamefont {Gebauer}}, \bibinfo {author} {\bibfnamefont
  {U.}~\bibnamefont {Gerstmann}}, \bibinfo {author} {\bibfnamefont
  {C.}~\bibnamefont {Gougoussis}}, \bibinfo {author} {\bibfnamefont
  {A.}~\bibnamefont {Kokalj}}, \bibinfo {author} {\bibfnamefont
  {M.}~\bibnamefont {Lazzeri}}, \bibinfo {author} {\bibfnamefont
  {L.}~\bibnamefont {Martin-Samos}}, \bibinfo {author} {\bibfnamefont
  {N.}~\bibnamefont {Marzari}}, \bibinfo {author} {\bibfnamefont
  {F.}~\bibnamefont {Mauri}}, \bibinfo {author} {\bibfnamefont
  {R.}~\bibnamefont {Mazzarello}}, \bibinfo {author} {\bibfnamefont
  {S.}~\bibnamefont {Paolini}}, \bibinfo {author} {\bibfnamefont
  {A.}~\bibnamefont {Pasquarello}}, \bibinfo {author} {\bibfnamefont
  {L.}~\bibnamefont {Paulatto}}, \bibinfo {author} {\bibfnamefont
  {C.}~\bibnamefont {Sbraccia}}, \bibinfo {author} {\bibfnamefont
  {S.}~\bibnamefont {Scandolo}}, \bibinfo {author} {\bibfnamefont
  {G.}~\bibnamefont {Sclauzero}}, \bibinfo {author} {\bibfnamefont {A.~P.}\
  \bibnamefont {Seitsonen}}, \bibinfo {author} {\bibfnamefont {A.}~\bibnamefont
  {Smogunov}}, \bibinfo {author} {\bibfnamefont {P.}~\bibnamefont {Umari}}, \
  and\ \bibinfo {author} {\bibfnamefont {R.~M.}\ \bibnamefont {Wentzcovitch}},\
  }\href {http://www.quantum-espresso.org} {\bibfield  {journal} {\bibinfo
  {journal} {Journal of Physics: Condensed Matter}\ }\textbf {\bibinfo {volume}
  {21}},\ \bibinfo {pages} {395502 (19pp)} (\bibinfo {year}
  {2009})}\BibitemShut {NoStop}%
\bibitem [{\citenamefont {Perdew}\ \emph {et~al.}(1996)\citenamefont {Perdew},
  \citenamefont {Burke},\ and\ \citenamefont {Ernzerhof}}]{PBE}%
  \BibitemOpen
  \bibfield  {author} {\bibinfo {author} {\bibfnamefont {J.~P.}\ \bibnamefont
  {Perdew}}, \bibinfo {author} {\bibfnamefont {K.}~\bibnamefont {Burke}}, \
  and\ \bibinfo {author} {\bibfnamefont {M.}~\bibnamefont {Ernzerhof}},\ }\href
  {\doibase 10.1103/PhysRevLett.77.3865} {\bibfield  {journal} {\bibinfo
  {journal} {Phys. Rev. Lett.}\ }\textbf {\bibinfo {volume} {77}},\ \bibinfo
  {pages} {3865} (\bibinfo {year} {1996})}\BibitemShut {NoStop}%
\bibitem [{Note2()}]{Note2}%
  \BibitemOpen
  \bibinfo {note} {\protect \url
  {http://www.qe-forge.org/gf/project/pslibrary}}\BibitemShut {NoStop}%
\bibitem [{\citenamefont {Derlet}\ \emph {et~al.}(2007)\citenamefont {Derlet},
  \citenamefont {Nguyen-Manh},\ and\ \citenamefont {Dudarev}}]{Dudarev_Derlet}%
  \BibitemOpen
  \bibfield  {author} {\bibinfo {author} {\bibfnamefont {P.~M.}\ \bibnamefont
  {Derlet}}, \bibinfo {author} {\bibfnamefont {D.}~\bibnamefont {Nguyen-Manh}},
  \ and\ \bibinfo {author} {\bibfnamefont {S.}~\bibnamefont {Dudarev}},\
  }\href@noop {} {\bibfield  {journal} {\bibinfo  {journal} {Physical Review
  B}\ }\textbf {\bibinfo {volume} {76}},\ \bibinfo {pages} {054107} (\bibinfo
  {year} {2007})}\BibitemShut {NoStop}%
\bibitem [{Note3()}]{Note3}%
  \BibitemOpen
  \bibinfo {note} {Ratio between cutoff on the density and cutoff on the
  wavefunction for the ultrasoft pseudopotential}\BibitemShut {NoStop}%
\bibitem [{\citenamefont {Marzari}\ \emph {et~al.}(1999)\citenamefont
  {Marzari}, \citenamefont {Vanderbilt}, \citenamefont {De~Vita},\ and\
  \citenamefont {Payne}}]{marzarismearing}%
  \BibitemOpen
  \bibfield  {author} {\bibinfo {author} {\bibfnamefont {N.}~\bibnamefont
  {Marzari}}, \bibinfo {author} {\bibfnamefont {D.}~\bibnamefont {Vanderbilt}},
  \bibinfo {author} {\bibfnamefont {A.}~\bibnamefont {De~Vita}}, \ and\
  \bibinfo {author} {\bibfnamefont {M.~C.}\ \bibnamefont {Payne}},\ }\href
  {\doibase 10.1103/PhysRevLett.82.3296} {\bibfield  {journal} {\bibinfo
  {journal} {Phys. Rev. Lett.}\ }\textbf {\bibinfo {volume} {82}},\ \bibinfo
  {pages} {3296} (\bibinfo {year} {1999})}\BibitemShut {NoStop}%
\bibitem [{\citenamefont {Pizzi}\ \emph {et~al.}(2016)\citenamefont {Pizzi},
  \citenamefont {Cepellotti}, \citenamefont {Sabatini}, \citenamefont
  {Marzari},\ and\ \citenamefont {Kozinsky}}]{Aiida}%
  \BibitemOpen
  \bibfield  {author} {\bibinfo {author} {\bibfnamefont {G.}~\bibnamefont
  {Pizzi}}, \bibinfo {author} {\bibfnamefont {A.}~\bibnamefont {Cepellotti}},
  \bibinfo {author} {\bibfnamefont {R.}~\bibnamefont {Sabatini}}, \bibinfo
  {author} {\bibfnamefont {N.}~\bibnamefont {Marzari}}, \ and\ \bibinfo
  {author} {\bibfnamefont {B.}~\bibnamefont {Kozinsky}},\ }\href@noop {}
  {\bibfield  {journal} {\bibinfo  {journal} {Computational Materials Science}\
  }\textbf {\bibinfo {volume} {111}},\ \bibinfo {pages} {218} (\bibinfo {year}
  {2016})}\BibitemShut {NoStop}%
\bibitem [{\citenamefont {Beeler~Jr}\ and\ \citenamefont
  {Johnson}(1967)}]{beelerVAC}%
  \BibitemOpen
  \bibfield  {author} {\bibinfo {author} {\bibfnamefont {J.}~\bibnamefont
  {Beeler~Jr}}\ and\ \bibinfo {author} {\bibfnamefont {R.}~\bibnamefont
  {Johnson}},\ }\href@noop {} {\bibfield  {journal} {\bibinfo  {journal}
  {Physical Review}\ }\textbf {\bibinfo {volume} {156}},\ \bibinfo {pages}
  {677} (\bibinfo {year} {1967})}\BibitemShut {NoStop}%
\bibitem [{Note4()}]{Note4}%
  \BibitemOpen
  \bibinfo {note} {A recent work based on a DFT+Gutzwiller approach (see
  Ref.~\protect \rev@citealpnum {ironGutzwiller}) has shown interesting
  improvements over standard DFT with respect to the agreement between
  experiments and theory for the description of the mechanical properties of
  $\alpha $-iron. One might therefore try to use such approach to generate a
  database for GAP models with predictions closer to experiments. At the
  moment, however, its computational cost remains an important
  limitation.}\BibitemShut {Stop}%
\bibitem [{Note5()}]{Note5}%
  \BibitemOpen
  \bibinfo {note} {This value is well within the DFT uncertainty and highlights
  the capabilities of our training procedure.}\BibitemShut {Stop}%
\bibitem [{\citenamefont {Lejaeghere}\ \emph {et~al.}(2014)\citenamefont
  {Lejaeghere}, \citenamefont {Van~Speybroeck}, \citenamefont {Van~Oost},\ and\
  \citenamefont {Cottenier}}]{Cottenier}%
  \BibitemOpen
  \bibfield  {author} {\bibinfo {author} {\bibfnamefont {K.}~\bibnamefont
  {Lejaeghere}}, \bibinfo {author} {\bibfnamefont {V.}~\bibnamefont
  {Van~Speybroeck}}, \bibinfo {author} {\bibfnamefont {G.}~\bibnamefont
  {Van~Oost}}, \ and\ \bibinfo {author} {\bibfnamefont {S.}~\bibnamefont
  {Cottenier}},\ }\href {\doibase 10.1080/10408436.2013.772503} {\bibfield
  {journal} {\bibinfo  {journal} {Critical Reviews in Solid State and Materials
  Sciences}\ }\textbf {\bibinfo {volume} {39}},\ \bibinfo {pages} {1} (\bibinfo
  {year} {2014})},\ \Eprint
  {http://arxiv.org/abs/http://dx.doi.org/10.1080/10408436.2013.772503}
  {http://dx.doi.org/10.1080/10408436.2013.772503} \BibitemShut {NoStop}%
\bibitem [{\citenamefont {Basinski}\ \emph {et~al.}(1955)\citenamefont
  {Basinski}, \citenamefont {Hume-Rothery},\ and\ \citenamefont
  {Sutton}}]{Basinski}%
  \BibitemOpen
  \bibfield  {author} {\bibinfo {author} {\bibfnamefont {Z.~S.}\ \bibnamefont
  {Basinski}}, \bibinfo {author} {\bibfnamefont {W.}~\bibnamefont
  {Hume-Rothery}}, \ and\ \bibinfo {author} {\bibfnamefont {A.~L.}\
  \bibnamefont {Sutton}},\ }\href {\doibase 10.1098/rspa.1955.0102} {\bibfield
  {journal} {\bibinfo  {journal} {Proceedings of the Royal Society of London.
  Series A. Mathematical and Physical Sciences}\ }\textbf {\bibinfo {volume}
  {229}},\ \bibinfo {pages} {459} (\bibinfo {year} {1955})}\BibitemShut
  {NoStop}%
\bibitem [{\citenamefont {Adams}\ \emph {et~al.}(2006)\citenamefont {Adams},
  \citenamefont {Agosta}, \citenamefont {Leisure},\ and\ \citenamefont
  {Ledbetter}}]{Adams}%
  \BibitemOpen
  \bibfield  {author} {\bibinfo {author} {\bibfnamefont {J.~J.}\ \bibnamefont
  {Adams}}, \bibinfo {author} {\bibfnamefont {D.~S.}\ \bibnamefont {Agosta}},
  \bibinfo {author} {\bibfnamefont {R.}~\bibnamefont {Leisure}}, \ and\
  \bibinfo {author} {\bibfnamefont {H.}~\bibnamefont {Ledbetter}},\ }\href
  {\doibase 10.1063/1.2365714} {\bibfield  {journal} {\bibinfo  {journal}
  {Journal of Applied Physics}\ }\textbf {\bibinfo {volume} {100}},\ \bibinfo
  {pages} {113530} (\bibinfo {year} {2006})}\BibitemShut {NoStop}%
\bibitem [{\citenamefont {Reith}\ \emph {et~al.}(2014)\citenamefont {Reith},
  \citenamefont {Podloucky}, \citenamefont {Marsman}, \citenamefont
  {Bedolla-Velazquez},\ and\ \citenamefont {Mohn}}]{Reith2014}%
  \BibitemOpen
  \bibfield  {author} {\bibinfo {author} {\bibfnamefont {D.}~\bibnamefont
  {Reith}}, \bibinfo {author} {\bibfnamefont {R.}~\bibnamefont {Podloucky}},
  \bibinfo {author} {\bibfnamefont {M.}~\bibnamefont {Marsman}}, \bibinfo
  {author} {\bibfnamefont {P.~O.}\ \bibnamefont {Bedolla-Velazquez}}, \ and\
  \bibinfo {author} {\bibfnamefont {P.}~\bibnamefont {Mohn}},\ }\href {\doibase
  10.1103/PhysRevB.90.014432} {\bibfield  {journal} {\bibinfo  {journal}
  {Physical Review B}\ }\textbf {\bibinfo {volume} {90}},\ \bibinfo {pages}
  {014432} (\bibinfo {year} {2014})},\ \Eprint {http://arxiv.org/abs/1312.4313}
  {arXiv:1312.4313} \BibitemShut {NoStop}%
\bibitem [{\citenamefont {Sch{\"{o}}necker}(2011)}]{Schonecker2011}%
  \BibitemOpen
  \bibfield  {author} {\bibinfo {author} {\bibfnamefont {S.}~\bibnamefont
  {Sch{\"{o}}necker}},\ }\emph {\bibinfo {title} {{Theoretical Studies of
  Epitaxial Bain Paths of Metals}}},\ \href
  {http://nbn-resolving.de/urn:nbn:de:bsz:14-qucosa-77263} {Ph.D. thesis},\
  \bibinfo  {school} {Technische Universit{\"{a}}t Dresden} (\bibinfo {year}
  {2011})\BibitemShut {NoStop}%
\bibitem [{\citenamefont {Fri{\'{a}}k}\ \emph {et~al.}(2001)\citenamefont
  {Fri{\'{a}}k}, \citenamefont {{\v{S}}ob},\ and\ \citenamefont
  {Vitek}}]{Friak2001}%
  \BibitemOpen
  \bibfield  {author} {\bibinfo {author} {\bibfnamefont {M.}~\bibnamefont
  {Fri{\'{a}}k}}, \bibinfo {author} {\bibfnamefont {M.}~\bibnamefont
  {{\v{S}}ob}}, \ and\ \bibinfo {author} {\bibfnamefont {V.}~\bibnamefont
  {Vitek}},\ }\href {\doibase 10.1103/PhysRevB.63.052405} {\bibfield  {journal}
  {\bibinfo  {journal} {Physical Review B}\ }\textbf {\bibinfo {volume} {63}},\
  \bibinfo {pages} {052405} (\bibinfo {year} {2001})}\BibitemShut {NoStop}%
\bibitem [{\citenamefont {Hoover}(1985)}]{Nose-Hoover}%
  \BibitemOpen
  \bibfield  {author} {\bibinfo {author} {\bibfnamefont {W.~G.}\ \bibnamefont
  {Hoover}},\ }\href {\doibase 10.1103/PhysRevA.31.1695} {\bibfield  {journal}
  {\bibinfo  {journal} {Phys. Rev. A}\ }\textbf {\bibinfo {volume} {31}},\
  \bibinfo {pages} {1695} (\bibinfo {year} {1985})}\BibitemShut {NoStop}%
\bibitem [{\citenamefont {Parrinello}\ and\ \citenamefont
  {Rahman}(1981)}]{Parrinello-Rahman}%
  \BibitemOpen
  \bibfield  {author} {\bibinfo {author} {\bibfnamefont {M.}~\bibnamefont
  {Parrinello}}\ and\ \bibinfo {author} {\bibfnamefont {A.}~\bibnamefont
  {Rahman}},\ }\href {\doibase http://dx.doi.org/10.1063/1.328693} {\bibfield
  {journal} {\bibinfo  {journal} {Journal of Applied Physics}\ }\textbf
  {\bibinfo {volume} {52}},\ \bibinfo {pages} {7182} (\bibinfo {year}
  {1981})}\BibitemShut {NoStop}%
\bibitem [{\citenamefont {Plimpton}(1995)}]{lammps}%
  \BibitemOpen
  \bibfield  {author} {\bibinfo {author} {\bibfnamefont {S.}~\bibnamefont
  {Plimpton}},\ }\href
  {http://citeseerx.ist.psu.edu/viewdoc/download?doi=10.1.1.35.6047&rep=rep1&type=pdf}
  {\bibfield  {journal} {\bibinfo  {journal} {Journal of Computational
  Physics}\ }\textbf {\bibinfo {volume} {117}},\ \bibinfo {pages} {1–19}
  (\bibinfo {year} {1995})}\BibitemShut {NoStop}%
\bibitem [{Note6()}]{Note6}%
  \BibitemOpen
  \bibinfo {note} {Note however that in our MD NPT runs we have allowed only
  for isotropic fluctuations of the starting cubic box. New runs where the
  initial box is allowed to fluctuate freely should be performed to confirm
  this observation.}\BibitemShut {Stop}%
\bibitem [{\citenamefont {K\"ormann}\ \emph {et~al.}(2010)\citenamefont
  {K\"ormann}, \citenamefont {Dick}, \citenamefont {Hickel},\ and\
  \citenamefont {Neugebauer}}]{koerman_specificheat}%
  \BibitemOpen
  \bibfield  {author} {\bibinfo {author} {\bibfnamefont {F.}~\bibnamefont
  {K\"ormann}}, \bibinfo {author} {\bibfnamefont {A.}~\bibnamefont {Dick}},
  \bibinfo {author} {\bibfnamefont {T.}~\bibnamefont {Hickel}}, \ and\ \bibinfo
  {author} {\bibfnamefont {J.}~\bibnamefont {Neugebauer}},\ }\href {\doibase
  10.1103/PhysRevB.81.134425} {\bibfield  {journal} {\bibinfo  {journal} {Phys.
  Rev. B}\ }\textbf {\bibinfo {volume} {81}},\ \bibinfo {pages} {134425}
  (\bibinfo {year} {2010})}\BibitemShut {NoStop}%
\bibitem [{\citenamefont {Lavrentiev}\ \emph {et~al.}(2010)\citenamefont
  {Lavrentiev}, \citenamefont {Nguyen-Manh},\ and\ \citenamefont
  {Dudarev}}]{dudarev_specificheat}%
  \BibitemOpen
  \bibfield  {author} {\bibinfo {author} {\bibfnamefont {M.~Y.}\ \bibnamefont
  {Lavrentiev}}, \bibinfo {author} {\bibfnamefont {D.}~\bibnamefont
  {Nguyen-Manh}}, \ and\ \bibinfo {author} {\bibfnamefont {S.~L.}\ \bibnamefont
  {Dudarev}},\ }\href {\doibase 10.1103/PhysRevB.81.184202} {\bibfield
  {journal} {\bibinfo  {journal} {Phys. Rev. B}\ }\textbf {\bibinfo {volume}
  {81}},\ \bibinfo {pages} {184202} (\bibinfo {year} {2010})}\BibitemShut
  {NoStop}%
\bibitem [{\citenamefont {Ridley}\ and\ \citenamefont {Stuart}(1968)}]{Ridley}%
  \BibitemOpen
  \bibfield  {author} {\bibinfo {author} {\bibfnamefont {N.}~\bibnamefont
  {Ridley}}\ and\ \bibinfo {author} {\bibfnamefont {H.}~\bibnamefont
  {Stuart}},\ }\href {\doibase 10.1088/0022-3727/1/10/308} {\bibfield
  {journal} {\bibinfo  {journal} {J. Phys. D: Appl. Phys.}\ }\textbf {\bibinfo
  {volume} {1}},\ \bibinfo {pages} {1291} (\bibinfo {year} {1968})}\BibitemShut
  {NoStop}%
\bibitem [{\citenamefont {Seki}\ and\ \citenamefont {Nagata}(2005)}]{Seki}%
  \BibitemOpen
  \bibfield  {author} {\bibinfo {author} {\bibfnamefont {I.}~\bibnamefont
  {Seki}}\ and\ \bibinfo {author} {\bibfnamefont {K.}~\bibnamefont {Nagata}},\
  }\href@noop {} {\bibfield  {journal} {\bibinfo  {journal} {ISIJ
  international}\ }\textbf {\bibinfo {volume} {45}},\ \bibinfo {pages} {1789}
  (\bibinfo {year} {2005})}\BibitemShut {NoStop}%
\bibitem [{\citenamefont {Desai}(1986)}]{Desai}%
  \BibitemOpen
  \bibfield  {author} {\bibinfo {author} {\bibfnamefont {P.~D.}\ \bibnamefont
  {Desai}},\ }\href {\doibase http://dx.doi.org/10.1063/1.555761} {\bibfield
  {journal} {\bibinfo  {journal} {Journal of Physical and Chemical Reference
  Data}\ }\textbf {\bibinfo {volume} {15}},\ \bibinfo {pages} {967} (\bibinfo
  {year} {1986})}\BibitemShut {NoStop}%
\bibitem [{\citenamefont {Wallace}\ \emph {et~al.}(1960)\citenamefont
  {Wallace}, \citenamefont {Sidles},\ and\ \citenamefont
  {Danielson}}]{Wallace2}%
  \BibitemOpen
  \bibfield  {author} {\bibinfo {author} {\bibfnamefont {D.~C.}\ \bibnamefont
  {Wallace}}, \bibinfo {author} {\bibfnamefont {P.~H.}\ \bibnamefont {Sidles}},
  \ and\ \bibinfo {author} {\bibfnamefont {G.~C.}\ \bibnamefont {Danielson}},\
  }\href {\doibase http://dx.doi.org/10.1063/1.1735393} {\bibfield  {journal}
  {\bibinfo  {journal} {Journal of Applied Physics}\ }\textbf {\bibinfo
  {volume} {31}},\ \bibinfo {pages} {168} (\bibinfo {year} {1960})}\BibitemShut
  {NoStop}%
\bibitem [{\citenamefont {Dever}(1972)}]{Dever}%
  \BibitemOpen
  \bibfield  {author} {\bibinfo {author} {\bibfnamefont {D.~J.}\ \bibnamefont
  {Dever}},\ }\href {http://dx.doi.org/10.1063/1.1661710} {\bibfield  {journal}
  {\bibinfo  {journal} {J. Appl. Phys.}\ }\textbf {\bibinfo {volume} {43}},\
  \bibinfo {pages} {3293} (\bibinfo {year} {1972})}\BibitemShut {NoStop}%
\bibitem [{Note7()}]{Note7}%
  \BibitemOpen
  \bibinfo {note} {The calculation of this quantity is performed at constant
  volume with relaxed atomic positions in a cubic cell containing 53/1999
  atoms.}\BibitemShut {Stop}%
\bibitem [{\citenamefont {Henkelman}\ and\ \citenamefont
  {J{\'o}nsson}(2000)}]{NEB}%
  \BibitemOpen
  \bibfield  {author} {\bibinfo {author} {\bibfnamefont {G.}~\bibnamefont
  {Henkelman}}\ and\ \bibinfo {author} {\bibfnamefont {H.}~\bibnamefont
  {J{\'o}nsson}},\ }\href@noop {} {\bibfield  {journal} {\bibinfo  {journal}
  {The Journal of chemical physics}\ }\textbf {\bibinfo {volume} {113}},\
  \bibinfo {pages} {9978} (\bibinfo {year} {2000})}\BibitemShut {NoStop}%
\bibitem [{\citenamefont {Djurabekova}\ \emph {et~al.}(2010)\citenamefont
  {Djurabekova}, \citenamefont {Malerba}, \citenamefont {Pasianot},
  \citenamefont {Olsson},\ and\ \citenamefont
  {Nordlund}}]{djurabekova2010kinetics}%
  \BibitemOpen
  \bibfield  {author} {\bibinfo {author} {\bibfnamefont {F.}~\bibnamefont
  {Djurabekova}}, \bibinfo {author} {\bibfnamefont {L.}~\bibnamefont
  {Malerba}}, \bibinfo {author} {\bibfnamefont {R.~C.}\ \bibnamefont
  {Pasianot}}, \bibinfo {author} {\bibfnamefont {P.}~\bibnamefont {Olsson}}, \
  and\ \bibinfo {author} {\bibfnamefont {K.}~\bibnamefont {Nordlund}},\
  }\href@noop {} {\bibfield  {journal} {\bibinfo  {journal} {Philosophical
  Magazine}\ }\textbf {\bibinfo {volume} {90}},\ \bibinfo {pages} {2585}
  (\bibinfo {year} {2010})}\BibitemShut {NoStop}%
\bibitem [{\citenamefont {Hayward}\ and\ \citenamefont
  {Fu}(2013)}]{hayward2013interplay}%
  \BibitemOpen
  \bibfield  {author} {\bibinfo {author} {\bibfnamefont {E.}~\bibnamefont
  {Hayward}}\ and\ \bibinfo {author} {\bibfnamefont {C.-C.}\ \bibnamefont
  {Fu}},\ }\href@noop {} {\bibfield  {journal} {\bibinfo  {journal} {Physical
  Review B}\ }\textbf {\bibinfo {volume} {87}},\ \bibinfo {pages} {174103}
  (\bibinfo {year} {2013})}\BibitemShut {NoStop}%
\bibitem [{\citenamefont {Fu}\ \emph {et~al.}(2005)\citenamefont {Fu},
  \citenamefont {Dalla~Torre}, \citenamefont {Willaime}, \citenamefont
  {Bocquet},\ and\ \citenamefont {Barbu}}]{nature_chuchun}%
  \BibitemOpen
  \bibfield  {author} {\bibinfo {author} {\bibfnamefont {C.-C.}\ \bibnamefont
  {Fu}}, \bibinfo {author} {\bibfnamefont {J.}~\bibnamefont {Dalla~Torre}},
  \bibinfo {author} {\bibfnamefont {F.}~\bibnamefont {Willaime}}, \bibinfo
  {author} {\bibfnamefont {J.-L.}\ \bibnamefont {Bocquet}}, \ and\ \bibinfo
  {author} {\bibfnamefont {A.}~\bibnamefont {Barbu}},\ }\href@noop {}
  {\bibfield  {journal} {\bibinfo  {journal} {Nature materials}\ }\textbf
  {\bibinfo {volume} {4}},\ \bibinfo {pages} {68} (\bibinfo {year}
  {2005})}\BibitemShut {NoStop}%
\bibitem [{\citenamefont {Becquart}\ and\ \citenamefont
  {Domain}(2003)}]{divacancy_DFT1}%
  \BibitemOpen
  \bibfield  {author} {\bibinfo {author} {\bibfnamefont {C.}~\bibnamefont
  {Becquart}}\ and\ \bibinfo {author} {\bibfnamefont {C.}~\bibnamefont
  {Domain}},\ }\href@noop {} {\bibfield  {journal} {\bibinfo  {journal}
  {Nuclear Instruments and Methods in Physics Research Section B: Beam
  Interactions with Materials and Atoms}\ }\textbf {\bibinfo {volume} {202}},\
  \bibinfo {pages} {44} (\bibinfo {year} {2003})}\BibitemShut {NoStop}%
\bibitem [{Note8()}]{Note8}%
  \BibitemOpen
  \bibinfo {note} {Performed at the constant equilibrium volume with atomic
  relaxation in a $3\times 3\times 3$ cubic cell}\BibitemShut {NoStop}%
\bibitem [{\citenamefont {Olsson}\ \emph {et~al.}(2007)\citenamefont {Olsson},
  \citenamefont {Domain},\ and\ \citenamefont {Wallenius}}]{Olsson}%
  \BibitemOpen
  \bibfield  {author} {\bibinfo {author} {\bibfnamefont {P.}~\bibnamefont
  {Olsson}}, \bibinfo {author} {\bibfnamefont {C.}~\bibnamefont {Domain}}, \
  and\ \bibinfo {author} {\bibfnamefont {J.}~\bibnamefont {Wallenius}},\
  }\href@noop {} {\bibfield  {journal} {\bibinfo  {journal} {Physical Review
  B}\ }\textbf {\bibinfo {volume} {75}},\ \bibinfo {pages} {014110} (\bibinfo
  {year} {2007})}\BibitemShut {NoStop}%
\bibitem [{\citenamefont {Fu}\ \emph {et~al.}(2004)\citenamefont {Fu},
  \citenamefont {Willaime},\ and\ \citenamefont {Ordej\'on}}]{Fu_SIA}%
  \BibitemOpen
  \bibfield  {author} {\bibinfo {author} {\bibfnamefont {C.-C.}\ \bibnamefont
  {Fu}}, \bibinfo {author} {\bibfnamefont {F.}~\bibnamefont {Willaime}}, \ and\
  \bibinfo {author} {\bibfnamefont {P.}~\bibnamefont {Ordej\'on}},\ }\href
  {\doibase 10.1103/PhysRevLett.92.175503} {\bibfield  {journal} {\bibinfo
  {journal} {Phys. Rev. Lett.}\ }\textbf {\bibinfo {volume} {92}},\ \bibinfo
  {pages} {175503} (\bibinfo {year} {2004})}\BibitemShut {NoStop}%
\bibitem [{\citenamefont {Spencer}\ \emph {et~al.}(2002)\citenamefont
  {Spencer}, \citenamefont {Hung}, \citenamefont {Snook},\ and\ \citenamefont
  {Yarovsky}}]{spencer2002}%
  \BibitemOpen
  \bibfield  {author} {\bibinfo {author} {\bibfnamefont {M.~J.}\ \bibnamefont
  {Spencer}}, \bibinfo {author} {\bibfnamefont {A.}~\bibnamefont {Hung}},
  \bibinfo {author} {\bibfnamefont {I.~K.}\ \bibnamefont {Snook}}, \ and\
  \bibinfo {author} {\bibfnamefont {I.}~\bibnamefont {Yarovsky}},\ }\href@noop
  {} {\bibfield  {journal} {\bibinfo  {journal} {Surface Science}\ }\textbf
  {\bibinfo {volume} {513}},\ \bibinfo {pages} {389} (\bibinfo {year}
  {2002})}\BibitemShut {NoStop}%
\bibitem [{\citenamefont {B{\l}o{\'n}ski}\ and\ \citenamefont
  {Kiejna}(2007)}]{blonski2007}%
  \BibitemOpen
  \bibfield  {author} {\bibinfo {author} {\bibfnamefont {P.}~\bibnamefont
  {B{\l}o{\'n}ski}}\ and\ \bibinfo {author} {\bibfnamefont {A.}~\bibnamefont
  {Kiejna}},\ }\href@noop {} {\bibfield  {journal} {\bibinfo  {journal}
  {Surface science}\ }\textbf {\bibinfo {volume} {601}},\ \bibinfo {pages}
  {123} (\bibinfo {year} {2007})}\BibitemShut {NoStop}%
\bibitem [{\citenamefont {Vitek}(1968)}]{vitek}%
  \BibitemOpen
  \bibfield  {author} {\bibinfo {author} {\bibfnamefont {V.}~\bibnamefont
  {Vitek}},\ }\href@noop {} {\bibfield  {journal} {\bibinfo  {journal}
  {Philosophical Magazine}\ }\textbf {\bibinfo {volume} {18}},\ \bibinfo
  {pages} {773} (\bibinfo {year} {1968})}\BibitemShut {NoStop}%
\bibitem [{\citenamefont {Itakura}\ \emph {et~al.}(2012)\citenamefont
  {Itakura}, \citenamefont {Kaburaki},\ and\ \citenamefont
  {Yamaguchi}}]{itakura2012first}%
  \BibitemOpen
  \bibfield  {author} {\bibinfo {author} {\bibfnamefont {M.}~\bibnamefont
  {Itakura}}, \bibinfo {author} {\bibfnamefont {H.}~\bibnamefont {Kaburaki}}, \
  and\ \bibinfo {author} {\bibfnamefont {M.}~\bibnamefont {Yamaguchi}},\
  }\href@noop {} {\bibfield  {journal} {\bibinfo  {journal} {Acta Materialia}\
  }\textbf {\bibinfo {volume} {60}},\ \bibinfo {pages} {3698} (\bibinfo {year}
  {2012})}\BibitemShut {NoStop}%
\bibitem [{\citenamefont {Vitek†}(2004)}]{vitek2004core}%
  \BibitemOpen
  \bibfield  {author} {\bibinfo {author} {\bibfnamefont {V.}~\bibnamefont
  {Vitek†}},\ }\href@noop {} {\bibfield  {journal} {\bibinfo  {journal}
  {Philosophical Magazine}\ }\textbf {\bibinfo {volume} {84}},\ \bibinfo
  {pages} {415} (\bibinfo {year} {2004})}\BibitemShut {NoStop}%
\bibitem [{\citenamefont {Frederiksen}\ and\ \citenamefont
  {Jacobsen}(2003)}]{frederiksen2003density}%
  \BibitemOpen
  \bibfield  {author} {\bibinfo {author} {\bibfnamefont {S.~L.}\ \bibnamefont
  {Frederiksen}}\ and\ \bibinfo {author} {\bibfnamefont {K.~W.}\ \bibnamefont
  {Jacobsen}},\ }\href@noop {} {\bibfield  {journal} {\bibinfo  {journal}
  {Philosophical magazine}\ }\textbf {\bibinfo {volume} {83}},\ \bibinfo
  {pages} {365} (\bibinfo {year} {2003})}\BibitemShut {NoStop}%
\bibitem [{\citenamefont {Domain}\ and\ \citenamefont
  {Monnet}(2005)}]{domain2005simulation}%
  \BibitemOpen
  \bibfield  {author} {\bibinfo {author} {\bibfnamefont {C.}~\bibnamefont
  {Domain}}\ and\ \bibinfo {author} {\bibfnamefont {G.}~\bibnamefont
  {Monnet}},\ }\href@noop {} {\bibfield  {journal} {\bibinfo  {journal}
  {Physical review letters}\ }\textbf {\bibinfo {volume} {95}},\ \bibinfo
  {pages} {215506} (\bibinfo {year} {2005})}\BibitemShut {NoStop}%
\bibitem [{\citenamefont {Bahn}\ and\ \citenamefont
  {Jacobsen}(2002)}]{bahn2002object}%
  \BibitemOpen
  \bibfield  {author} {\bibinfo {author} {\bibfnamefont {S.~R.}\ \bibnamefont
  {Bahn}}\ and\ \bibinfo {author} {\bibfnamefont {K.~W.}\ \bibnamefont
  {Jacobsen}},\ }\href@noop {} {\bibfield  {journal} {\bibinfo  {journal}
  {Computing in Science \& Engineering}\ }\textbf {\bibinfo {volume} {4}},\
  \bibinfo {pages} {56} (\bibinfo {year} {2002})}\BibitemShut {NoStop}%
\bibitem [{\citenamefont {Kolsbjerg}\ \emph {et~al.}(2016)\citenamefont
  {Kolsbjerg}, \citenamefont {Groves},\ and\ \citenamefont
  {Hammer}}]{kolsbjerg2016automated}%
  \BibitemOpen
  \bibfield  {author} {\bibinfo {author} {\bibfnamefont {E.~L.}\ \bibnamefont
  {Kolsbjerg}}, \bibinfo {author} {\bibfnamefont {M.~N.}\ \bibnamefont
  {Groves}}, \ and\ \bibinfo {author} {\bibfnamefont {B.}~\bibnamefont
  {Hammer}},\ }\href@noop {} {\bibfield  {journal} {\bibinfo  {journal} {The
  Journal of Chemical Physics}\ }\textbf {\bibinfo {volume} {145}},\ \bibinfo
  {pages} {094107} (\bibinfo {year} {2016})}\BibitemShut {NoStop}%
\bibitem [{\citenamefont {Ventelon}\ \emph {et~al.}(2013)\citenamefont
  {Ventelon}, \citenamefont {Willaime}, \citenamefont {Clouet},\ and\
  \citenamefont {Rodney}}]{Ventelon2013}%
  \BibitemOpen
  \bibfield  {author} {\bibinfo {author} {\bibfnamefont {L.}~\bibnamefont
  {Ventelon}}, \bibinfo {author} {\bibfnamefont {F.}~\bibnamefont {Willaime}},
  \bibinfo {author} {\bibfnamefont {E.}~\bibnamefont {Clouet}}, \ and\ \bibinfo
  {author} {\bibfnamefont {D.}~\bibnamefont {Rodney}},\ }\href {\doibase
  10.1016/j.actamat.2013.03.012} {\bibfield  {journal} {\bibinfo  {journal}
  {Acta Materialia}\ }\textbf {\bibinfo {volume} {61}},\ \bibinfo {pages}
  {3973} (\bibinfo {year} {2013})}\BibitemShut {NoStop}%
\bibitem [{\citenamefont {Dezerald}\ \emph {et~al.}(2014)\citenamefont
  {Dezerald}, \citenamefont {Ventelon}, \citenamefont {Clouet}, \citenamefont
  {Denoual}, \citenamefont {Rodney},\ and\ \citenamefont
  {Willaime}}]{Dezerald2014AbMetals}%
  \BibitemOpen
  \bibfield  {author} {\bibinfo {author} {\bibfnamefont {L.}~\bibnamefont
  {Dezerald}}, \bibinfo {author} {\bibfnamefont {L.}~\bibnamefont {Ventelon}},
  \bibinfo {author} {\bibfnamefont {E.}~\bibnamefont {Clouet}}, \bibinfo
  {author} {\bibfnamefont {C.}~\bibnamefont {Denoual}}, \bibinfo {author}
  {\bibfnamefont {D.}~\bibnamefont {Rodney}}, \ and\ \bibinfo {author}
  {\bibfnamefont {F.}~\bibnamefont {Willaime}},\ }\href {\doibase
  10.1103/PhysRevB.89.024104} {\bibfield  {journal} {\bibinfo  {journal} {Phys.
  Rev. B}\ }\textbf {\bibinfo {volume} {89}},\ \bibinfo {pages} {024104}
  (\bibinfo {year} {2014})}\BibitemShut {NoStop}%
\bibitem [{\citenamefont {Tateyama}\ and\ \citenamefont
  {Ohno}(2003)}]{tateyama2003HFe}%
  \BibitemOpen
  \bibfield  {author} {\bibinfo {author} {\bibfnamefont {Y.}~\bibnamefont
  {Tateyama}}\ and\ \bibinfo {author} {\bibfnamefont {T.}~\bibnamefont
  {Ohno}},\ }\href@noop {} {\bibfield  {journal} {\bibinfo  {journal} {Phys.
  Rev. B}\ }\textbf {\bibinfo {volume} {67}},\ \bibinfo {pages} {174105}
  (\bibinfo {year} {2003})}\BibitemShut {NoStop}%
\bibitem [{\citenamefont {Matter}\ \emph {et~al.}(1979)\citenamefont {Matter},
  \citenamefont {Winter},\ and\ \citenamefont
  {Triftsh{\"a}user}}]{monovac_posannih1}%
  \BibitemOpen
  \bibfield  {author} {\bibinfo {author} {\bibfnamefont {H.}~\bibnamefont
  {Matter}}, \bibinfo {author} {\bibfnamefont {J.}~\bibnamefont {Winter}}, \
  and\ \bibinfo {author} {\bibfnamefont {W.}~\bibnamefont {Triftsh{\"a}user}},\
  }\href@noop {} {\bibfield  {journal} {\bibinfo  {journal} {Applied physics}\
  }\textbf {\bibinfo {volume} {20}},\ \bibinfo {pages} {135} (\bibinfo {year}
  {1979})}\BibitemShut {NoStop}%
\bibitem [{\citenamefont {De~Schepper}\ \emph {et~al.}(1983)\citenamefont
  {De~Schepper}, \citenamefont {Segers}, \citenamefont {Dorikens-Vanpraet},
  \citenamefont {Dorikens}, \citenamefont {Knuyt}, \citenamefont {Stals},\ and\
  \citenamefont {Moser}}]{de1983positron}%
  \BibitemOpen
  \bibfield  {author} {\bibinfo {author} {\bibfnamefont {L.}~\bibnamefont
  {De~Schepper}}, \bibinfo {author} {\bibfnamefont {D.}~\bibnamefont {Segers}},
  \bibinfo {author} {\bibfnamefont {L.}~\bibnamefont {Dorikens-Vanpraet}},
  \bibinfo {author} {\bibfnamefont {M.}~\bibnamefont {Dorikens}}, \bibinfo
  {author} {\bibfnamefont {G.}~\bibnamefont {Knuyt}}, \bibinfo {author}
  {\bibfnamefont {L.}~\bibnamefont {Stals}}, \ and\ \bibinfo {author}
  {\bibfnamefont {P.}~\bibnamefont {Moser}},\ }\href@noop {} {\bibfield
  {journal} {\bibinfo  {journal} {Phys. Rev. B}\ }\textbf {\bibinfo {volume}
  {27}},\ \bibinfo {pages} {5257} (\bibinfo {year} {1983})}\BibitemShut
  {NoStop}%
\bibitem [{\citenamefont {Christian}(2002)}]{christian2002theory}%
  \BibitemOpen
  \bibfield  {author} {\bibinfo {author} {\bibfnamefont {J.}~\bibnamefont
  {Christian}},\ }\href {https://books.google.ch/books?id=kiWjakQeUSAC} {\emph
  {\bibinfo {title} {The Theory of Transformations in Metals and Alloys}}},\
  The Theory of Transformations in Metals and Alloys\ (\bibinfo  {publisher}
  {Elsevier Science},\ \bibinfo {year} {2002})\BibitemShut {NoStop}%
\bibitem [{\citenamefont {Dudarev}(2013)}]{dudarev2013density}%
  \BibitemOpen
  \bibfield  {author} {\bibinfo {author} {\bibfnamefont {S.}~\bibnamefont
  {Dudarev}},\ }\href@noop {} {\bibfield  {journal} {\bibinfo  {journal}
  {Materials Research}\ }\textbf {\bibinfo {volume} {43}},\ \bibinfo {pages}
  {35} (\bibinfo {year} {2013})}\BibitemShut {NoStop}%
\bibitem [{\citenamefont {B{\l}o{\'n}ski}\ and\ \citenamefont
  {Kiejna}(2004)}]{blonski2004}%
  \BibitemOpen
  \bibfield  {author} {\bibinfo {author} {\bibfnamefont {P.}~\bibnamefont
  {B{\l}o{\'n}ski}}\ and\ \bibinfo {author} {\bibfnamefont {A.}~\bibnamefont
  {Kiejna}},\ }\href@noop {} {\bibfield  {journal} {\bibinfo  {journal}
  {Vacuum}\ }\textbf {\bibinfo {volume} {74}},\ \bibinfo {pages} {179}
  (\bibinfo {year} {2004})}\BibitemShut {NoStop}%
\bibitem [{\citenamefont {Alling}\ \emph {et~al.}(2016)\citenamefont {Alling},
  \citenamefont {K\"ormann}, \citenamefont {Grabowski}, \citenamefont {Glensk},
  \citenamefont {Abrikosov},\ and\ \citenamefont {Neugebauer}}]{neuge2016}%
  \BibitemOpen
  \bibfield  {author} {\bibinfo {author} {\bibfnamefont {B.}~\bibnamefont
  {Alling}}, \bibinfo {author} {\bibfnamefont {F.}~\bibnamefont {K\"ormann}},
  \bibinfo {author} {\bibfnamefont {B.}~\bibnamefont {Grabowski}}, \bibinfo
  {author} {\bibfnamefont {A.}~\bibnamefont {Glensk}}, \bibinfo {author}
  {\bibfnamefont {I.~A.}\ \bibnamefont {Abrikosov}}, \ and\ \bibinfo {author}
  {\bibfnamefont {J.}~\bibnamefont {Neugebauer}},\ }\href {\doibase
  10.1103/PhysRevB.93.224411} {\bibfield  {journal} {\bibinfo  {journal} {Phys.
  Rev. B}\ }\textbf {\bibinfo {volume} {93}},\ \bibinfo {pages} {224411}
  (\bibinfo {year} {2016})}\BibitemShut {NoStop}%
\bibitem [{mca()}]{mca}%
  \BibitemOpen
  \href@noop {} {}\bibinfo {howpublished}
  {\url{http://www.materialscloud.org/archive/2017.0006/v1/}}\BibitemShut
  {NoStop}%
\bibitem [{\citenamefont {Schickling}\ \emph {et~al.}(2016)\citenamefont
  {Schickling}, \citenamefont {B\"unemann}, \citenamefont {Gebhard},\ and\
  \citenamefont {Boeri}}]{ironGutzwiller}%
  \BibitemOpen
  \bibfield  {author} {\bibinfo {author} {\bibfnamefont {T.}~\bibnamefont
  {Schickling}}, \bibinfo {author} {\bibfnamefont {J.}~\bibnamefont
  {B\"unemann}}, \bibinfo {author} {\bibfnamefont {F.}~\bibnamefont {Gebhard}},
  \ and\ \bibinfo {author} {\bibfnamefont {L.}~\bibnamefont {Boeri}},\ }\href
  {\doibase 10.1103/PhysRevB.93.205151} {\bibfield  {journal} {\bibinfo
  {journal} {Phys. Rev. B}\ }\textbf {\bibinfo {volume} {93}},\ \bibinfo
  {pages} {205151} (\bibinfo {year} {2016})}\BibitemShut {NoStop}%
\end{thebibliography}%

\clearpage
\end{document}